\documentclass[journal, onecolumn]{IEEEtran}

\usepackage[T1]{fontenc}
\usepackage{amsmath,amssymb,amsfonts,amsthm,mathtools}
\usepackage{bm}
\usepackage{booktabs}
\usepackage{cite}
\usepackage{enumitem}
\usepackage{graphicx}
\usepackage{microtype}
\usepackage[caption=false,font=footnotesize]{subfig}
\usepackage[hidelinks,final]{hyperref}
\usepackage[capitalize,noabbrev]{cleveref}
\usepackage{xcolor}

\newtheorem{theorem}{Theorem}
\newtheorem{lemma}{Lemma}
\newtheorem{proposition}{Proposition}
\newtheorem{corollary}{Corollary}
\theoremstyle{definition}
\newtheorem{definition}{Definition}
\newtheorem{remark}{Remark}

\newcommand{\cX}{\mathcal{X}}
\newcommand{\cY}{\mathcal{Y}}
\newcommand{\cM}{\mathcal{M}}
\newcommand{\cS}{\mathcal{S}}
\newcommand{\cC}{\mathcal{C}}
\newcommand{\cN}{\mathcal{N}}
\newcommand{\cA}{\mathcal{A}}
\newcommand{\cE}{\mathcal{E}}
\newcommand{\ind}{\mathbf{1}}
\newcommand{\EE}{\mathbb{E}}
\newcommand{\PP}{\mathbb{P}}
\newcommand{\eps}{\varepsilon}
\newcommand{\hb}{h_2}
\newcommand{\other}[1]{[J]\setminus\{#1\}}
\newcommand{\ins}[3]{{#2}\,\sqcup_{#1}\,{#3}}
\newcommand{\fallfac}[2]{(#1)_{#2}}
\DeclareMathOperator*{\argmax}{arg\,max}

\allowdisplaybreaks

\begin{document}

\title{{Symmetry Breaking over Multiway Channels}}

\author{Raymond Zhang, Richard Combes, and Sheng Yang%
	\thanks{R. Zhang is with INRIA, France. R. Combes and S. Yang are with CentraleSupélec, Paris-Saclay University, France. This paper is supported by the CHIST-ERA Wireless AI 2022 call MLDR project (ANR-23-CHR4-0005), partially funded by AEI and NCN under projects PCI2023-145958-2 and 2023/05/Y/ST7/00004, respectively. This paper was presented in part at the 2025 IEEE Information Theory Workshop (ITW), Sydney, Australia, September--October 2025, and at the 2026 IEEE International Symposium on Information Theory (ISIT), Guangzhou, China, June--July 2026.
	}
}

\maketitle

\begin{abstract}
Identical users running the same algorithm must often break their symmetry and agree on distinct roles, as in leader election, naming, and resource allocation. We study anonymous symmetry breaking over a discrete memoryless multiway channel invariant under permutations of its $J$ users: each user observes only a private output and must choose a number in $\{1,\dots,J\}$, and the symmetry is broken if the $J$ numbers are distinct. Using a R\'enyi change of measure, we derive a finite-blocklength converse on the error probability of every symmetric causal protocol, including adaptive and randomized ones. For achievability, block coding with random names reduces the problem to all-source multicast, for which we derive a Gallager random-coding exponent for every fixed $J$, and a cut-set bound that limits the exponent of the probability that two names coincide or are decoded wrongly. Interactive role controllers instead let the users contend for identifiers directly: we give their exact error probability, an achievable exponent given by a spectral radius when some outputs certify a role, and a floor for finite-state controllers when every output has positive probability, and we introduce generalized musical chairs. On a noisy bandit channel motivated by multiplayer multi-armed bandits, with noiseless collisions and arms of a common mean below one, interactive schemes provably exceed the limit of name decoding, and numerical estimates suggest the same with noisy collisions. For two users, noiseless collisions, one informative arm and a silent arm, the optimal error probability is determined up to a factor of two.
\end{abstract}


\section{Introduction}
\label{sec:introduction}

Anonymous agents must often agree on distinct roles before any higher-level distributed task can begin. This symmetry-breaking step appears as leader election, naming, rendezvous, and channel or resource allocation. The agents may be radio devices or identical software programs (for instance, copies of the same AI model started from the same prompt) that must accomplish a task jointly without a coordinator. These problems tend to be harder when every agent starts from the same prior information, executes the same algorithm, and observes only a private noisy response to the joint actions. Shared randomness does not by itself break the symmetry, since all users see the same random value. Independent private randomness creates different tentative identifiers, but two users may still draw the same one, and they have to find out through the channel whether or not this has happened.

Our main motivation is the decentralized multiplayer multi-armed bandit problem. In this model, $J$ players repeatedly choose among $K$ arms and observe private rewards. The players label the arms consistently, and they act in synchronized time slots, so that they share a common clock but no common identity.
Simultaneous use of the same arm causes a collision and suppresses the corresponding rewards. Algorithms with strong regret guarantees use collisions to assign roles and exchange statistics \cite{boursier_sic-mmab_2019,shi_decentralized_2020,wang_optimal_2020,huang22a}; see \cite{boursier_survey_2024} for a recent survey. State-of-the-art algorithms typically start with an initialization phase in which the players obtain distinct ranks, and the simplest rule to accomplish this is called musical chairs \cite{rosenski_multi-player_nodate}: each player selects a random arm until they receive a nonzero reward, and then keeps selecting that arm, since without noise a nonzero reward proves that nobody else pulled it. Random ranks are also used to break the symmetry in \cite{Anandkumar2011}, and initialization without collision information is studied in \cite{lugosi2022multiplayer}. The resulting channel model was formalized as the \emph{bandit channel} in \cite{zhang2025}. Unlike a conventional multiple-access channel, each player is simultaneously a transmitter and a receiver, its own action is known as receiver side information, and interference is useful because it carries information about the actions of the other players.

Two architectures come to mind for symmetry breaking over such a channel. In the first, each user draws a long random name and transmits the codeword of a common codebook indexed by its name; every user then decodes the names of the others, and ranking the names gives distinct identifiers unless two names coincide---an event of exponentially small probability in the number of bits used to encode the name---or decoding fails. This turns symmetry breaking into all-source multicast over a multiway channel. The two-user case is naturally related to Shannon's two-way channel \cite{shannon_two-way_1961}, for which a complete single-letter capacity characterization is unavailable in general \cite{han1984general,dueck1979capacity}. With more than two users, each receiver must recover every other user's message. A tractable inner bound is nevertheless obtained by independent random coding and simultaneous decoding: at a fixed receiver, the other $J-1$ codewords form a multiple-access channel with the receiver's own codeword as side information.
In \cite{zhang2025}, two players communicate over the bandit channel with a common codebook and inputs that do not depend on the outputs: the random-coding exponent and the achievable rate are derived, and Shannon's outer bound for the two-way channel is evaluated as an upper bound on the capacity. In \cite{zhang2026}, block coding with random names breaks the symmetry between two players on that channel, with codebooks that time-share between input laws; its error exponent is derived under maximum-likelihood and universal decoding, and the regret of the resulting initialization phase of a bandit algorithm is bounded, with a minimax lower bound. The two-user, noiseless case of our block-coding results with a single input law is therefore in \cite{zhang2025,zhang2026} (see \Cref{subsec:bandit-block}), and so is the value of our name-decoding ceiling for two users and arms with equal means (\Cref{prop:bandit-name-cap}). The change-of-measure converses, block coding and the cut-set bound for $J\geq3$ users, the ceiling on the name-decoding exponent, the interactive role controllers and noisy collisions are not in \cite{zhang2025,zhang2026}, and universal decoding and regret are not treated here.
In the second architecture, the users contend directly for the identifiers in $[J]:=\{1,\ldots,J\}$. A user keeps its tentative identifier as long as its outputs suggest that nobody else holds it, and draws a new one otherwise, as in musical chairs. No name is decoded: the channel itself resolves the conflicts.

In this paper, we consider the problem of determining how fast the probability of error of symmetry breaking can decay with the number of channel uses, for every strategy. For those two architectures we quantify their error exponents, and we show that, on the noisy bandit channel below, the second architecture can exceed every name-decoding exponent of the first, that is, every exponent of the probability that two names coincide or are decoded wrongly: provably with noiseless collisions, and according to numerical estimates with noisy ones.

We consider a memoryless channel whose law is unchanged when the users are permuted: permuting the inputs of the users permutes their outputs in the same way, and the outputs may be arbitrarily correlated given the inputs. All users apply the same strategy: the input of a user may depend on its own past outputs, the users may use common and private randomness, and each user outputs a number in $[J]$.
The symmetry is broken if the $J$ numbers are distinct. \Cref{sec:model} gives the formal definition.

We then specialize to a \emph{noisy bandit channel}, in which arm $x$ yields a Bernoulli reward with probability $\mu_x$ when it is pulled by a single user and with probability $\eps_x$ when it is shared. The case $\eps_x=0$ for every arm is the bandit channel of \cite{zhang2025}, where a reward certifies that the user was alone on its arm. Allowing $\eps_x>0$ is not meant as a detailed model of collisions, although a collision does not need to suppress every reward, for instance with capture in wireless systems or imperfect collision detection, and nonzero rewards on collisions are also considered in multiplayer bandits \cite{magesh2022decentralized}. It is the simplest departure from certified outputs: the outputs stay binary and conditionally independent, the channel stays symmetric, and only a collision can change the law of an output---but a reward no longer proves that the user was alone. For $\eps_x=0$, a single observation can certify a role, musical chairs and its generalization apply, and, for two users, the converse can be tight. For $0<\mu_x<1$ and $0<\eps_x<1$, every output has positive probability, no observation is conclusive, the users have to accumulate evidence over many uses, and a gap remains between the converses and the schemes. When $\eps_x=\mu_x$ for every $x$, collisions are invisible and symmetry breaking is impossible.

\subsection{Related Work}
\label{subsec:related-symmetry-breaking}

The objective of assigning distinct names to initially indistinguishable processes is classical in distributed computing \cite{itai1990symmetry,lipton1990processor}. Under the names \emph{naming} and \emph{initialization}, randomized wireless networking protocols assign identifiers to anonymous stations in single- and multichannel collision models \cite{hayashi1999initialization,nakano2000initialization}. More recent work treats anonymous beeping networks, where a listening node learns only whether at least one neighbor beeped \cite{chlebus2017naming,aldawsari2020multichannel}, also with noisy beeps \cite{ashkenazi2022noisy}. That literature primarily optimizes rounds, energy, random bits, and success probability for particular distributed models. Indistinguishability arguments give lower bounds for contention resolution over collision channels \cite{willard1986,newport2014radio}, and optimal protocols are known for two parties \cite{wang2021contention}; the collision channel without feedback \cite{massey1985collision} is an information-theoretic model of uncoordinated access in which the users have distinct protocol sequences. Our symmetry-breaking criterion is the same underlying task, but the object of study is its reliability as a function of the number of channel uses $n$ over a general noisy multiway channel.

The closely related leader-election task asks each user to output a bit, with success when exactly one user outputs $1$ and becomes the leader while other users become followers; stronger formulations additionally require identifying or communicating with the elected leader \cite{ghaffari2016leader}. A symmetry-breaking protocol also solves the task of leader election if the user whose identifier is $1$ elects itself as the leader by outputting $1$,
so that an election failure is contained in a symmetry-breaking failure.
For $J=2$ the tasks are exactly equivalent, since an election bit $L_j$ is mapped back to the identifier $2-L_j$ without additional observations. For $J>2$, symmetry breaking requires all $J$ identifiers to be different, whereas election may leave the followers indistinguishable. Neither criterion additionally requires a certificate or common knowledge of success.

Both tasks are special cases of a population version of symmetry breaking. Given $L$ states and a population vector $\bm N=(N_1,\ldots,N_L)$ with $N_1+\cdots+N_L=J$, the users succeed when exactly $N_l$ of them choose state $l$, for every $l\in[L]$. Our task is $\bm N=(1,\ldots,1)$, and leader election is $\bm N=(1,J-1)$. Our task is the finest of these: if every user maps its identifier through a fixed map from $[J]$ to $[L]$ that sends exactly $N_l$ identifiers to state $l$, for instance one that splits the ranks into consecutive blocks of sizes $N_1,\ldots,N_L$, a protocol for $\bm N=(1,\ldots,1)$ becomes a protocol for $\bm N$ with no larger error probability. Every achievability result of this paper therefore holds for every population vector, and so do the converses of \Cref{sec:meta-converse}, for every population vector with at least two positive entries (\Cref{prop:population-converse}). Apart from this, the rest of the paper treats $\bm N=(1,\ldots,1)$ only.

The differences between symmetry breaking and two related information-theoretic problems are worth mentioning. Coordination capacity includes task assignment without duplication, for which the target law is uniform over the permutations of the tasks \cite{cuff2010coordination}. Its terminals, however, have fixed labels and use terminal-dependent mappings over prescribed communication links. Unsourced random access instead gives statistically identical transmitters a common codebook and asks one central receiver to recover an unordered message list \cite{polyanskiy2017perspective}. Here there is no central observer: every anonymous user is simultaneously a transmitter and a receiver, sees a different private output, and must infer its own place in a common ordering. The distinction is already visible before communication. If $C$ is common randomness and $\Pi_C$ is a random permutation, labeled terminals may set their values to $\Pi_C(j)$, which is collision-free, whereas identical anonymous maps of $C$ produce the same value for every user. Private randomness can create different tentative names, but the channel must let the users determine their relative order consistently.

Our converse is a change-of-measure argument in the spirit of the meta-converse \cite{polyanskiy2010channel} and of its R\'enyi form \cite{polyanskiy2010arimoto}; with noninteracting references, its exponent has the form of Haroutunian's sphere-packing bound for channels with feedback \cite{haroutunian1977}. Change-of-measure lower bounds that hold for adaptive strategies are also standard in bandit problems \cite{kaufmann2016complexity,garivier2019explore}.

Finally, when there are more arms than players, the goal of the players should be to cycle through the arms in a fair fashion while avoiding collisions. This task is sometimes called orthogonalization \cite{boursier_sic-mmab_2019} and can be reduced to symmetry breaking as well.

To summarize, the symmetry-breaking objective is ubiquitous in distributed systems; our emphasis is its information-theoretic reliability. To the best of our knowledge, the present work is the first Shannon-theoretic treatment of anonymous symmetry breaking over a general noisy symmetric multiway channel, with private receiver observations, identical causal strategies, general converses, reliability exponents, and both coding and interactive achievability.

\subsection{Contributions}

The main contributions are summarized as follows.

\begin{enumerate}

  \item \emph{Formulation.} We cast anonymous symmetry breaking as a Shannon-theoretic problem for $J$ users over a general symmetric memoryless channel, with the error probability of the best strategy and its reliability exponent as the quantities of interest (\Cref{sec:model}).

  \item \emph{A converse for every strategy.} We prove finite-blocklength converses on the probability of error for every symmetric causal strategy, including adaptive ones and those using arbitrary common and private randomness (\Cref{sec:meta-converse}). We compare the channel with a reference under which the identifiers of $m$ users are conditionally i.i.d.\ given a latent variable, so that they collide with probability at least that of i.i.d.\ uniform draws. A R\'enyi change of measure transfers this probability to the true channel, at a cost that is the $n$-th power of a one-step quantity of a nonnegative kernel defining the reference. The bound is optimized over the R\'enyi order and the kernel. Two instances are explicit: a noninteracting reference, whose exponent reduces to a KL-radius formula, and a common-output reference, which keeps only the trajectories on which the selected users receive identical outputs and is effective when collisions are detected without noise. The converses extend to every population vector with at least two positive entries. In the examples we evaluate kernels with $m=2$ selected users only, and the exponents of the resulting bounds do not depend on $J$ beyond the distinction between $J=2$ and $J\geq3$.

  \item \emph{Block coding and the limit of name decoding.} A shared random codebook and independent private names reduce achievability to all-source multicast, and the error probability is at most the probability that two names coincide, plus the multicast decoding error.
    For arbitrary fixed $J$, we prove a Gallager random-coding exponent for the error probability of symmetry breaking,
    which for i.i.d.\ inputs is positive if and only if $I(X_{-j};Y_j\mid X_j)>0$, and is at most $\frac1J I(X_{-j};Y_j\mid X_j)$. A multiway cut-set bound shows that no name-decoding scheme, block or interactive, has a name-decoding exponent above $\log|\cY|/(J-1)$. Symmetry breaking itself is not subject to this limit: for $J=2$, a channel whose output pair is uniform on $\{(0,1),(1,0)\}$ for any input breaks the symmetry in a single use, although it carries no information~(\Cref{sec:block-coding}).

  \item \emph{Interactive schemes.} We introduce role controllers (\Cref{sec:interactive-coding}), in which each user runs a probabilistic automaton, in the spirit of the finite-memory automata studied for hypothesis testing and the two-armed bandit \cite{cover1969,hellman1970,cover1970}: it holds a tentative identifier and a state updated from its own outputs. Their error probability is the exact probability of a failure state of a Markov chain on the joint states, and every symmetric causal strategy, block coding in particular, is a role controller. When some output certifies that a role is not shared, a finite-state controller achieves an error exponent of at least $-\log$ of a spectral radius, with equality under a condition that holds in the cases we evaluate. When every output has positive probability, no time-invariant finite-state controller has a vanishing error probability, as time-invariant finite memory keeps the error of hypothesis testing away from zero \cite{hellman1970}. On the noisy bandit channel with $J$ arms of common parameters, a state set of bounded size with maps that vary with time does give a vanishing error probability, by a device of Cover \cite{cover1969}, but whether it can give a positive exponent is open. We analyze confidence controllers, which accumulate evidence in a counter whose range grows with $n$. We also introduce \emph{generalized musical chairs}, a finite-state scheme in which a user avoids the roles on which it recently failed. It contains musical chairs as the special case of no exclusion.

  \item \emph{The noisy bandit channel.} We define a noisy bandit channel with one pair of reward probabilities per arm
($\mu_x$ for a user alone on arm $x$, $\eps_x$ for each user sharing it), for which we obtain explicit converses, in closed form for arms with equal parameters, the Gallager function of block coding, and the exponents of the interactive schemes (\Cref{sec:erasure}). For two users, one informative arm and a silent arm (an arm that never yields a reward, $\mu_2=\eps_2=0$), and noiseless collisions $\eps_1 = 0$, the converse and one-arm musical chairs match, and the error exponent is $-\log(1-\mu_1/2)$; for two users and two noiseless arms of mean one, musical chairs matches the converse as well.
  With noiseless collisions, musical chairs provably exceeds every name-decoding exponent, for every number $J$ of users and $J$ arms of a common mean below one. With noisy collisions, our numerical estimates indicate that the confidence controllers also exceed it, but we have no proof that their exponent is positive~(\Cref{tab:bandit-comparison}). Generalized musical chairs goes further: for three users and three noiseless arms of mean $\mu=1$, its exponent is more than twice that of musical chairs. How the optimal exponent depends on $J$ is left open.

\end{enumerate}

The remainder of the paper is organized as follows. \Cref{sec:model} introduces the general channel and the symmetry-breaking problem, and \Cref{sec:meta-converse} proves the converses. \Cref{sec:block-coding} develops block coding and the limits of name decoding, and \Cref{sec:interactive-coding} the interactive schemes. \Cref{sec:erasure} applies the results to the noisy bandit channel and compares the schemes with the converses, and \Cref{sec:conclusion} concludes. Proofs that would interrupt the main flow are presented in the appendices.

\section{Symmetric Channels and the {Symmetry-Breaking} Problem}
\label{sec:model}

\subsection{Notation}

For a positive integer $m$, let $[m]:=\{1,\ldots,m\}$. For $\cS\subseteq[J]$, write $X_{\cS}:=(X_j:j\in\cS)$ and $\cS^c:=[J]\setminus\cS$. For $j\in[J]$, write $X_{-j}:=X_{\other{j}}\in\cX^{J-1}$, with the indices in increasing order. {For a finite set $\cA$, $\cS\subseteq[J]$, $u\in\cA^{|\cS|}$ and $v\in\cA^{|\cS^c|}$, let $\ins{\cS}{u}{v}\in\cA^J$ denote the vector $z$ with $z_\cS=u$ and $z_{\cS^c}=v$, the components of $u$ and $v$ being read in increasing index order. Thus $x_{[J]}=\ins{\cS}{x_\cS}{x_{\cS^c}}$ for every $x_{[J]}\in\cX^J$. For a singleton, we abbreviate $\sqcup_{\{j\}}$ as $\sqcup_j$, so that $\ins{j}{u}{v}=(v_1,\ldots,v_{j-1},u,v_j,\ldots,v_{J-1})$ for $u\in\cA$ and $v\in\cA^{J-1}$, $x_{[J]}=\ins{j}{x_j}{x_{-j}}$, and $(u,v)\mapsto\ins{j}{u}{v}$ is a bijection $\cA\times\cA^{J-1}\to\cA^J$.} A tuple written with a comma, such as $(u,v)$ or $(x_j,x_{-j})$, always denotes the plain concatenation of its entries.
A superscript $n$ denotes a length-$n$ sequence and $X^n[i]$ {denotes} its $i${-}th entry. All alphabets are finite, all logarithms are natural, and rates are measured in nats per channel use. For a finite set $\cA$, $\mathcal P(\cA)$ denotes the set of probability distributions on $\cA$. The binary entropy is denoted
\begin{equation*}
  \hb(u) := -u\log u-(1-u)\log(1-u),\qquad u\in[0,1]\, .
\end{equation*}
{For integers $1\leq k\leq N$, let $\fallfac{N}{k}:=N(N-1)\cdots(N-k+1)$ denote the falling factorial.}

\subsection{Symmetric Memoryless Channel}

At time $i$, user $j$ selects $X_j[i]\in\cX$ and observes $Y_j[i]\in\cY$ at the end of the channel use. The memoryless channel has the joint transition law
\begin{equation}
  P_{Y_{[J]}\mid X_{[J]}}(y_{[J]}\mid x_{[J]}).
  \label{eq:general-joint-channel}
\end{equation}
Conditioned on the input chosen in each slot, successive output vectors are generated independently according to this law. {When no confusion arises we abbreviate $P_{Y_{[J]}\mid X_{[J]}}$ as $P$.}

The users are anonymous, and the physical channel does not depend on how they are labeled.
\begin{definition}[Symmetric channel]
\label{def:symmetric-channel}
For a permutation $\sigma$ of $[J]$, write $x_\sigma:=(x_{\sigma(1)},\ldots,x_{\sigma(J)})$, and define $y_\sigma$ analogously. The memoryless channel $P$ is \emph{symmetric}~(or \emph{permutation invariant}) if, for every permutation $\sigma$ of $[J]$, every $x_{[J]}\in\cX^J$ and every $y_{[J]}\in\cY^J$,
\begin{equation}
  P(y_\sigma\mid x_\sigma)=P(y_{[J]}\mid x_{[J]}) \,.
  \label{eq:user-permutation-symmetry}
\end{equation}
\end{definition}
Throughout, the channel is symmetric.  Note that this is a symmetry under relabeling of the users, not the symmetry of a single-user channel in the sense of \cite{gallager1968}: no symmetry is imposed on the input or the output alphabet.

For $j\in[J]$, the channel observed by receiver $j$ is the marginal
\begin{equation}
  W_j(u\mid x_{[J]}) := \sum_{v\in\cY^{J-1}} {P}(\ins{j}{u}{v}\mid x_{[J]}),\qquad u\in\cY,
  \label{eq:receiver-marginal-channel}
\end{equation}
obtained by summing over the outputs $y_{-j}=v$ of the other users while the output $y_j=u$ of user $j$ is held fixed.
The symmetry \eqref{eq:user-permutation-symmetry} implies that the marginals yield a single local law. For $b\in\cX^{J-1}$ and a permutation $\tau$ of $[J-1]${,} write $b_\tau:=(b_{\tau(1)},\ldots,b_{\tau(J-1)})$. Define
\begin{equation}
  W(y\mid a,b):=W_1(y\mid \ins{1}{a}{b}),\qquad a\in\cX,\ b\in\cX^{J-1},\ y\in\cY.
  \label{eq:single-local-law}
\end{equation}
{The next lemma shows that every user sees the same local law $W$, which is invariant under permutations of the other users' inputs.}
\begin{lemma}[Single local law]
\label{lem:single-local-law}
Under \eqref{eq:user-permutation-symmetry}, for every $j\in[J]$, $x_{[J]}\in\cX^J$, $y\in\cY$, and every permutation $\sigma$ of $[J]$ and $\tau$ of $[J-1]$,
\begin{enumerate}[label=(\roman*)]
  \item $W_j(y\mid x_\sigma)=W_{\sigma(j)}(y\mid x_{[J]})$;
  \item $W_j(y\mid x_{[J]})=W(y\mid x_j,x_{-j})$;
  \item $W(y\mid x_j,(x_{-j})_\tau)=W(y\mid x_j,x_{-j})$.
\end{enumerate}
\end{lemma}
\begin{proof}
(i) Since $y\mapsto y_\sigma$ is a bijection of $\cY^J$ with $(y_\sigma)_j=y_{\sigma(j)}$, summing \eqref{eq:user-permutation-symmetry} over the $y$ with $y_{\sigma(j)}=u$ gives $W_j(u\mid x_\sigma)=W_{\sigma(j)}(u\mid x_{[J]})$, by \eqref{eq:receiver-marginal-channel}. (ii) follows from (i) and \eqref{eq:single-local-law} for the permutation $\sigma$ with $\sigma(1)=j$ that keeps the order of the other indices, for which $x_\sigma=\ins{1}{x_j}{x_{-j}}$. (iii) follows from (i) for the permutation that fixes $1$ and permutes the other positions by $\tau$.
\end{proof}

\begin{definition}[Anonymous symmetry breaking]
  \label{def:admissible-causal-strategy}

Let $\Pi_n$ denote the class of all admissible $n$-round anonymous symmetric causal strategies (defined below), where $n$ is the number of channel uses.  We call $n$ the blocklength even though no block code is involved in general. The private seeds
$S_1,\ldots,S_J$ are i.i.d.\ with common law $P_S$ and are independent of the common randomness $C$, whose law is $P_C$.
The laws $P_C$ and $P_S$ are part of the strategy, and, given the inputs, the channel outputs are independent of $(C,S_{[J]})$. Under our convention, all private randomization is absorbed into the private seeds, and a \emph{strategy} $\pi_n\in\Pi_n$ is specified by a sequence of common causal deterministic encoding maps $\{f_i\}_{i=1}^n$ such that
\begin{equation}
 X_j[i]=f_i(C,S_j,Y_j^{i-1}), \quad i\in[n], j\in[J],
\end{equation}%
and a common terminal map $g_n$ that produces the identifier $Z_j\in[J]$ of user $j$,
\begin{equation}
  Z_j=g_n(C,S_j,Y_j^n) \in [J], \quad j\in[J].
\end{equation}%
We call a strategy \emph{nonadaptive} if the encoding maps $f_i$ do not depend on the past channel outputs $Y_j^{i-1}$, i.e., $X_j[i]=f_i(C,S_j)$ for all $i\in[n]$ and $j\in[J]$. A strategy is also called a \emph{protocol}. We define the \emph{symmetry-breaking error event} as
\begin{equation}
\mathcal E
:=
\{\exists\,j\neq k: Z_j=Z_k\},
\end{equation}
and its probability under strategy $\pi_n$ and channel $P$ is denoted by
\begin{equation}
P_{\mathrm{b}}^{(n)} := P_{\mathrm{b}}^{(n)}(\pi_n;P)\,.
\end{equation}
\emph{Symmetry breaking} is \emph{reliable} if there exists a sequence
{$\{\pi_n\}_{n\ge1}$ of strategies}
with $\pi_n\in\Pi_n$ such that $P_{\mathrm{b}}^{(n)}\to0$.
\end{definition}

Common randomness cannot lower the optimal error probability, since $P_{\mathrm b}^{(n)}$ averages the conditional error probabilities over $C$; we keep $C$ because shared random codebooks are naturally described with it.

\paragraph*{Collisions of i.i.d.~identifiers}
Any strategy whose terminal map ignores the channel outputs cannot be reliable, because under such a strategy the identifiers $Z_1,\ldots,Z_J$ are i.i.d.~given $C$. For integers $2\leq m\leq N$, let
\begin{equation}
 \gamma_{m,N}:=1-\frac{\fallfac{N}{m}}{N^m}
 \label{eq:gamma-mj}
\end{equation}
be the probability that $m$ i.i.d.~uniform draws from $[N]$ are not all distinct.

\begin{lemma}[Collision among i.i.d.~identifiers]
\label{lem:iid-identifier-collision}
Let $2\leq m\leq J$ and let $Z_1,\ldots,Z_m$ be i.i.d.\ random variables taking values in $[J]$. Then
\begin{equation}
 \PP\{\exists j\neq k:Z_j=Z_k\}\geq\gamma_{m,J}.
 \label{eq:iid-identifier-collision}
\end{equation}
\end{lemma}

\begin{proof}
If $Z_1,\ldots,Z_m$ are i.i.d.~with distribution $(p_z)_{z\in[J]}$, then
\begin{equation}
 \PP\{Z_1,\ldots,Z_m\ \text{all distinct}\} =m!\,e_m(p_1,\ldots,p_J) \leq\frac{\fallfac{J}{m}}{J^m},
 \label{eq:m-iid-distinct-bound}
\end{equation}
where $e_m(p_1,\ldots,p_J):=\sum_{i_1<\cdots< i_m} p_{i_1}\cdots p_{i_m}$ is the $m${-}th elementary symmetric polynomial and the inequality is Maclaurin's inequality~\cite[Th.~52]{hardy1952inequalities}. The uniform law attains equality. Taking the complement proves the claim.
\end{proof}

Thus $\gamma_{m,J}$ is the smallest collision probability among $m$ i.i.d.\ draws from $[J]$. In particular,
\begin{equation}
 \gamma_J:=\gamma_{J,J}=1-\frac{J!}{J^J}, \qquad \gamma_{2,J}=\frac1J.
 \label{eq:gamma-j}
\end{equation}

If the identifiers $Z_1,\ldots,Z_J$ were i.i.d.\ given the common randomness, for instance drawn from the private seeds without using the channel, then $P_{\mathrm{b}}^{(n)}\geq\gamma_J$ for every $n$. Reliable symmetry breaking therefore requires the channel to make the users' observations dependent. The converse of \Cref{sec:meta-converse} bounds how fast the error probability can decay, and the collision term $P_{\mathrm{col}}(M,J)=\gamma_{J,M}$ of the ranking scheme below, \eqref{eq:birthday-bound}, is another instance of \eqref{eq:gamma-mj}.

Private randomness is essential for the coding construction below: users independently draw tentative names. The codebook is common but can be fixed deterministically after the random-coding analysis.

\subsection{Random Names, Multicast Decoding, and Ranking}
\label{subsec:ranking-reduction}

We propose a nonadaptive scheme that uses a common codebook and private randomization to generate tentative names. {It is one particular strategy in $\Pi_n$; many other strategies exist, including the interactive ones of \Cref{sec:interactive-coding}, and the converse of \Cref{sec:meta-converse} applies to all of them.} Fix a codebook size $M$ and a common codebook
\begin{equation}
  \cC:=(x^n(m))_{m\in[M]} \in (\cX^n)^M.
  \label{eq:common-codebook}
\end{equation}
Each user independently draws
\begin{equation}
  S_j = M_j\sim\operatorname{Unif}[M]
  \label{eq:private-random-message}
\end{equation}
and transmits $X_j^n=x^n(M_j)$. {We call $M_j$ the \emph{name} of user $j$. It plays the role of the message in the multicast code, and $R:=\frac1n\log M$ is the \emph{name rate}.} User $j$ knows $(M_j,X_j^n,Y_j^n)$ and decodes an unordered list~(a multiset) $\widehat{\cM}_{-j}$ of $J-1$ foreign names. The unordered formulation is important: anonymous users need not attach physical user labels to the decoded codewords.

User $j$ forms the multiset $\{M_j\}\cup\widehat{\cM}_{-j}$ and chooses its rank, the number of decoded names below its own plus one,
\begin{equation}
  Z_j:=1+\sum_{m\in\widehat{\cM}_{-j}}\ind\{m<M_j\} \in [J] \,.
  \label{eq:rank-rule}
\end{equation}
If all names are distinct and every decoder is correct, the $J$ ranks $Z_1,\ldots,Z_J$ are a permutation of $[J]$, so the symmetry-breaking error event $\mathcal E$ does not occur.
Define the decoding error probability given that the names are distinct,
\begin{equation}
  P_{e,j}^{(n)} := \PP\{\widehat{\cM}_{-j}\neq\{M_k:k\neq j\} \mid M_1,\ldots,M_J\text{ distinct}\},
  \label{eq:individual-decoding-error}
\end{equation}
and the collision probability, i.e., the probability that the names are not distinct,
\begin{align}
  P_{\mathrm{col}}(M,J) &:= 1-\frac{\fallfac{M}{J}}{M^J}=\gamma_{J,M},\qquad M\geq J.
  \label{eq:birthday-bound}
\end{align}
{By the union bound over the $J$ receivers,} the ranking scheme \eqref{eq:rank-rule} gives
\begin{align}
  P_{\mathrm{b}}^{(n)} &\leq P_{\mathrm{col}}(M,J)+\sum_{j=1}^J P_{e,j}^{(n)}.
  \label{eq:sync-decomposition}
\end{align}

{By the union bound over the $\binom{J}{2}$ pairs of names, $P_{\mathrm{col}}(M,J)\leq\binom J2\frac1M$, and $P_{\mathrm{col}}(M,J)=\binom J2\frac1M+O(M^{-2})$ as $M\to\infty$.} For fixed $J$ and $R$, and $M=\lceil e^{nR}\rceil$, the tentative-name collision term $P_{\mathrm{col}}(M,J)$ has exponent $R$.
\Cref{sec:block-coding} bounds the decoding error probability $P_{e,j}^{(n)}$ in \eqref{eq:sync-decomposition} when the common codebook is drawn from the i.i.d.~random-coding ensemble: independently for every name $m$, generate
\begin{equation}
\PP_{\cC}\{X^n(m)=x^n\} = \prod_{i=1}^n Q_X(x[i]).
 \label{eq:iid-codebook-law}
\end{equation}
Conditioned on distinct names, the $J$ transmitted codewords are independent draws from $Q_X^n$, exactly as required by the random-coding analysis.

\section{Change-of-Measure Converses}
\label{sec:meta-converse}

In this section, we derive converses, i.e., lower bounds on the error probability of every anonymous symmetric causal strategy in $\Pi_n$, independently of any particular coding architecture. To do so, we compare the physical experiment with references under which the identifiers of $m$ users are conditionally i.i.d.\ given a latent variable, using a change-of-measure argument.
For a channel $P$, define the optimal error probability and the upper reliability exponent, respectively,
\begin{align}
 \epsilon_n^\star(P)&:=\inf_{\pi_n\in\Pi_n} P_{\mathrm{b}}^{(n)}(\pi_n;P),
 \label{eq:optimal-sync-error}\\
 \overline E_{\mathrm{b}}(P)&:=\limsup_{n\to\infty}-\frac1n\log\epsilon_n^\star(P).
 \label{eq:optimal-sync-exponent}
\end{align}
The lower reliability exponent $\underline E_{\mathrm b}(P)$ is defined with $\liminf$ in place of $\limsup$; converses bound $\overline E_{\mathrm b}(P)$ from above, and schemes bound $\underline E_{\mathrm b}(P)$ from below.
Unlike the cut-set bounds of \Cref{sec:multi-user}, which constrain only name decoding, these converses hold for every strategy.

\subsection{Law of the Complete Local Views}\label{subsec:local-views}

By definition, given a protocol $\pi_n\in\Pi_n$ and a realization $C=c$ of the common randomness, the identifier chosen by user $j$, $Z_j=g_n(c,S_j,Y_j^n)$, is a function of
\begin{equation}
 H_j:=(S_j,Y_j^n),
 \label{eq:complete-local-view}
\end{equation}
which we call the \emph{complete local view} of user $j$, and our first step is to derive its distribution.
The private seeds $S_1,\ldots,S_J$ are i.i.d.\ with law $P_S$, hence the distribution of the complete local views of all users $H_{[J]}$ is
\begin{equation}
 R_P^{\pi_n,c}(ds_{[J]},dy_{[J]}^n):=\prod_{j=1}^JP_S(ds_j)\prod_{i=1}^n P\bigl(y_{[J]}[i]\bigm| x_{[J]}[i]\bigr),
 \qquad x_j[i]=f_i(c,s_j,y_j^{i-1}),
 \label{eq:probability-trajectory-measure}
\end{equation}
and this distribution depends on the protocol $\pi_n$ through $f_i$. We also write $P^{\pi_n,c}:=R_P^{\pi_n,c}$.

\subsection{Symmetric References}\label{subsec:symmetric-references}

Consider $2 \le m \le J$, and an alternative distribution $Q$ for local views $H_{[J]}$ where the local views of the first $m$ users $H_{[m]}$ are i.i.d. Under this alternative distribution, the identifiers $Z_{[m]}$ are i.i.d.~as well, so that they collide with probability at least $\gamma_{m,J}$, as shown in Lemma~\ref{lem:iid-identifier-collision}. By linearity, this is also true when considering a mixture of such $Q$, which leads to the following definition and lemma.

\begin{definition}[Symmetric reference]
\label{def:symmetric-reference}
A probability law $Q$ on the complete local views $H_{[J]}$ is an \emph{$m$-symmetric reference} for $(\pi_n,c)$ if $Q\ll P^{\pi_n,c}$ and there exists a latent random variable $\Lambda$ such that, under $Q$, the views $H_1,\ldots,H_m$ are conditionally i.i.d.\ given $\Lambda$; formally, $Q$ is the marginal on $H_{[J]}$ of a law of $(\Lambda,H_{[J]})$ under which $H_1,\ldots,H_m$ are conditionally i.i.d.\ given $\Lambda$. The set of such laws is denoted by $\mathfrak L_{m,J}^{(n)}(\pi_n,c;P)$.
\end{definition}
Note that exchangeability of $H_1,\ldots,H_m$ would not suffice: on the channel of \Cref{subsec:name-decoding-limits} whose output pair is uniform on $\{(0,1),(1,0)\}$, the two views are exchangeable, and yet the identifiers $1+Y_j[1]$ never collide.

\begin{lemma}[Failure under a symmetric reference]
\label{lem:reference-failure}
Every $Q\in\mathfrak L_{m,J}^{(n)}(\pi_n,c;P)$ satisfies
\begin{equation}
 Q(\cE)\geq\gamma_{m,J},
 \qquad \cE:=\{\exists j\neq k:Z_j=Z_k\}.
 \label{eq:latent-reference-failure}
\end{equation}
\end{lemma}
\begin{proof}
Conditionally on $\Lambda$, the identifiers $Z_j=g_n(c,S_j,Y_j^n)$, $j\in[m]$, are i.i.d., since all users apply the same map $g_n$. By \Cref{lem:iid-identifier-collision} they collide with conditional probability at least $\gamma_{m,J}$. A collision among the selected users is a failure; averaging over $\Lambda$ proves the claim.
\end{proof}

\subsection{R\'enyi Divergences}\label{subsec:renyi-divergences}

Before performing a change of measure, we first recall a few notions about R\'enyi divergences, which are critical for our analysis. For probability laws $Q\ll P$ with likelihood ratio $L=dQ/dP$ and for $\alpha>1$, the R\'enyi divergence is
\begin{equation}
 D_\alpha(Q\Vert P):=\frac{1}{\alpha-1}\log\EE_P[L^\alpha],
 \label{eq:renyi-definition}
\end{equation}
and $D_\alpha(Q\Vert P):=+\infty$ if $Q\not\ll P$. On a finite set, $\EE_P[L^\alpha]=\sum_\omega Q(\omega)^\alpha P(\omega)^{1-\alpha}$. Moreover,
\begin{align}
 \lim_{\alpha\downarrow1}D_\alpha(Q\Vert P) &=D_{\mathrm{KL}}(Q\Vert P), \label{eq:renyi-kl-limit}\\
 D_\infty(Q\Vert P) &:=\log\operatorname*{ess\,sup}_{P}L;
 \label{eq:renyi-infinity-definition}
\end{align}
see \cite[Ths.~5 and~6]{vanerven2014}; \eqref{eq:renyi-kl-limit} holds since $D_\alpha(Q\Vert P)<\infty$ on a finite set with $Q\ll P$. Set $b_\alpha:=\alpha/(\alpha-1)$ for $1<\alpha<\infty$ and $b_\infty:=1$.

\begin{lemma}[R\'enyi change of measure]
\label{lem:renyi-event}
For every event $\cA$ and every $\alpha\in(1,\infty]$,
\begin{equation}
 P(\cA)\geq Q(\cA)^{b_\alpha}\exp\{-D_\alpha(Q\Vert P)\}.
 \label{eq:renyi-event}
\end{equation}
\end{lemma}

\begin{proof}
For $\alpha<\infty$, H\"older's inequality gives
\begin{equation}
 Q(\cA)=\EE_P[L\ind_{\cA}]
 \leq \bigl(\EE_P[L^\alpha]\bigr)^{1/\alpha} P(\cA)^{(\alpha-1)/\alpha}
 =e^{\frac{\alpha-1}{\alpha}D_\alpha(Q\Vert P)} P(\cA)^{(\alpha-1)/\alpha},
 \label{eq:holder-renyi-proof}
\end{equation}
and rearranging proves the claim. For $\alpha=\infty$, $Q(\cA)=\int_{\cA}L\,dP\leq e^{D_\infty(Q\Vert P)}P(\cA)$. If $Q\not\ll P$ the bound is trivial.
\end{proof}
The bound also follows from the data-processing inequality for R\'enyi divergences, as in the Arimoto converse of \cite{polyanskiy2010arimoto}.

\subsection{Change of Measure}\label{subsec:change-of-measure}

Equipped with the above, we then use a change-of-measure argument between the distribution $P^{\pi_n,c}$ and the closest symmetric reference distribution $Q \in \mathfrak L_{m,J}^{(n)}(\pi_n,c;P)$ to lower bound the error probability, where the closeness between $P^{\pi_n,c}$ and $Q$ is quantified by the R\'enyi divergence. Namely, combining \Cref{lem:reference-failure} with \Cref{lem:renyi-event} for $\cA=\cE$ gives the basic converse of Proposition~\ref{prop:trajectory-converse}.

\begin{proposition}[Trajectory converse]
\label{prop:trajectory-converse}
For every protocol $\pi_n$, every realization $c$ of the common randomness, every $2\leq m\leq J$ and every $\alpha\in(1,\infty]$,
\begin{equation}
 P_{\mathrm{b}}^{(n)}(\pi_n;P\mid C=c)\geq\gamma_{m,J}^{b_\alpha}\,e^{-\mathcal D_{\alpha,m}^{(n)}(\pi_n,c;P)},
 \label{eq:unified-trajectory-converse}
\end{equation}
where the \emph{trajectory radius} is
\begin{equation}
 \mathcal D_{\alpha,m}^{(n)}(\pi_n,c;P):=\inf_{Q\in\mathfrak L_{m,J}^{(n)}(\pi_n,c;P)}D_\alpha(Q\Vert P^{\pi_n,c}).
 \label{eq:unified-trajectory-radius}
\end{equation}
\end{proposition}

The trajectory radius $\mathcal D_{\alpha,m}^{(n)}(\pi_n,c;P)$ quantifies the closeness between the law $P^{\pi_n,c}$ and the closest symmetric reference $Q\in\mathfrak L_{m,J}^{(n)}(\pi_n,c;P)$, and it depends on the blocklength $n$, the channel $P$ and the protocol $\pi_n$. For a fixed protocol $\pi_n$, computing $\mathcal D_{\alpha,m}^{(n)}(\pi_n,c;P)$ does not seem feasible, mostly because of the complex structure of $\mathfrak L_{m,J}^{(n)}(\pi_n,c;P)$. Even worse, a converse that holds for all protocols would require maximizing $\mathcal D_{\alpha,m}^{(n)}(\pi_n,c;P)$ over the protocols, which seems even more involved. We now provide a tractable bound that involves only single-letter quantities.

\subsection{Product-Form Symmetric References}\label{subsec:product-form-references}

To obtain a tractable converse, we consider a subset of the set of symmetric references $\mathfrak L_{m,J}^{(n)}(\pi_n,c;P)$ which have a special product form, which allows us to derive a single-letter bound.
Consider distributions $Q_{B}^{\pi_n,c}$ of the form
\begin{equation}
Q_{B}^{\pi_n,c}:=R_B^{\pi_n,c}/\beta_{B}^{\pi_n,c},
 \label{eq:factorized-reference-law}
\end{equation}
where
\begin{equation}
 R_B^{\pi_n,c}(ds_{[J]},dy_{[J]}^n):= \prod_{j=1}^JP_S(ds_j)\prod_{i=1}^n B\bigl(y_{[J]}[i]\bigm| x_{[J]}[i]\bigr),
 \qquad x_j[i]=f_i(c,s_j,y_j^{i-1}),
 \label{eq:kernel-trajectory-measure}
\end{equation}
and $\beta_{B}^{\pi_n,c}$ is the total mass of $R_B^{\pi_n,c}$, which depends on the kernel\footnote{We call $B$ a ``kernel''; however, this does not imply that $B$ is a transition kernel, because we do not assume $B$ to be normalized, i.e., we can have $\sum_{y_{[J]}} B(y_{[J]}\mid x_{[J]}) \ne 1$.} $B$, of the form
\begin{equation}
 B(y_{[J]}\mid x_{[J]})=\sum_{r\in\mathcal R}\left[\prod_{j=1}^m a_r(x_j,y_j)\right]c_r\bigl(y_{m+1:J}\bigm| x_{[J]},y_{[m]}\bigr),
 \label{eq:latent-factorized-kernel}
\end{equation}
with $\mathcal R$ a finite set of component indices and all factors $a_r,c_r$ chosen nonnegative, and with $B\ll P$, which means that $B(y\mid x)=0$ whenever $P(y\mid x)=0$. The same one-user factor $a_r$ is used by every selected user. For $m=J$ the last factor is absent.
We say that a kernel $B$ is \emph{admissible} if the corresponding $a_r$ and $c_r$ satisfy the following.
\begin{definition}[Admissible kernels]
\label{def:admissible-kernel}
A kernel of the form \eqref{eq:latent-factorized-kernel} is \emph{admissible} if one of the following holds.
\begin{enumerate}
	\item[(I)] \emph{Independent completion:} every $c_r(y_{m+1:J}\mid x_{[J]},y_{[m]})$ does not depend on the inputs and outputs $(x_{[m]},y_{[m]})$ of the selected users; we then write it $c_r(x_{m+1:J},y_{m+1:J})$.
 \item[(N)] \emph{Normalized completion:} for every $r$, the sum $\sum_{y_{m+1:J}} c_r(y_{m+1:J}\mid x_{[J]},y_{[m]})$ is a constant, independent of $x_{[J]}$ and $y_{[m]}$. Since the constant can be absorbed into $a_r$, we normalize it to $1$ without loss of generality.
\end{enumerate}
The set of admissible kernels is denoted by $\mathfrak A_{m,J}(P)$. For $m=J$ both conditions are void.
\end{definition}
Expanding the product in \eqref{eq:kernel-trajectory-measure} writes $R_B^{\pi_n,c}$ as a sum over component sequences $r^n$. For a fixed $r^n$, the selected users run the protocol with the same one-user weights $a_{r_i}$, so they are i.i.d.\ unless the factors $c_{r_i}$ couple them. Under (I), $c_r$ ignores the selected users; under (N), it may depend on them but has total mass one, so that summing out the unselected outputs, from the last use backward, removes it without favoring any selected trajectory. The following lemma makes this precise.
\begin{lemma}[Product-form references]
\label{lem:product-form}
Let $B\in\mathfrak A_{m,J}(P)$. For every protocol $\pi_n$ and every $c$ such that $\beta_{B}^{\pi_n,c}>0$: $Q_{B}^{\pi_n,c}\in\mathfrak L_{m,J}^{(n)}(\pi_n,c;P)$.
\end{lemma}
\begin{proof}[Proof sketch]
The full proof is in Appendix~\ref{app:kernel-reference}.
Fix a component sequence $r^n$ and sum out the unselected users. Under (I) they factor off. Under (N), each factor $c_{r_i}$ sums to one, working backward from $i=n$. What remains is a product of $m$ identical one-user measures.
\end{proof}
\begin{definition}[Normalized kernels]
\label{def:normalized-kernel}
A kernel of the form \eqref{eq:latent-factorized-kernel} is \emph{normalized} if $\sum_{y_{[J]}} B(y_{[J]}\mid x_{[J]}) = 1$ for all $x_{[J]}$.
\end{definition}
Two remarks are in order: i)~We do not assume that $B$ is normalized, which allows for a better converse as we shall see. In turn, when $B$ is not normalized, $Q_{B}^{\pi_n,c}$ is in general not the law of the protocol run over a channel, so that computing  $\beta_{B}^{\pi_n,c}$ in closed form seems difficult. We will show how to obtain bounds on $\beta_{B}^{\pi_n,c}$. ii)~In the special case where $B$ is normalized, one may readily check that $\beta_{B}^{\pi_n,c}=1$, and $B$ can be interpreted as an alternative channel to the actual channel $P$; furthermore, the R\'enyi divergence quantifies the difficulty of distinguishing $P$ from $B$ based on the local views $H_{[J]}$.

\subsection{R\'enyi Divergence of Product-Form References}\label{subsec:renyi-product-form}

We now show how to bound the R\'enyi divergence between $Q_{B}^{\pi_n,c}$ and $P^{\pi_n,c}$ by exploiting the product form, to evaluate the converse. We first define the following quantities. In the remainder of this section, to simplify notation, we write $u:=x_{m+1:J}$ and $v:=y_{m+1:J}$ for the inputs and outputs of the unselected users.
The likelihood ratio between $Q_{B}^{\pi_n,c}$ and $P^{\pi_n,c}$ is given by the product
\begin{equation}
	\frac{d Q_B^{\pi_n,c}}{dP^{\pi_n,c}}= \frac{1}{\beta_B^{\pi_n,c}} \prod_{i=1}^n\frac{B(Y_{[J]}[i]\mid X_{[J]}[i])}{P(Y_{[J]}[i]\mid X_{[J]}[i])}.
 \label{eq:causal-trajectory-likelihood-ratio}
\end{equation}

\paragraph{Persistence coefficient}
Define $A_r(x):=\sum_y a_r(x,y)$, and $C_r(u):=\sum_v c_r(u,v)$ under (I), or $C_r \equiv 1$ under (N). The one-step mass of $B$ is then
\begin{equation}
b_B(x_{[J]}):=\sum_{y_{[J]}}B(y_{[J]}\mid x_{[J]}) =\sum_r\left[\prod_{j\leq m}A_r(x_j)\right] \,C_r(u).
\end{equation}
In turn, define the minimal expected one-step mass when the selected users draw their inputs i.i.d.\ and the unselected users play arbitrary inputs, namely
\begin{align}
	\kappa_m(B):=\min_{\substack{p\in\mathcal P(\cX)\\ u\in\cX^{J-m}}}\ \sum_{x_{[m]}}\left[\prod_{j=1}^mp(x_j)\right]b_B(x_{[m]},u)
 	=\min_{p,u}\sum_r\left[\sum_xp(x)A_r(x)\right]^mC_r(u),
 \label{eq:partial-persistence-kernel-form}
\end{align}
omitting $u$ when $m=J$.
It is noted that $\kappa_m(B)$ only depends on single-letter quantities.
We will use it to lower bound the normalization constant $\beta_{B}^{\pi_n,c}$.
Further, $\kappa_m(B)$ is readily computable by convex minimization over the simplex, since $p \mapsto \sum_r\left[\sum_xp(x)A_r(x)\right]^mC_r(u)$ is convex for any fixed $u$, and $u$ ranges over a finite set.
\paragraph{Likelihood-ratio moment and reference radius}
For $1<\alpha<\infty$ and $\kappa_m(B)>0$, define the worst-case likelihood-ratio moment and the one-step reference radius
\begin{align}
 \Psi_{\alpha,m}(B;P)&:=\max_{x\in\cX^J}\sum_{y:P(y\mid x)>0}B(y\mid x)^\alpha P(y\mid x)^{1-\alpha},
 \label{eq:partial-renyi-moment}\\
 d_{\alpha,m}^{\mathrm{ref}}(B\Vert P)&:=\frac{\log\Psi_{\alpha,m}(B;P)-\alpha\log\kappa_m(B)}{\alpha-1},
 \label{eq:partial-renyi-ref-divergence}
\end{align}
and the effective persistence
\begin{align}
 K_{\alpha,m}(B;P)&:=e^{-d_{\alpha,m}^{\mathrm{ref}}(B\Vert P)}=\left[\frac{\kappa_m(B)^\alpha}{\Psi_{\alpha,m}(B;P)}\right]^{1/(\alpha-1)}.
 \label{eq:partial-renyi-effective-coefficient}
\end{align}
At order infinity put
\begin{align}
 \rho_{\infty,m}(B;P)&:=\max_{x,y:B(y\mid x)>0}\frac{B(y\mid x)}{P(y\mid x)},\notag\\
 d_{\infty,m}^{\mathrm{ref}}(B\Vert P)&:=\log\frac{\rho_{\infty,m}(B;P)}{\kappa_m(B)},\qquad
 K_{\infty,m}(B;P):=\frac{\kappa_m(B)}{\rho_{\infty,m}(B;P)}.
 \label{eq:partial-factorized-infinity-limit}
\end{align}
In the special case where $B$ is normalized we have $b_B(x_{[J]}) = 1$, so that $\kappa_m(B) = 1$, and $\frac1{\alpha-1}\log\Psi_{\alpha,m}(B;P)$ is the maximal R\'enyi divergence between the conditional distributions $B(\cdot\mid x_{[J]})$ and $P(\cdot\mid x_{[J]})$, when maximizing over the input $x_{[J]}$. In turn, the maximizer can be interpreted as the ``most informative'' input to differentiate between the channels $P$ and $B$. We now provide a tractable bound on the R\'enyi divergence $D_\alpha(Q_B^{\pi_n,c}\Vert P^{\pi_n,c})$ for any given protocol $\pi_n$, which makes the converse of \Cref{prop:trajectory-converse} tractable for every protocol $\pi_n$, every $c$ and every product-form reference $Q_B^{\pi_n,c}$.

\begin{lemma}[Kernel references]
\label{lem:factorized-reference-structure}
\label{lem:factorized-reference-renyi-cost}
Let $B\in\mathfrak A_{m,J}(P)$ with $\kappa_m(B)>0$. For every protocol $\pi_n$ and every $c$:
\begin{enumerate}
\item $\beta_{B}^{\pi_n,c}\geq\kappa_m(B)^n$;
\item $D_\alpha(Q_{B}^{\pi_n,c}\Vert P^{\pi_n,c})\leq n\,d_{\alpha,m}^{\mathrm{ref}}(B\Vert P)$ for every $\alpha\in(1,\infty]$.
\end{enumerate}
\end{lemma}
\begin{proof}[Proof sketch]
The full proof is in Appendix~\ref{app:kernel-reference}.
\begin{enumerate}
 \item Summing out the unselected users as in the proof of \Cref{lem:product-form}, now on prefixes of the component sequence, shows that the branches that extend a given branch by one channel use have a total mass equal to the mass of that branch times the factor $\sum_r[\EE_pA_r]^m\,\EE_qC_r$. Here $p$ is the next-input law of a selected user and $q$ that of the unselected users on this branch, so the factor is at least $\kappa_m(B)$.
\item Conditioning on the past, the $\alpha$-th moment of the product in \eqref{eq:causal-trajectory-likelihood-ratio}, that is, of $dR_B^{\pi_n,c}/dP^{\pi_n,c}$, grows by at most $\Psi_{\alpha,m}(B;P)$ per channel use. Dividing by $(\beta_{B}^{\pi_n,c})^\alpha$ and using the first item gives the claim.
\end{enumerate}
\end{proof}

\subsection{The Kernel Converse}
\label{subsec:kernel-converse}

The above provides a lower bound on the error probability for any fixed protocol $\pi_n$ and any value of $c$; however, this bound depends on $\pi_n$ and $c$. To obtain a lower bound which applies to all protocols, we average over $C$, and combine \Cref{prop:trajectory-converse} with \Cref{lem:factorized-reference-structure}, which gives the main converse. This converse only involves single-letter quantities defined above.

\begin{theorem}[Kernel converse]
\label{thm:latent-symmetry-meta-converse}
\label{thm:finite-alpha-factorized-converse}
For every $2\leq m\leq J$, every $B\in\mathfrak A_{m,J}(P)$ with $\kappa_m(B)>0$, every symmetric causal protocol $\pi_n$, and every $\alpha\in(1,\infty]$,
\begin{equation}
 P_{\mathrm{b}}^{(n)}(\pi_n;P)\geq\gamma_{m,J}^{b_\alpha}K_{\alpha,m}(B;P)^n.
 \label{eq:finite-alpha-factorized-converse}
\end{equation}
\end{theorem}

\begin{corollary}[Optimized kernel converse]
\label{thm:combined-protocol-converse}
\label{subsec:converse-complementarity}
Let
\begin{equation}
 K_{\alpha,m}^{\mathrm{ref}}(P):=\sup_{\substack{B\in\mathfrak A_{m,J}(P)\\\kappa_m(B)>0}}K_{\alpha,m}(B;P),
 \label{eq:optimized-unified-coefficient}
\end{equation}
with the convention that an empty supremum is zero. Then
\begin{align}
 \epsilon_n^\star(P)&\geq\sup_{\substack{2\leq m\leq J\\\alpha\in(1,\infty]}}\gamma_{m,J}^{b_\alpha}[K_{\alpha,m}^{\mathrm{ref}}(P)]^n,
 \label{eq:combined-protocol-converse}\\
 \overline E_{\mathrm{b}}(P)&\leq\inf_{\substack{2\leq m\leq J\\\alpha\in(1,\infty]}}[-\log K_{\alpha,m}^{\mathrm{ref}}(P)].
 \label{eq:combined-protocol-exponent}
\end{align}
\end{corollary}
\begin{proof}[Proof of \Cref{thm:latent-symmetry-meta-converse} and \Cref{thm:combined-protocol-converse}]
Fix $c$. By the first item of \Cref{lem:factorized-reference-structure}, $\beta_B^{\pi_n,c}\geq\kappa_m(B)^n>0$, so that $Q_B^{\pi_n,c}$ is defined, and it belongs to $\mathfrak L_{m,J}^{(n)}(\pi_n,c;P)$ by \Cref{lem:product-form}. \Cref{prop:trajectory-converse} and the second item of \Cref{lem:factorized-reference-structure} give $P_{\mathrm b}^{(n)}(\pi_n;P\mid C=c)\geq\gamma_{m,J}^{b_\alpha}e^{-nd_{\alpha,m}^{\mathrm{ref}}(B\Vert P)}=\gamma_{m,J}^{b_\alpha}K_{\alpha,m}(B;P)^n$, and averaging over $C$ proves \eqref{eq:finite-alpha-factorized-converse}. Optimizing over $B$, $m$ and $\alpha$ proves \Cref{thm:combined-protocol-converse}.
\end{proof}
Note that the symmetry \eqref{eq:user-permutation-symmetry} of the channel is not used in this section. Therefore, the converses hold for every memoryless channel, provided that all users apply the same strategy.

\paragraph{Computing the coefficient}
For $s>0$, $\kappa_m(sB)=s\kappa_m(B)$ and $\Psi_{\alpha,m}(sB;P)=s^\alpha\Psi_{\alpha,m}(B;P)$, so that $K_{\alpha,m}(B;P)$ is scale invariant, and, for $1<\alpha<\infty$, $K_{\alpha,m}^{\mathrm{ref}}(P)=(\Psi^\star)^{-1/(\alpha-1)}$, where $\Psi^\star$ is the infimum of $\Psi_{\alpha,m}(B;P)$ over $B\in\mathfrak A_{m,J}(P)$ with $\kappa_m(B)\geq1$.
Under each of the completions (I) and (N), the admissible kernels form a convex cone (a sum merges the components), $\Psi_{\alpha,m}(\cdot;P)$ is convex and $\kappa_m$ is concave, so that the minimization is a convex program over each of the two cones, although neither has an explicit description in terms of the $|\cX|^J|\cY|^J$ entries of $B$. For $m<J$ one solves it over both cones and keeps the better value; for $m=J$ there are no unselected users, and the two cones coincide. Convexity does not make it easy: for $m=J=2$, the product form says that $B$, as a matrix indexed by $(x_1,y_1)$ and $(x_2,y_2)$, is completely positive, and membership in that cone is NP-hard \cite{dickinson2014}, as the order of the matrices grows. In practice we optimize over a fixed number of components, a nonconvex problem; this affects only optimality, since every admissible $B$ with $\kappa_m(B)>0$ gives a valid bound.
Also, at order infinity,
\begin{equation}
 K_{\infty,m}^{\mathrm{ref}}(P)=\sup\{\kappa_m(B):B\in\mathfrak A_{m,J}(P),\ B\leq P\}.
 \label{eq:latent-infinity-optimization}
\end{equation}

\paragraph{{Choice of the order}}
For a normalized kernel, $d_{\alpha,m}^{\mathrm{ref}}(B\Vert P)$ is a maximal R\'enyi divergence, nondecreasing in $\alpha$ \cite{vanerven2014}, so that $\alpha\downarrow1$ gives the best exponent, while at finite blocklength the prefactor $\gamma_{m,J}^{b_\alpha}$ favors larger orders. If $b_B$ is not constant, however, $\Psi_{\alpha,m}(B;P)\to\max_xb_B(x)>\kappa_m(B)$ as $\alpha\downarrow1$, so that $d_{\alpha,m}^{\mathrm{ref}}(B\Vert P)\to+\infty$; the best order is then finite or $\infty$, as for the common-output bounds and the optimized references of \Cref{tab:bandit-comparison}.

\paragraph{Support bound}
The effective persistence of a kernel is at most the physical probability of its support. Let $S_B(x):=\sum_{y:B(y\mid x)>0}P(y\mid x)$ be the physical probability of the support of $B$ and $t=(\alpha-1)/\alpha$. H\"older's inequality at each input gives $b_B(x)\leq\Psi_{\alpha,m}(B;P)^{1/\alpha}S_B(x)^t$, and $b_B(x)\leq\rho_{\infty,m}(B;P)S_B(x)$ at order infinity. Averaging, minimizing, and using the concavity of $s\mapsto s^t$ yields
\begin{align}
 K_{\alpha,m}(B;P)
 &\leq\left[\min_{p,u}\EE_{p^{\otimes m}}S_B(X_{[m]},u)^t\right]^{1/t}
 \leq\min_{p,u}\EE_{p^{\otimes m}}S_B(X_{[m]},u),\quad 1<\alpha<\infty,
 \notag\\
 K_{\infty,m}(B;P)
 &\leq\min_{p,u}\EE_{p^{\otimes m}}S_B(X_{[m]},u).
 \label{eq:upper-bound-K-all}
\end{align}
Thus a reference cannot retain more mass than the channel places on the outputs it keeps. At order infinity, the bound holds with equality for every admissible restriction of the channel, $B(y\mid x)=P(y\mid x)\ind\{y\in\mathcal S(x)\}$ with $\mathcal S(x)\subseteq\cY^J$, since then $\rho_{\infty,m}(B;P)=1$ and $b_B=S_B$.

\subsection{Noninteracting and Common-Output References}
\label{subsec:reference-instances}
\label{subsec:noninteraction-converse}

We now evaluate \Cref{thm:latent-symmetry-meta-converse} for two explicit families of admissible kernels: noninteracting references, under which the output of each selected user depends only on its own input, and common-output references, under which the selected users always receive the same output.

\paragraph{Noninteracting references}
Let $V$ be a one-user channel from $\cX$ to $\cY$ and write
\begin{equation}
 q_{m,V}(y_{[m]}\mid x_{[m]}):=\prod_{j=1}^mV(y_j\mid x_j),
 \qquad
 P_m(y_{[m]}\mid x_{[J]}):=\sum_{v}P(y_{[m]},v\mid x_{[J]}).
 \label{eq:physical-m-user-marginal}
\end{equation}
Complete the selected outputs with the physical conditional law of the unselected outputs:
\begin{equation}
 B_V(y_{[J]}\mid x_{[J]}):=q_{m,V}(y_{[m]}\mid x_{[m]})\,P(v\mid x_{[J]},y_{[m]}),
 \label{eq:canonical-partial-reference}
\end{equation}
with an arbitrary probability law in place of $P(\cdot\mid x_{[J]},y_{[m]})$ where $P_m(y_{[m]}\mid x_{[J]})=0$. Under $B_V$ each selected user behaves as if it were alone on the channel $V$, so the selected users do not interact. This kernel has one component ($|\mathcal{R}| = 1$, $a=V$) and a normalized completion, so $B_V$ satisfies (N) and $b_{B_V}\equiv1$, hence $\kappa_m(B_V)=1$. It satisfies $B_V\ll P$ if and only if $q_{m,V}(\cdot\mid x_{[m]})\ll P_m(\cdot\mid x_{[J]})$ for every $x_{[J]}$. In that case the completion cancels in $B_V/P$, and $\Psi_{\alpha,m}(B_V;P)=e^{(\alpha-1)d_{\alpha,m}(V\Vert P)}$ with the one-step radius
\begin{equation}
 d_{\alpha,m}(V\Vert P):=\max_{x_{[J]}}D_\alpha\bigl(q_{m,V}(\cdot\mid x_{[m]})\bigm\Vert P_m(\cdot\mid x_{[J]})\bigr).
 \label{eq:partial-renyi-radius}
\end{equation}
At $\alpha=1$ the R\'enyi divergence is the KL divergence, and at $\alpha=\infty$ it is the logarithm of the largest likelihood ratio. We set $d_{\alpha,m}(V\Vert P)=\infty$ if absolute continuity fails.

\begin{corollary}[Noninteraction bound]
\label{thm:j-meta-converse}
For every one-user channel $V$, every $2\leq m\leq J$, every symmetric causal protocol $\pi_n$, and every $\alpha\in(1,\infty]$,
\begin{equation}
 P_{\mathrm{b}}^{(n)}(\pi_n;P)\geq\gamma_{m,J}^{b_\alpha}e^{-n d_{\alpha,m}(V\Vert P)},
 \label{eq:j-meta-converse-finite}
\end{equation}
and hence
\begin{equation}
 \overline E_{\mathrm{b}}(P)\leq\inf_V d_{1,2}(V\Vert P).
 \label{eq:j-meta-converse-exponent}
\end{equation}
\end{corollary}
\begin{proof}
The finite-blocklength bound is \Cref{thm:latent-symmetry-meta-converse} for $B=B_V$, since $K_{\alpha,m}(B_V;P)=e^{-d_{\alpha,m}(V\Vert P)}$. For fixed $V$ and finite alphabets, $d_{\alpha,m}(V\Vert P)$ is nondecreasing in $\alpha$ and tends to $d_{1,m}(V\Vert P)$ as $\alpha\downarrow1$. Marginalizing one selected output gives $d_{\alpha,m}\leq d_{\alpha,m+1}$ by data processing. Hence $m=2$ and $\alpha\downarrow1$ give the best exponent. At finite blocklength, $m$ still trades the cost against $\gamma_{m,J}$.
\end{proof}

For noninteracting references the best exponent is obtained as $\alpha\downarrow1$, and at finite blocklength the order can be optimized for each $n$: if $d_{\alpha,m}(V\Vert P)=d_{1,m}(V\Vert P)+(\alpha-1)\vartheta+o(\alpha-1)$ with $\vartheta>0$, the order $\alpha^\star=1+\sqrt{\log(1/\gamma_{m,J})/(n\vartheta)}$ gives $P_{\mathrm b}^{(n)}\geq\exp\{-n\,d_{1,m}(V\Vert P)-2\sqrt{n\vartheta\log(1/\gamma_{m,J})}+o(\sqrt n)\}$.

\begin{remark}[Other normalized completions]
\label{rem:physical-completion}
The physical completion in \eqref{eq:canonical-partial-reference} can be combined with any selected factor: the kernel $B=\sum_r\prod_{j\leq m}a_r(x_j,y_j)\,P(v\mid x_{[J]},y_{[m]})$ satisfies condition (N), $\kappa_m(B)=\min_p\sum_r[\sum_xp(x)A_r(x)]^m$, and, whenever $B\ll P$, $\Psi_{\alpha,m}(B;P)$ involves only the $m$-user marginal $P_m$. Noninteracting references are the normalized one-component case.
\end{remark}

\paragraph{Common-output references}

Let $\mathcal O_m:=\{y\in\cY^J:y_1=\cdots=y_m\}$ be the set of output vectors on which the selected users agree. Let $\mathfrak C_{m,J}(P)$ be the set of kernels $B$ that satisfy
\begin{equation}
 B(y_{[J]}\mid x_{[J]})=\sum_{r\in\mathcal R}\ind\{y_{[J]}=\bar y_r\}\left[\prod_{j=1}^ma_r(x_j)\right]c_r(x_{m+1:J}) \,,
 \label{eq:partial-factorized-kernel}
\end{equation}
with $\mathcal R$ finite, an output vector $\bar y_r\in\mathcal O_m$ attached to each component $r$, nonnegative factors $a_r$ on $\cX$ and $c_r$ on $\cX^{J-m}$, and $B\ll P$. Such a kernel is supported on $\mathcal O_m$. Writing $\bar y_r=(\bar y_{r,1},\ldots,\bar y_{r,J})$, it has the form \eqref{eq:latent-factorized-kernel} with selected factor $a_r(x)\ind\{y=\bar y_{r,1}\}$ and unselected factor $c_r(u)\ind\{v=\bar y_{r,m+1:J}\}$. The unselected factor depends only on the unselected users, so the kernel satisfies (I). The physical completion of \Cref{rem:physical-completion} gives a second family, with the same selected factors and the unselected factor $P(v\mid x_{[J]},y_{[m]})$, which satisfies (N); the kernel of \Cref{prop:bandit-co-converse} is of this kind. For $m=J$ the two families coincide, and we call the kernels of both families \emph{common-output} kernels.

Under such a reference the selected users always receive identical outputs, so their views differ only through their private seeds. The reference retains only trajectories on which the channel gives the selected users no distinguishing information, and $\kappa_m$ measures how much physical probability these trajectories keep, uniformly over protocols.

Common-output kernels are admissible, so that they give at most $K_{\alpha,m}^{\mathrm{ref}}(P)$, and the inequality can be strict: for a noninteracting channel, $P(y_{[J]}\mid x_{[J]})=\prod_jV(y_j\mid x_j)$ with no $V(\cdot\mid x)$ a point mass, the reference $B_V=P$ gives $K_{\alpha,m}^{\mathrm{ref}}(P)=1$, whereas by \eqref{eq:upper-bound-K-all} every common-output kernel has $K_{\alpha,m}(B;P)\leq\max_{x_{[J]}}P(Y_1=\cdots=Y_m\mid x_{[J]})<1$.

\begin{remark}[Extension to leader election]
\label{rem:leader-election-converse}
Let each user output a bit $L_j$, and let $\cE_{\mathrm{LE}}:=\{\sum_jL_j\neq1\}$, whose probability under $\pi_n$ is $P_{\mathrm{LE}}^{(n)}(\pi_n;P)$; let $\overline E_{\mathrm{LE}}(P)$ be the upper exponent of $\inf_{\pi_n}P_{\mathrm{LE}}^{(n)}(\pi_n;P)$. Take $m=J$. Under any $J$-symmetric reference, the terminal bits are conditionally i.i.d.\ Bernoulli given $\Lambda$. Hence
\begin{equation}
 Q(\cE_{\mathrm{LE}})\geq
 \eta_J:=1-\max_{p\in[0,1]}Jp(1-p)^{J-1}
 =1-\left(1-\frac1J\right)^{J-1}>0.
 \label{eq:leader-election-reference-failure}
\end{equation}
The same proof gives, for every $B\in\mathfrak A_{J,J}(P)$ with $\kappa_J(B)>0$,
\begin{equation}
 P_{\mathrm{LE}}^{(n)}(\pi_n;P)\geq\eta_J^{b_\alpha}K_{\alpha,J}(B;P)^n,
 \qquad
 \overline E_{\mathrm{LE}}(P)\leq\inf_{\alpha\in(1,\infty]}[-\log K_{\alpha,J}^{\mathrm{ref}}(P)].
 \label{eq:leader-election-reference-converses}
\end{equation}
Unlike for symmetry breaking, where any $m\geq2$ selected users suffice, the reference must make all $J$ users conditionally i.i.d.: a proper subset can consist entirely of followers while an unselected user is the leader. For $J=2$, $\eta_2=\gamma_2=1/2$, and symmetry breaking and basic leader election are equivalent.
\end{remark}

\begin{proposition}[General population vectors]
\label{prop:population-converse}
Consider the population version of the task mentioned in \Cref{subsec:related-symmetry-breaking}: each user outputs $Z_j\in[L]$ through a common terminal map, and the task fails unless exactly $N_l$ users output $l$, for every $l\in[L]$. Let $P_{\bm N}^{(n)}(\pi_n;P)$ be the probability of failure, $N_{\max}:=\max_lN_l$, and, for $N_{\max}<m\leq J$,
\begin{equation}
 \gamma_m(\bm N):=\min_{p\in\mathcal P([L])}\PP_{p^{\otimes m}}\bigl\{\exists\,l:\ |\{j\leq m:Z_j=l\}|>N_l\bigr\}\geq L^{1-m}.
 \label{eq:population-prefactor}
\end{equation}
Then, for every such $m$, every $B\in\mathfrak A_{m,J}(P)$ with $\kappa_m(B)>0$, every strategy $\pi_n$ and every $\alpha\in(1,\infty]$,
\begin{equation}
 P_{\bm N}^{(n)}(\pi_n;P)\geq\gamma_m(\bm N)^{b_\alpha}K_{\alpha,m}(B;P)^n.
 \label{eq:population-converse}
\end{equation}
\end{proposition}
\begin{proof}
Under an $m$-symmetric reference, $Z_1,\ldots,Z_m$ are conditionally i.i.d.\ given $\Lambda$, with some law $p$ on $[L]$. Success requires that at most $N_l$ of them output $l$, for every $l$, so that the reference fails with probability at least $\gamma_m(\bm N)$. The event $Z_1=\cdots=Z_m$ has conditional probability $\sum_lp_l^m\geq L^{1-m}$ and implies failure, since $m>N_{\max}$, which proves the bound in \eqref{eq:population-prefactor}. The proofs of \Cref{prop:trajectory-converse} and \Cref{thm:latent-symmetry-meta-converse} then apply verbatim, with $\gamma_m(\bm N)$ in place of $\gamma_{m,J}$.
\end{proof}
The range of $m$ is sharp: for $m\leq N_{\max}$, the law that puts all its mass on a state $l$ with $N_l=N_{\max}$ gives zero in \eqref{eq:population-prefactor}, so the bound is vacuous. For $\bm N=(1,\ldots,1)$, $\gamma_m(\bm N)=\gamma_{m,J}$, which is \Cref{lem:reference-failure}, and for $\bm N=(1,J-1)$, only $m=J$ is allowed and $\gamma_J(\bm N)=\eta_J$, which is \Cref{rem:leader-election-converse}. The exponents of these bounds are those of symmetry breaking with $m>N_{\max}$ selected users, so that $\bm N$ affects only the prefactor and the range of $m$; with noninteracting references, for instance, the best exponent bound is $\inf_Vd_{1,N_{\max}+1}(V\Vert P)$, since $d_{1,m}$ is nondecreasing in $m$.

\section{{Achievability with Block Coding}}
\label{sec:block-coding}
\label{sec:multi-user}

In a \emph{block code}, each user transmits the codeword of the common codebook indexed by its private name and ignores its outputs until the end of the block. The outputs are used only in the terminal decision: user $j$ decodes the unordered set of the other names from its own codeword and its outputs, and ranks its own name as in \eqref{eq:rank-rule}.

\subsection{{The Block-Coding Exponent}}
\label{subsec:block-exponent}

The codebook is drawn from the i.i.d.\ ensemble \eqref{eq:iid-codebook-law} with input law $Q_X$. For $x\in\cX^k$ write $Q_X^k(x):=\prod_{i=1}^kQ_X(x_i)$.

\paragraph{Decoder}
Receiver $j$ knows its name $M_j$, hence its codeword $X_j^n=X^n(M_j)$. For an unordered set $\cA\subseteq[M]\setminus\{M_j\}$ of $J-1$ names, listed as $a_1,\ldots,a_{J-1}$, let
\begin{equation}
 L_j(\cA):=\prod_{i=1}^nW\bigl(Y_j[i]\bigm| X_j[i],(X[i](a_1),\ldots,X[i](a_{J-1}))\bigr).
 \label{eq:unordered-ml-likelihood}
\end{equation}
By \Cref{lem:single-local-law}, $L_j(\cA)$ does not depend on the order in which the elements of $\cA$ are listed, so it is a function of the unordered set. The receiver outputs $\widehat\cM_{-j}\in\argmax_\cA L_j(\cA)$, chosen by a fixed rule if the maximizer is not unique. The bounds below are for the event that some set other than the true one has a likelihood at least that of the true set, which contains the decoding error event. No physical user labels are attached to the decoded names.

\paragraph{{Error events}}
{A decoding error replaces the names of some nonempty set of foreign users while keeping the others (Appendix~\ref{app:multi-user-gallager} shows that every error is of this form).} For a receiver $j$, a subset $\cS\subseteq[J]\setminus\{j\}$ of size $s$, and $\rho\in[0,1]$, define
{\begin{equation}
	E_{0,s}(\rho,Q_X):=-\log\sum_{x_{\cS^c},\,y_j}Q_X^{J-s}(x_{\cS^c})
	\Biggl[\sum_{x_\cS}Q_X^s(x_\cS)\,W_j(y_j\mid x_{[J]})^{\frac1{1+\rho}}\Biggr]^{1+\rho},
	\label{eq:multi-user-ezero}
\end{equation}
where $\cS^c=[J]\setminus\cS$ contains the receiver $j$. By \Cref{lem:single-local-law}, the right-hand side} depends on $s$ but not on $j$ or the particular subset{. It is Gallager's function for the channel from $X_\cS$ to $(X_{\cS^c},Y_j)$ with i.i.d.~inputs: the receiver's own codeword and the correctly decoded names act as side information}.

Define
\begin{equation}
 E_{\mathrm r,s}(R,Q_X) := \max_{0\leq\rho\leq1} \{E_{0,s}(\rho,Q_X)-\rho sR\}.
 \label{eq:multi-user-random-exponent}
\end{equation}

\begin{theorem}[{Block-coding exponent}]
\label{thm:multi-user-gallager}
{Fix $J\geq2$ and $Q_X$.}
\begin{enumerate}[label=(\roman*)]
 \item For every receiver $j$, the ensemble-average {decoding error probability, conditioned on distinct names,} satisfies
 \begin{equation}
  \EE_{\cC}[P_{e,j}^{(n)}(\cC)] \leq \sum_{s=1}^{J-1}\binom{J-1}{s} \inf_{0\leq\rho\leq1} (M-J)^{s\rho}e^{-nE_{0,s}(\rho,Q_X)}.
  \label{eq:multi-user-gallager-bound}
 \end{equation}
 \item For $M=\lceil e^{nR}\rceil$, there exists a deterministic common codebook such that
 \begin{equation}
  P_{\mathrm{b}}^{(n)} \leq \binom{J}{2}e^{-nR+o(n)}+J\sum_{s=1}^{J-1}\binom{J-1}{s} e^{-nE_{\mathrm r,s}(R,Q_X)+o(n)}.
  \label{eq:multi-user-sync-bound}
 \end{equation}
 \item {Consequently, $\underline E_{\mathrm b}(P)\geq E_{\mathrm{bc}}(Q_X;P)$, where
 \begin{equation}
  E_{\mathrm{bc}}(Q_X;P):=\sup_{R>0}\min\Bigl\{R,\ \min_{1\leq s\leq J-1}E_{\mathrm r,s}(R,Q_X)\Bigr\}.
  \label{eq:multi-user-sync-exponent}
 \end{equation}}
\end{enumerate}
\end{theorem}

The proof is given in Appendix~\ref{app:multi-user-gallager}. {In \eqref{eq:multi-user-gallager-bound},} the factor $\binom{J-1}{s}$ chooses the {users whose names are replaced}, while $(M-J)^s$ upper bounds the number of {competing unordered sets. The exponent \eqref{eq:multi-user-sync-exponent} is a minimum of two competing terms. Duplicate names occur with probability of order $e^{-nR}$, which favors a large name rate. Longer names are harder to decode, since $E_{\mathrm r,s}(R,Q_X)$ decreases in $R$, which favors a small one. The name rate is therefore a design parameter to be optimized.} As in the random-coding exponents of the multiple-access channel \cite{gallager1985perspective,nazari2014error}, \eqref{eq:multi-user-gallager-bound} is a union bound over the sets of wrongly decoded users; here the receiver decodes an unordered set of names, with its own codeword as side information. The exponent $E_{\mathrm{bc}}(Q_X;P)$ is achieved by one ensemble with one decoder; it is not an upper bound for block codes.

For $J=2$ the only error event is $s=1$, and with $\{j,j'\}=[2]$, part (i) is the usual Gallager bound $\EE_\cC[P_{e,j}^{(n)}(\cC)]\leq(M-2)^\rho e^{-nE_{0,1}(\rho,Q_X)}$, with the receiver's own input as side information; the derivative of $E_{0,1}$ at $\rho=0$ is $I_Q(X_{j'};Y_j\mid X_j)$, the rate of Shannon's independent-input inner bound for the two-way channel \cite{shannon_two-way_1961}.

\subsection{{Condition for a Positive Block-Coding Exponent}}
\label{subsec:rate-threshold}

For a nonempty subset $\cS\subseteq[J]\setminus\{j\}$ {with $|\cS|=s$ and i.i.d.\ inputs $X_1,\ldots,X_J$ with law $Q_X$}, define
\begin{equation}
 {I_s(Q_X)} := I_Q(X_{\cS};Y_j\mid X_{\cS^c}),
 \label{eq:subset-mutual-information}
\end{equation}
where $\cS^c:=[J]\setminus\cS$ includes the receiver input $X_j$. {By user symmetry,} the value depends only on $s$.

Differentiating \eqref{eq:multi-user-ezero} gives
\begin{equation}
 \left.\frac{\partial E_{0,s}(\rho,Q_X)}{\partial\rho}\right|_{\rho=0} =I_s(Q_X),
 \label{eq:multi-ezero-derivative}
\end{equation}
{and $E_{0,s}(\cdot,Q_X)$ is concave with $E_{0,s}(0,Q_X)=0$ \cite{gallager1968}. Hence $E_{\mathrm r,s}(R,Q_X)>0$ if and only if $sR<I_s(Q_X)$. The name rates at which every decoding exponent is positive are therefore those below
\begin{equation}
 R_{\mathrm{th}}(Q_X):=\min_{1\leq s\leq J-1}\frac{I_s(Q_X)}{s}.
 \label{eq:multi-user-symmetric-inner}
\end{equation}
This is also the symmetric rate of the all-source multicast inner bound: the constraints $sR<I_s(Q_X)$ are those of a multiple-access channel with the receiver's input as side information.} The following lemma eliminates the minimization over subset sizes.

\begin{lemma}[Normalized subset information is nonincreasing]
\label{lem:normalized-information-decreasing}
Suppose $X_1,\ldots,X_m$ are i.i.d.\ and independent of $T$, and the conditional law of $Y$ given $(X_1,\ldots,X_m,T)$ is invariant under permutations of the $m$ inputs. Then
\begin{equation}
 \frac1s I(X_{[s]};Y\mid X_{[m]\setminus[s]},T)
 \label{eq:normalized-information-sequence}
\end{equation}
is nonincreasing in $s=1,\ldots,m$.
\end{lemma}

A complete entropy proof is given in Appendix~\ref{app:normalized-information}.

\begin{proposition}[{Rate threshold of block coding}]
\label{thm:multi-user-inner}
\label{cor:full-set-bottleneck}
For every symmetric channel and every $Q_X$,
\begin{equation}
 {R_{\mathrm{th}}(Q_X)} = \frac{1}{J-1} I_Q(X_{-j};Y_j\mid X_j),
 \label{eq:full-set-bottleneck}
\end{equation}
{and
\begin{equation}
 E_{\mathrm{bc}}(Q_X;P)=\min_{1\leq s\leq J-1}\max_{0\leq\rho\leq1}\frac{E_{0,s}(\rho,Q_X)}{1+s\rho}\leq\frac1J I_Q(X_{-j};Y_j\mid X_j).
 \label{eq:block-exponent-threshold}
\end{equation}
Moreover, $E_{\mathrm{bc}}(Q_X;P)>0$ if and only if $I_Q(X_{-j};Y_j\mid X_j)>0$.}
\end{proposition}
\begin{proof}
The equality \eqref{eq:full-set-bottleneck} is \Cref{lem:normalized-information-decreasing} with $m=J-1$ and $T=X_j$. For each $s$, $E_{\mathrm r,s}(R,Q_X)\geq R$ if and only if $R(1+s\rho)\leq E_{0,s}(\rho,Q_X)$ for some $\rho\in[0,1]$, that is, if and only if $R\leq R_s:=\max_\rho E_{0,s}(\rho,Q_X)/(1+s\rho)$; moreover $E_{\mathrm r,s}(R_s,Q_X)=R_s$ by continuity, and $E_{\mathrm r,s}(\cdot,Q_X)$ is nonincreasing. Hence $\min\{R,\min_sE_{\mathrm r,s}(R,Q_X)\}$ equals $R$ for $R\leq\min_sR_s$ and is at most $\min_sR_s$ otherwise, which gives the equality in \eqref{eq:block-exponent-threshold}. Concavity gives $E_{0,s}(\rho,Q_X)\leq\rho I_s(Q_X)$, so that $R_s\leq I_s(Q_X)/(1+s)$, and $s=J-1$ gives the inequality. Conversely, suppose $R_{\mathrm{th}}(Q_X)>0$ and fix $0<R<R_{\mathrm{th}}(Q_X)$. Every function $\rho\mapsto E_{0,s}(\rho,Q_X)-\rho sR$ vanishes at $\rho=0$ with slope $I_s(Q_X)-sR>0$. Hence every $E_{\mathrm r,s}(R,Q_X)$ is positive, and so is the minimum.
\end{proof}

This is the role of the symmetric rate in symmetry breaking: it decides whether block coding with random codes breaks the symmetry at an exponential rate, and it caps the resulting exponent, which is even at most $\frac{J-1}{J}R_{\mathrm{th}}(Q_X)$. Since $I_Q(X_{-j};Y_j\mid X_j)\leq\log|\cY|$, the random-coding exponent is at most $\log|\cY|/J$; on the bandit channel of \Cref{sec:erasure} with two users, two arms of mean one and $\bm\eps=\bm0$, both sides equal $\frac12\log2$. The full set dominates the rate threshold, but not necessarily the exponent at the optimized rate: the minimizing $s$ in \eqref{eq:block-exponent-threshold} can be smaller than $J-1$; for three users and three arms of mean one with the uniform law, $s=1$ gives $0.176$ and $s=2$ gives $0.227$ (the former is the entry of \Cref{tab:gmc-users}).

The proof also applies to unequal message rates and independent user-specific input laws. The resulting familiar multiple-access constraints are $\sum_{k\in\cS}R_k<I(X_{\cS};Y_j\mid X_{\cS^c})$ for every $j$ and every nonempty $\cS\subseteq[J]\setminus\{j\}$. In the anonymous shared-codebook construction, anonymity, that is, the use of identical strategies, requires every user to employ the same name set and the same codeword ensemble. We therefore specialize to $R_k=R$ and a common marginal input law $Q_{X_k}=Q_X$ for all $k$. This is a restriction of the present random-coding construction, not a claim that every anonymous protocol must have independent or identically distributed inputs; more general exchangeable input processes remain possible.

\subsection{{Limits of Name Decoding}}
\label{subsec:name-decoding-limits}

The ranking scheme \eqref{eq:rank-rule}, as implemented in this section, has every user recover all other names. A cut-set argument bounds the name rate of any such scheme, even an adaptive one.

\begin{proposition}[Multiway cut-set bound]
\label{prop:multi-user-cutset}
Suppose each user draws an independent uniform name from a set of size $\lceil e^{nR}\rceil$ and, using an arbitrary causal protocol, every user recovers the multiset of all names with an error probability that tends to zero as $n\to\infty$. Then there is a joint input law $P_{X_{[J]}}$ on $\cX^J$ such that, for every nonempty proper subset $\cS\subset[J]$,
\begin{equation}
 |\cS|R \leq I(X_{\cS};Y_{\cS^c}\mid X_{\cS^c}),
 \label{eq:multiway-cutset}
\end{equation}
where $Y_{[J]}$ is the output of the channel for the input $X_{[J]}\sim P_{X_{[J]}}$. In particular, choosing $\cS=[J]\setminus\{j\}$ yields
\begin{equation}
 (J-1)R \leq I(X_{-j};Y_j\mid X_j) \leq H(Y_j).
 \label{eq:single-receiver-cutset}
\end{equation}
Since $H(Y_j)\leq\log|\cY|$,
\begin{equation}
 R\leq\frac{\log|\cY|}{J-1}.
 \label{eq:alphabet-cutset}
\end{equation}
\end{proposition}

The bound \eqref{eq:multiway-cutset} is the standard cut-set bound for memoryless networks \cite{el_gamal_network_2012}; a proof specialized to the present model, with common and private randomness and the decoding of a multiset, is included in Appendix~\ref{app:cutset}. For binary outputs, \eqref{eq:alphabet-cutset} is $R\leq\log2/(J-1)$.

Call a scheme \emph{name-decoding at rate $R$} if every user draws an independent uniform name from a set of size~$M:=\lceil e^{nR}\rceil$, forms by an arbitrary causal protocol an estimate of the multiset of the $J-1$ other names, and outputs one plus the number of names in this estimate that are smaller than its own, as in \eqref{eq:rank-rule}. Let $\cE_{\mathrm{col}}$ be the event that two names coincide, $\cE_{\mathrm{dec}}$ the event that the estimate of some user is wrong, and $\cN_n:=\cE_{\mathrm{col}}\cup\cE_{\mathrm{dec}}$. When every estimate is correct, the ranks are a permutation of $[J]$ if the names are distinct, and two users with the same name obtain the same rank otherwise. Hence the symmetry-breaking error event $\cE$ satisfies
\begin{equation}
 \cE_{\mathrm{col}}\cap\cE_{\mathrm{dec}}^{\mathrm c}\subseteq\cE\subseteq\cE_{\mathrm{col}}\cup\cE_{\mathrm{dec}}=\cN_n,
 \label{eq:name-decoding-sandwich}
\end{equation}
and in particular $P_{\mathrm b}^{(n)}\leq\PP(\cN_n)$. We call $\liminf_{n\to\infty}-\frac1n\log\PP(\cN_n)$ the \emph{name-decoding exponent} of the scheme. It is a lower bound on the error exponent of the scheme, and it is the exponent that an analysis through $\cN_n$ can prove; for block coding, the right-hand side of \eqref{eq:sync-decomposition} is an upper bound on $\PP(\cN_n)$.
\begin{corollary}[Ceiling on the name-decoding exponent]
\label{cor:name-decoding-cap}
For every sequence of name-decoding schemes at rate $R$,
\begin{equation}
 \limsup_{n\to\infty}-\frac1n\log\PP(\cN_n)\leq\frac1{J-1}\max_{P_{X_{[J]}}}I(X_{-j};Y_j\mid X_j)\leq\frac{\log|\cY|}{J-1}.
 \label{eq:name-decoding-cap}
\end{equation}
\end{corollary}
\begin{proof}
Two names coincide with probability $P_{\mathrm{col}}(M,J)=\gamma_{J,M}\geq1/M$, so that $-\frac1n\log\PP(\cN_n)\leq\frac1n\log\lceil e^{nR}\rceil\to R$. If the left-hand side of \eqref{eq:name-decoding-cap} is positive, then $\PP(\cN_n)\to0$ along a subsequence, along which every user recovers the multiset of all names with vanishing error probability; \eqref{eq:single-receiver-cutset}, applied to this subsequence, then gives $(J-1)R\leq\max_{P_{X_{[J]}}}I(X_{-j};Y_j\mid X_j)$, and the left-hand side is at most $R$.
\end{proof}
The two inclusions in \eqref{eq:name-decoding-sandwich} can be far apart: the error probability $P_{\mathrm b}^{(n)}$ of a name-decoding scheme can be much smaller than $\PP(\cN_n)$, since two users with the same name need not receive the same rank when their estimates are wrong. 

However, for channels with conditionally independent outputs, we have
\begin{proposition}[{Nonadaptive strategies with conditionally independent outputs}]
\label{prop:name-collision}
Suppose that the outputs are conditionally independent given the inputs,
\begin{equation}
 P(y_{[J]}\mid x_{[J]})=\prod_{j=1}^JW(y_j\mid x_j,x_{-j}) \,.
 \label{eq:cond-indep-outputs}
\end{equation}
Then every nonadaptive strategy satisfies
\begin{equation}
 P_{\mathrm b}^{(n)}\geq\gamma_{2,J}\,\PP\{\exists\,j\neq k:X_j^n=X_k^n\} \,.
 \label{eq:codeword-collision-bound}
\end{equation}
In particular, every block code with $M$ uniform names, with a fixed or random codebook and any terminal map, satisfies $P_{\mathrm b}^{(n)}\geq P_{\mathrm{col}}(M,J)/J\geq1/(JM)$, and, for $M=\lceil e^{nR}\rceil$, $\limsup_{n\to\infty}-\frac1n\log P_{\mathrm b}^{(n)}\leq R$.
\end{proposition}
\begin{proof}[Proof sketch]
Two users that send the same codeword have equal inputs in every slot. Under \eqref{eq:cond-indep-outputs} they then receive i.i.d.\ outputs, by \Cref{lem:single-local-law}, and their seeds are i.i.d.\ given the codeword. Their identifiers are thus i.i.d., and they coincide with probability at least $\gamma_{2,J}$ by \Cref{lem:iid-identifier-collision}. Equal names give equal codewords, and two names coincide with probability $P_{\mathrm{col}}(M,J)\geq1/M$, see \eqref{eq:birthday-bound}.
\end{proof}
The noisy bandit channel of \Cref{def:noisy-bandit} satisfies \eqref{eq:cond-indep-outputs}, so that on it the error exponent of a block code at rate $R$ is at most $R$, whatever its decoder, and at most the right-hand side of \eqref{eq:name-decoding-cap} if $\PP(\cN_n)\to0$, by \Cref{prop:multi-user-cutset}.

Symmetry breaking itself is not limited in this way. For $J=2$, a channel whose output pair is uniform on $\{(0,1),(1,0)\}$, whatever the inputs, breaks the symmetry in a single use although it carries no information, even if all users send the same codeword, so that \Cref{prop:name-collision} needs conditional independence.

\begin{remark}[Relation to noisy network coding]
For the corresponding labeled network, noisy network coding can exploit causal relaying and may yield a stronger interactive inner bound \cite{lim2011noisy}, but its standard construction uses node-specific codebooks and compression indices, whereas our users share one encoding rule and decode unordered sets of names. Adapting it to the anonymous setting is left for future work. Also note that \Cref{prop:name-collision} only holds for nonadaptive codes.
\end{remark}

\section{Achievability with Interactive Role Controllers}
\label{sec:interactive-coding}
\label{sec:role-controllers}

Block coding gives every user a name from a very large space at time zero. Conflicts between names are never detected; they are only made rare, at the price of the collision term $e^{-nR}$, and the names are then ordered by decoding. An alternative approach lets the users contend directly for the identifiers in $[J]$ and resolves conflicts through the channel. A user keeps its tentative identifier while its observations suggest that nobody else holds it, and draws a new one otherwise; musical chairs is the simplest rule of this kind. This section develops the second approach: \Cref{subsec:role-controllers,subsec:confidence-general} define role controllers and confidence controllers, \Cref{subsec:certified-completion} bounds the exponent of finite-state controllers that use certifying outputs, and \Cref{subsec:gmc} introduces generalized musical chairs.

\subsection{Role Controllers}
\label{subsec:role-controllers}
\label{subsec:persistent-state-coding}

\begin{definition}[Role controller]
\label{def:role-controller}

A \emph{role controller} consists of a countable state set $ \mathcal T$, an initial state $t_0 \in \mathcal T$, a law $\omega$ on a finite set $\mathcal U$, maps $\phi_i:\mathcal T\times\mathcal U\to\cX$ and $\psi_i: \mathcal T\times\mathcal U\times\cY\to\mathcal T$ for $i\in[n]$, and a terminal map $d_n: \mathcal T\to[J]$. User $j$ draws $U_j[1],\ldots,U_j[n]$ i.i.d.\ with law $\omega$, independently of the other users; that is, its private seed is $S_j=(U_j[1],\ldots,U_j[n])$. It sets $ T_j[0]= t_0$ and applies
\begin{equation}
 X_j[i]=\phi_i(T_j[i-1],U_j[i]),\qquad
T_j[i]=\psi_i(T_j[i-1],U_j[i],Y_j[i]),\qquad
 Z_j=d_n(T_j[n]).
 \label{eq:persistent-state-controller}
\end{equation}
\end{definition}

All the elements of the definition may depend on $n$. More generally, every user may draw its initial state from a law $\pi_0$ on $\mathcal T$ with its private randomness; then $\mathbf e_{\bm t_0}^{\mathsf T}$ is replaced by $\pi_0^{\otimes J}$ below. Since the randomization is private and every user applies the same maps, a role controller is an anonymous symmetric causal strategy in $\Pi_n$. We call $d_n(T_j[i])\in[J]$ the \emph{tentative role} of user $j$ at time $i$: the identifier it would output if the block ended then. The state contains the tentative role together with whatever else the user remembers. The final identifiers collide exactly when two users hold the same role at time $n$. For example, musical chairs, which is treated in \Cref{sec:erasure}, is a role controller in which a user that has not yet found a free role draws a role uniformly at random in every channel use and keeps it for good once it finds it free; the identifier of a user that is still searching at time $n$ is the last role it drew.

Because each $U_j[i]$ is used only at time $i$ and the channel is memoryless, the joint state $\bm T[i]=(T_1[i],\ldots,T_J[i])$ is a Markov chain with transition matrix
\begin{equation}
 K_i(\bm t,\bm t')
 =\sum_{u_{[J]}}\left(\prod_{j=1}^J\omega(u_j)\right)
 \sum_{y_{[J]}}P\bigl(y_{[J]}\bigm|\phi_i(t_1,u_1),\ldots,\phi_i(t_J,u_J)\bigr)
 \prod_{j=1}^J\ind\{\psi_i(t_j,u_j,y_j)=t_j'\}.
 \label{eq:persistent-state-transition}
\end{equation}
Let
\begin{equation}
 \mathcal F:=\{\bm t\in\mathcal T^J:d_n(t_j)=d_n(t_k)\text{ for some }j\neq k\}
\end{equation}
be the set of failure states.

\begin{proposition}[Exact error probability]
\label{prop:role-controller-exact}
For every role controller,
\begin{equation}
 P_{\mathrm{b}}^{(n)}=\mathbf e_{\bm t_0}^{\mathsf T}K_1K_2\cdots K_n\mathbf 1_{\mathcal F},
 \label{eq:role-controller-exact}
\end{equation}
where $\bm t_0=(t_0,\ldots,t_0)$ and $\mathbf 1_{\mathcal F}$ is the indicator vector of $\mathcal F$.
\end{proposition}

\begin{proposition}[{Symmetric causal strategies are role controllers}]
\label{prop:block-coding-controller}
For every strategy $\pi_n\in\Pi_n$ and every value $c$ of the common randomness, there is a role controller whose inputs, outputs and identifiers have the same joint law as those of $\pi_n$ given $C=c$. In particular, the infimum of $P_{\mathrm b}^{(n)}$ over role controllers is $\epsilon_n^\star(P)$, and every block code of \Cref{sec:block-coding} is a role controller, so that, for every $Q_X$, role controllers achieve the exponent $E_{\mathrm{bc}}(Q_X;P)$ of \eqref{eq:multi-user-sync-exponent}, and this exponent is positive whenever $I_Q(X_{-j};Y_j\mid X_j)>0$.
\end{proposition}
\begin{proof}
Given $C=c$, a seed $s$ enters only through the maps $f_1(c,s,\cdot),\ldots,f_n(c,s,\cdot)$ and $g_n(c,s,\cdot)$, which range over a finite set since the alphabets are finite, so that we may assume that the seed takes values in a finite set $\mathcal S$. Take $\mathcal U=\mathcal S$ with $\omega=P_S$, of which only $U[1]$ is used, as the seed of the user, and $\mathcal T=\{t_0\}\cup\bigcup_{i\leq n}\bigl(\mathcal S\times\cY^i\bigr)$. Put $\phi_1(t_0,u)=f_1(c,u)$ and $\psi_1(t_0,u,y)=(u,y)$; for $i\geq2$, put $\phi_i((s,y^{i-1}),u)=f_i(c,s,y^{i-1})$ and $\psi_i((s,y^{i-1}),u,y)=(s,(y^{i-1},y))$; and let $d_n(s,y^n)=g_n(c,s,y^n)$. The joint law is then that of $\pi_n$ given $C=c$, and since $P_{\mathrm b}^{(n)}(\pi_n;P)$ averages over $C$, some $c$ does at least as well as $\pi_n$. A block code is the strategy whose seed is the name $m$, with $f_i(c,m,y^{i-1})=x[i](m)$ and the rank rule \eqref{eq:rank-rule} as terminal map. The exponent is \Cref{thm:multi-user-gallager}, which provides a deterministic common codebook, and its positivity is \Cref{thm:multi-user-inner}.
\end{proof}

The state of this role controller contains the whole output sequence, and by \Cref{prop:finite-controller-floor} below, a finite-state controller cannot match it when every output has positive probability. The mutual-information condition in \Cref{prop:block-coding-controller} is sufficient but not necessary for a positive exponent: on the channel of \Cref{subsec:name-decoding-limits} that carries no information, a role controller that outputs one plus its first output bit as its identifier has zero error.

\begin{definition}[Stationary and finite-state controllers]
\label{def:finite-state-controller}
A role controller is \emph{stationary} if $\phi_i=\phi$ and $\psi_i=\psi$ for all $i\in[n]$. Its transition matrix is then a single matrix $K$, which may depend on $n$, and \eqref{eq:role-controller-exact} reads $P_{\mathrm b}^{(n)}=\mathbf e_{\bm t_0}^{\mathsf T}K^n\mathbf 1_{\mathcal F}$. A \emph{finite-state controller} is a stationary controller with a finite state set $\mathcal T$ such that $\mathcal T$, $t_0$ (or the initial law $\pi_0$), $\omega$, $\phi$, $\psi$ and $d_n=d$ do not depend on $n$: the same controller, with the same matrix $K$, is run at every blocklength.
\end{definition}

\begin{remark}[Time in the state]
\label{rem:clock-in-state}
Stationarity alone does not restrict the class: any role controller becomes stationary once the time index is stored in its state, with $\mathcal T\times\{0,\ldots,n\}$ in place of $\mathcal T$. In block coding, for instance, the clock is already in the state $(m,y^i)$, since $i$ is the length of $y^i$. What a finite-state controller excludes is this clock: a finite state set that does not depend on $n$ cannot count the channel uses. For the same reason, the finite-memory theory of hypothesis testing and of the two-armed bandit is restricted to time-invariant machines~\cite{hellman1970,cover1970}: with maps that vary with time, the time index serves as a free clock, and four states then suffice for an error probability that tends to zero in hypothesis testing~\cite{cover1969}, as $4J$ states do on the noisy bandit channel~(\Cref{app:bounded-memory}); see \cite{berg2023survey} for a recent survey. The memory $\log|\mathcal T|$ of a user grows linearly in $n$ for block coding, logarithmically for confidence controllers with thresholds of order $n$ (\Cref{subsec:confidence-general}), and stays bounded for generalized musical chairs (\Cref{subsec:gmc}).
\end{remark}

If every output has positive probability, or more generally if the possible outputs of a user do not depend on what the others play or observe, no finite-state controller has a vanishing error probability (\Cref{prop:finite-controller-floor}). To drive the error probability to zero, a finite-state controller needs an output that breaks this condition, such as a certificate, an observation that is impossible when the role is shared (\Cref{subsec:certified-completion}).

\begin{proposition}[Floor for finite-state controllers]
\label{prop:finite-controller-floor}
Let the controller be a finite-state controller, let $\mathcal X_0$ be the set of inputs that it plays with positive probability, $\mathcal X_0=\{\phi(t,u):t \in \mathcal T,\,\omega(u)>0\}$, and suppose that there are sets $\mathcal S(x)\subseteq\cY$, $x\in\mathcal X_0$, such that for all $x_{[J]}\in\mathcal X_0^J$, $P(y_{[J]}\mid x_{[J]})>0$ if and only if $y_j\in\mathcal S(x_j)$ for every $j$; this holds in particular if $P(y_{[J]}\mid x_{[J]})>0$ for all $x_{[J]}\in\mathcal X_0^J$ and all $y_{[J]}$.
Then the controller satisfies $\liminf_{n\to\infty}P_{\mathrm{b}}^{(n)}>0$.
\end{proposition}

\begin{proof}
If the users draw their initial state from a law $\pi_0$, they all start at a common state $t_0$ with $\pi_0(t_0)>0$ with probability $\pi_0(t_0)^J>0$, and we argue from there. Call $t'$ a successor of $t$ if $t'=\psi(t,u,y)$ with $\omega(u)>0$ and $y\in\mathcal S(\phi(t,u))$. Since the seeds are independent and the supports have product form, a joint move has positive probability if and only if every user makes a successor step; hence the joint states reachable in $k$ steps are the products of the local states reachable in $k$ steps, and from $(t,\ldots,t)$ all users make the same move with probability at least some $\eta>0$, which leads to a failure state. Let $a$ be reachable from $t_0$ with the smallest set of reachable states, counting $a$ itself, which is reachable from itself in zero steps; by minimality, one can show that $a$ is reachable from every state reachable from $a$. If $(b_1,\ldots,b_J)$ is reachable from $(a,\ldots,a)$ in $k$ steps and $b_i$ reaches $a$ in $\ell_i$ steps, then every $b_j$ reaches $a$ in exactly $L=\sum_i\ell_i+(J-1)k$ steps, through the loops $a\to b_i\to a$, $i\neq j$. Hence, from every joint state reachable from $(a,\ldots,a)$, the users return to $(a,\ldots,a)$ within some $L^\star$ steps with probability at least some $p>0$.

Since $a$ is reachable from $t_0$, say in $k_0$ steps, the users are at $(a,\ldots,a)$ at time $k_0$ with probability at least some $q>0$, by following this path together. Let $n\geq k_0+L^\star$. On this event, the joint state at time $n-L^\star\geq k_0$ is reachable from $(a,\ldots,a)$. From there, the users return to $(a,\ldots,a)$ within $L^\star$ steps with probability at least $p$, and then make identical moves until time $n$, at most $L^\star$ of them, so that the joint state at time $n$ is a failure state with probability at least $\eta^{L^\star}$. Hence $P_{\mathrm b}^{(n)}\geq qp\,\eta^{L^\star}$ for all $n\geq k_0+L^\star$.
\end{proof}

\Cref{prop:finite-controller-floor} uses the finiteness of $\mathcal T$, stationarity, and the independence from $n$.
Without stationarity, a state set of bounded size can suffice for a vanishing error probability: \Cref{app:bounded-memory} gives such a controller with $4J$ states on the noisy bandit channel, by a device of Cover~\cite{cover1969}, but its error probability decays only subexponentially. Whether such a controller can achieve a positive exponent when every output has positive probability is open.

\subsection{Confidence Controllers}
\label{subsec:confidence-general}

By \Cref{prop:finite-controller-floor}, on a channel on which every output has positive probability, a controller with a vanishing error probability is not a finite-state controller: its state set is infinite, its maps vary with time, or it depends on $n$. Block coding stores the whole output sequence. A more economical summary of the observations is a user's confidence in its current role: a counter that goes up when the observations suggest that the role is held alone, and down when they suggest that it is shared.

\begin{definition}[Confidence controller]
\label{def:confidence-controller}
A confidence controller is specified by the following data.
\begin{itemize}
 \item For every role $a\in[J]$: an input law $Q_a$ on $\cX$ for a user that is still searching, an input law $Q_a^\star$ on $\cX$ for a user that has committed (by default $Q_a^\star=Q_a$), a score $s_a(x,y)\in\mathbb Z\cup\{ \pm \infty\}$ of an input-output pair, and thresholds $\xi_+^a\in\{0,1,2,\ldots\}\cup\{\infty\}$ and $\xi_-^a\in\{1,2,\ldots\}$.
 \item A redraw kernel $R(a,\cdot)$ on $[J]$ for every $a\in[J]$, and an initial law $\pi_0$ on $[J]$.
\end{itemize}
The state of a user is its role $a\in[J]$ together with an integer counter $\xi\in\{-\xi_-^a+1,\ldots,\xi_+^a\}$; initially $a\sim\pi_0$ and $\xi=0$. In every channel use, a user in state $(a,\xi)$ proceeds as follows, all draws being made with fresh private randomness.
\begin{enumerate}
 \item It draws $X\sim Q_a$ if it is not committed and $X\sim Q_a^\star$ if it is, transmits $X$ and observes $Y$.
 \item If $\xi<\xi_+^a$, it updates $\xi\leftarrow\min\{\xi+s_a(X,Y),\,\xi_+^a\}$, with $\xi+\infty:=\infty$ and $\xi-\infty:=-\infty$. A user with $\xi=\xi_+^a$ is \emph{committed}: it keeps its state until time $n$, and transmits with the law $Q_a^\star$.
 \item If $\xi\leq-\xi_-^a$, it \emph{releases} its role: it draws a new role $a'\sim R(a,\cdot)$ and resets $\xi=0$.
\end{enumerate}
After $n$ channel uses, the user outputs its role $a$ as its identifier.
\end{definition}
Since every draw uses fresh private randomness, a confidence controller is a role controller in the sense of \Cref{def:role-controller}, with state $(a,\xi)$ and tentative role $d_n(a,\xi)=a$, and with the initial role drawn from $\pi_0$, as allowed after \Cref{def:role-controller}. Let $N\bigl(\cdot\mid(a,\xi),x,y\bigr)$ be the law of the next state of a user in state $(a,\xi)$ that transmitted $x$ and observed $y$. If $\xi=\xi_+^a$, the state does not change. Otherwise, put $\xi^+:=\min\{\xi+s_a(x,y),\xi_+^a\}$: the next state is $(a,\xi^+)$ if $\xi^+>-\xi_-^a$, and $(a',0)$ with probability $R(a,a')$ if $\xi^+\leq-\xi_-^a$. A user in state $(a,\xi)$ transmits with the law $Q_a^{\langle\xi\rangle}$, where $Q_a^{\langle\xi\rangle}:=Q_a$ if $\xi<\xi_+^a$ and $Q_a^{\langle\xi\rangle}:=Q_a^\star$ if $\xi=\xi_+^a$.

\begin{proposition}[Exact error probability of a confidence controller]
\label{prop:confidence-exact}
The joint state $\bm t=(t_1,\ldots,t_J)$, $t_j=(a_j,\xi_j)$, of the $J$ users is a Markov chain with initial law $\mu_0=(\pi_0\otimes\delta_0)^{\otimes J}$ and transition matrix
\begin{equation}
 K(\bm t,\bm t')=\sum_{x_{[J]},y_{[J]}}\prod_{j=1}^JQ_{a_j}^{\langle\xi_j\rangle}(x_j)\;P\bigl(y_{[J]}\bigm|x_{[J]}\bigr)\prod_{j=1}^JN\bigl(t_j'\bigm|t_j,x_j,y_j\bigr).
 \label{eq:confidence-transition}
\end{equation}
With $\mathcal F$ the set of joint states in which two users hold the same role,
\begin{equation}
 P_{\mathrm{b}}^{(n)}=\mu_0K^n\mathbf 1_{\mathcal F}.
 \label{eq:confidence-exact}
\end{equation}
\end{proposition}
\begin{proof}
This is \Cref{prop:role-controller-exact}
for the stationary role controller with the states $(a,\xi)$, with the initial state drawn from $\pi_0\otimes\delta_0$.
\end{proof}

For $J=2$ and thresholds of order $n$, the chain has $O(n^2)$ states, and the transitions of the counters are sparse, so \eqref{eq:confidence-exact} can be evaluated exactly for moderate $n$. Its transitions are simple: the counters move by the scores, and a release redraws a role. The state of a user takes $O(\log n)$ bits whenever the thresholds are $O(n)$.

A score $+\infty$ commits the user at once. It is meant for outputs that cannot occur when the role is shared, that is, for a certificate; combined with finite scores on the other outputs, it lets a controller use a certificate and soft evidence at the same time. The score is meant to be positive on average when the role is held alone and negative on average when it is shared, so that the counter of a user in a shared role drifts down until it releases, and the counter of a user in an unshared role drifts up until it commits. A typical score is a quantized log-likelihood ratio between the two situations. The input law $Q_a$ makes the role informative: a user in role $a$ must send something whose effect on its own output depends on whether another user holds the same role.

The class contains the following schemes.
\begin{itemize}
 \item \emph{Uniform confidence controllers.} Fixed inputs $Q_a=\delta_{\chi_a}$ with $\chi_a\in\cX$, the same parameters for every role, the uniform kernel $R(a,\cdot)=\mathrm{Unif}([J])$ and $\pi_0=\mathrm{Unif}([J])$, with large thresholds; \Cref{sec:erasure} evaluates such controllers, together with optimized ones that use other kernels and initial laws.
 \item \emph{Musical chairs} and its one-arm variant are also confidence controllers~(\Cref{sec:erasure}).
\end{itemize}

The thresholds determine the behavior.
\begin{itemize}
 \item With $\xi_+^a$ finite and independent of $n$, this is a finite-state controller, subject to \Cref{prop:finite-controller-floor}. If some output certifies that a role is not shared, a finite-state controller can nevertheless work, and \Cref{prop:persistent-state-exponent} gives an achievable exponent, namely that of the survival probability; musical chairs is of this kind.
 \item With $\xi_+^a$ growing linearly in $n$, the controller is stationary but not finite-state: the counter needs only $O(\log n)$ bits and is not subject to the floor; \Cref{sec:erasure} evaluates such controllers on the noisy bandit channel.
 \item With $\xi_+^a=\infty$, the controller is stationary and does not depend on $n$, but its state set is infinite, so that \Cref{prop:finite-controller-floor} does not apply either.
\end{itemize}

\subsection{Certified Completion}
\label{subsec:certified-completion}

Some channels have \emph{certifying} outputs, which are impossible when the role is shared. A user that receives one knows that it holds its role alone and can keep it for good.
A finite-state controller then completes with an error exponent of at least $-\log$ of a spectral radius.

\begin{proposition}[Certified completion]
\label{prop:persistent-state-exponent}
Let the controller be a finite-state controller; with an initial law $\pi_0$, read $\pi_0^{\otimes J}$ for $\mathbf e_{\bm t_0}^{\mathsf T}$ and the support of $\pi_0^{\otimes J}$ for $\bm t_0$ below. Let $\mathcal G\subseteq \mathcal T^J\setminus\mathcal F$ be absorbing, that is, $K(\bm t,\mathcal G)=1$ for $\bm t\in\mathcal G$, with $\bm t_0\notin\mathcal G$. Let $\mathcal B$ be the set of states reachable from $\bm t_0$ that are not in $\mathcal G$, and $K_{\mathcal B}$ the corresponding substochastic submatrix. Then
\begin{equation}
 P_{\mathrm{b}}^{(n)}\leq\mathbf e_{\bm t_0}^{\mathsf T}K_{\mathcal B}^{\,n}\mathbf 1.
 \label{eq:persistent-state-error-bound}
\end{equation}
If $0<\varrho:=\rho(K_{\mathcal B})<1$, where $\rho$ denotes the spectral radius, the right-hand side decays with exponent exactly $-\log\varrho$:
\begin{equation}
 \lim_{n\to\infty}-\frac1n\log\left(\mathbf e_{\bm t_0}^{\mathsf T}K_{\mathcal B}^{\,n}\mathbf 1\right)=-\log\varrho.
 \label{eq:persistent-state-spectral-exponent}
\end{equation}
If, moreover, a strongly connected class of $\mathcal B$~(a maximal set of joint states that all reach one another inside $\mathcal B$) whose submatrix has spectral radius $\varrho$ contains a failure state in each of its cyclic classes, then $P_{\mathrm b}^{(n)}\geq c\varrho^n$ for some $c>0$ and all large $n$, so that $P_{\mathrm b}^{(n)}$ itself decays with exponent exactly $-\log\varrho$.
\end{proposition}
\begin{proof}
Since $\mathcal G$ is absorbing and disjoint from $\mathcal F$, failure at time $n$ implies that $\mathcal G$ has not been entered, whose probability is the right-hand side of \eqref{eq:persistent-state-error-bound}. Write $s_n$ for this right-hand side. Since $s_n\leq\|K_{\mathcal B}^{\,n}\|_\infty$, the largest row sum of $K_{\mathcal B}^{\,n}$, and $\|K_{\mathcal B}^{\,n}\|_\infty^{1/n}\to\varrho$ by Gelfand's formula, $\liminf_{n\to\infty}-\frac1n\log s_n\geq-\log\varrho$. More precisely, the Jordan form of $K_{\mathcal B}$ gives $s_n\leq C\,n^k\varrho^n$ for a constant $C$, where $k+1$ is the size of the largest Jordan block of an eigenvalue of modulus $\varrho$, so that $k\leq|\mathcal B|-1$. For the lower bound, a strongly connected class of $\mathcal B$ attaining $\varrho$ is reachable from $\bm t_0$ by definition of $\mathcal B$. Following a fixed path to it and then using a positive Perron eigenvector of its irreducible submatrix bounds $s_n$ below by a positive constant times $\varrho^n$, without an aperiodicity assumption. For the last claim, let $\mathcal B_0\subseteq\mathcal B$ be such a class, of period $d$, entered at a state $\bm e$ with positive probability. By the Perron--Frobenius theorem for irreducible matrices, for every $\bm f\in\mathcal B_0$, $(K_{\mathcal B_0}^{\,k})_{\bm e\bm f}$ is at least a positive constant times $\varrho^k$ for all large $k$ in the residue class modulo $d$ of the cyclic class of $\bm f$. Taking $\bm f$ among the failure states of the $d$ cyclic classes shows that the probability of a failure state at time $n$ is at least $c\varrho^n$ for all large $n$.
\end{proof}
When $\varrho<1$, the error probability vanishes, so that, by \Cref{prop:finite-controller-floor}, the possible outputs of a user must depend on what the others play or observe, as they do with certificates. Generalized musical chairs, below, is a role controller of this kind.

\subsection{Generalized Musical Chairs}
\label{subsec:gmc}

Musical chairs redraws its role uniformly after a failure, although a failure may mean that the role is held by a committed user. Generalized musical chairs lets a user remember the roles on which it failed and avoid them. It applies when every role $a\in[J]$ has an input $\chi_a\in\cX$, the inputs being distinct, and a set $\mathcal C_a\subseteq\cY$ of \emph{certifying} outputs, which have probability zero whenever another user also plays $\chi_a$.

\begin{definition}[Generalized musical chairs]
\label{def:gmc}
Fix an exclusion probability $\beta\in[0,1]$. A user is either \emph{committed} to a role or \emph{searching} with an excluded set $E\subseteq[J]$, initially $E=\emptyset$. In every channel use, a searching user proceeds as follows.
\begin{enumerate}
 \item If $E=[J]$, it sets $E=\emptyset$.
 \item It draws a role $a$ uniformly in $[J]\setminus E$, transmits $\chi_a$ and observes $y$.
 \item If $y\in\mathcal C_a$, it commits to $a$. Otherwise, with probability $\beta$, it sets $E\leftarrow E\cup\{a\}$.
\end{enumerate}
A committed user transmits $\chi_a$ until time $n$. The identifier of a user is its committed role, or, if it is still searching, the last role it drew (the bound below counts every such case as a failure).
\end{definition}

For $\beta=0$ the excluded set stays empty, and this is musical chairs: a user commits at its first certificate and otherwise redraws a role uniformly. For $\beta>0$ a searching user avoids the roles on which it recently failed. A failure is not conclusive: the role may be held by a committed user, but it may also be a collision with another searching user or, when the certifying outputs are not certain under a lone user, a missed reward. The exclusion is therefore a heuristic, and $\beta$ sets how much to trust it; the reset of $E$ when it covers $[J]$ prevents a permanent lock-out, even for $\beta=1$. The exponent is discontinuous at $\beta=0$, as \Cref{sec:erasure} shows.

\begin{corollary}[Exponent of generalized musical chairs]
\label{cor:gmc-exponent}

Generalized musical chairs is a finite-state controller. Let $\mathcal G$ be the set of joint states in which all users are committed to distinct roles, and $K_{\mathcal B}$ the substochastic matrix of \Cref{prop:persistent-state-exponent}. Then $P_{\mathrm b}^{(n)}\leq\mathbf e_{\bm t_0}^{\mathsf T}K_{\mathcal B}^{\,n}\mathbf 1$, and the error exponent is at least $-\log\rho(K_{\mathcal B})$, with equality under the last condition of \Cref{prop:persistent-state-exponent}. Maximized over $\beta$, this bound is at least the exponent of musical chairs, which is the case $\beta=0$.
\end{corollary}
\begin{proof}
The state of a searching user is its excluded set together with the last role it drew, which determines its identifier, and that of a committed user is its role; this state set is finite, the maps do not depend on $n$, and the randomization is private. A user commits at a channel use only if it received a certifying output, which means that no other user played the same input at that use. Committed users keep playing their input, so a later user cannot obtain a certificate for their roles. The committed roles are therefore distinct at all times, so every reachable joint state in which all users are committed lies in $\mathcal G$. The set $\mathcal G$ is disjoint from the failure states, and it is absorbing because committed users do not change. \Cref{prop:persistent-state-exponent} gives the claims.
\end{proof}
The exponent is positive, that is, $\rho(K_{\mathcal B})<1$, when, for every $a\in[J]$, a user alone on $\chi_a$ receives an output in $\mathcal C_a$ with positive probability, whatever the others play. Indeed, since commitments are permanent, it suffices that some searching user $j$ commits within $J+1$ uses with positive probability. Failing and excluding the roles it draws, $j$ empties its excluded set within $J$ uses unless it commits. It then draws, with positive probability, a role that no committed user holds and that no other searching user is forced to draw~(one exists, since the other searching users are fewer than such roles), while the others draw other roles, and it receives a certificate with positive probability. Hence from every reachable joint state in which some user is still searching, all users are committed within some $m$ steps with probability at least some $p>0$. This implies that every row sum of $K_{\mathcal B}^{\,m}$ is at most $1-p$, and hence $\rho(K_{\mathcal B})\leq(1-p)^{1/m}<1$.

The survival probability does not depend on the last roles drawn and is invariant under permutations of the users, so that the chain that computes it has one state per assignment of committed roles and excluded sets, up to a permutation of the users, and the spectral radius is computed numerically in \Cref{sec:erasure}. With $\beta>0$, generalized musical chairs is not a confidence controller, since the excluded set is needed, and it requires certifying outputs~(\Cref{prop:finite-controller-floor}).

\section{Application: The Noisy Bandit Channel}
\label{sec:erasure}

We now apply the general results to the channel that motivated the work: $J$ players choose among $K$ arms and observe a Bernoulli reward, but a collision suppresses the reward of the players involved. We define a \emph{noisy} bandit channel with one pair of parameters per arm, so that the channel of \cite{zhang2025} is the noiseless case.

\subsection{The Channel}
\label{subsec:bandit-channel}

\begin{definition}[Noisy bandit channel]
\label{def:noisy-bandit}
Let $\cX=[K]$ and $\cY=\{0,1\}$; the inputs are the arms. Each arm has a \emph{solo} reward probability $\mu_x\in[0,1]$ and a \emph{collision} reward probability $\eps_x\in[0,1]$. Given the inputs $x_{[J]}$, the outputs are independent, and user $j$ observes $Y_j=1$ with probability
\begin{equation}
 W(1\mid x_j,x_{-j})=
 \begin{cases}
 \mu_{x_j} &\text{if } x_k\neq x_j\text{ for all }k\neq j,\\
 \eps_{x_j} &\text{if } x_k=x_j\text{ for some }k\neq j,
 \end{cases}
 \qquad W(0\mid\cdot)=1-W(1\mid\cdot).
 \label{eq:noisy-bandit-law}
\end{equation}
\end{definition}

A user that is alone on arm $x$ is rewarded with probability $\mu_x$. Users that share an arm are each rewarded, independently, with probability $\eps_x$, whatever the number of colliders. An arm with $\mu_x=\eps_x=0$ is never rewarded, whether or not it is shared, so it is a \emph{silent} input. The channel is symmetric (\Cref{def:symmetric-channel}), and the outputs are conditionally independent, so that the per-receiver statements of \Cref{sec:block-coding} apply.

Below are some notable special cases.
\begin{itemize}
 \item $\bm\eps=\bm 0$ is the original bandit channel: a collision is an erasure whose output is indistinguishable from a failed reward. A positive output is then a \emph{certificate} that the user was alone, and no collision is ever reported as a reward.
 \item If $0<\mu_x<1$ and $0<\eps_x<1$ for the arms in use, every output has positive probability, no output certifies a role, and by \Cref{prop:finite-controller-floor} no finite-state controller has a vanishing error probability, even if silent arms are also used. With maps that vary with time and $J$ arms with the same parameters $\eps<\mu$, a bounded state set suffices for a vanishing error probability (cf. \Cref{prop:bounded-memory-clock}).
 \item If $\eps_x=\mu_x$ for every $x$, collisions are invisible, the channel is noninteracting, and symmetry breaking is impossible (cf.~\Cref{cor:bandit-dichotomy}).
\end{itemize}

\subsection{The Converse}
\label{subsec:bandit-converse}

The converse of \Cref{sec:meta-converse} bounds the error probability of every symmetric causal protocol. We apply \Cref{thm:latent-symmetry-meta-converse} with $m=2$: users $1$ and $2$ are selected, and the other $J-2$ users are unselected. Since the outputs are independent given the inputs, the selected pair sees the channel through the two local laws $W(\cdot\mid x_1,x_{-1})$ and $W(\cdot\mid x_2,x_{-2})$. For $p,q\in[0,1]$ write $D_\alpha(p\Vert q):=D_\alpha(\mathrm{Bern}(p)\Vert\mathrm{Bern}(q))$, with $D_1$ the Kullback--Leibler divergence.

\paragraph{A noninteracting reference}
A noninteracting reference gives every selected user a private reward law $v_x\in[0,1]$ for arm $x$, whatever the other users do. In each channel use, a selected user on arm $x$ contributes to the cost of the change of measure the R\'enyi divergence between $\mathrm{Bern}(v_x)$ and its true reward law, which is $\mathrm{Bern}(\mu_x)$ if it is alone on the arm and $\mathrm{Bern}(\eps_x)$ if it shares it:
\begin{equation}
 {\mathsf a_x}:=D_\alpha(v_x\Vert\mu_x),\qquad
 {\mathsf s_x}:=D_\alpha(v_x\Vert\eps_x).
 \label{eq:bandit-alone-shared-cost}
\end{equation}
On a silent arm both laws are $\delta_0$, so that ${\mathsf a_x}={\mathsf s_x}=0$ if $v_x=0$ and ${\mathsf a_x}={\mathsf s_x}=\infty$ otherwise. For $J\geq3$, the unselected users can make a selected user alone or in collision. Let ${\bar{\mathsf a}_x}:={\mathsf a_x}$ if $J=2$ and ${\bar{\mathsf a}_x}:=\max\{{\mathsf a_x},{\mathsf s_x}\}$ if $J\geq3$, and put
\begin{equation}
 \Delta_\alpha(\bm v):=\max\Bigl\{\ \max_{x\in[K]}2{\mathsf s_x},\ \ \max_{1\leq x<x'\leq K}({\bar{\mathsf a}_x}+{\bar{\mathsf a}_{x'}})\Bigr\}.
 \label{eq:bandit-ni-radius}
\end{equation}

\begin{proposition}[Noninteraction converse]
\label{prop:bandit-ni-converse}
For every $\bm v\in[0,1]^K$, every $\alpha\in(1,\infty]$, every blocklength $n$ and every symmetric causal protocol,
\begin{equation}
 P_{\mathrm b}^{(n)}\geq\gamma_{2,J}^{b_\alpha}\,e^{-n\Delta_\alpha(\bm v)}.
 \label{eq:bandit-ni-finite}
\end{equation}
Consequently $\overline E_{\mathrm b}(P)\leq\inf_{\bm v}\Delta_1(\bm v)$.
\end{proposition}
\begin{proof}
Let $V(1\mid x)=v_x$. Since $P_2$ is a product of the two local laws and $D_\alpha$ is additive on products, the radius \eqref{eq:partial-renyi-radius} is
\[
 d_{\alpha,2}(V\Vert P)=\max_{x_{[J]}}\sum_{j=1}^2D_\alpha\bigl(V(\cdot\mid x_j)\bigm\Vert W(\cdot\mid x_j,x_{-j})\bigr).
\]
The term of user $j$ is ${\mathsf a_{x_j}}$ if it is alone and ${\mathsf s_{x_j}}$ if it shares its arm.  If $x_1=x_2=x$, both users share their arm and the sum is $2{\mathsf s_x}$. If $x_1\neq x_2$, then for $J=2$ both users are alone, and for $J\geq3$ the unselected users can make each selected user alone or in collision, except that for $J=3$ the two selected users cannot both be in collision, and for $K=2$ they cannot both be alone. Hence $d_{\alpha,2}(V\Vert P)\leq\Delta_\alpha(\bm v)$, with equality if $J=2$ or $K\geq3$, since ${\mathsf s_x}+{\mathsf s_{x'}}\leq\max\{2{\mathsf s_x},2{\mathsf s_{x'}}\}$. \Cref{thm:j-meta-converse} gives the bound, and its second statement the exponent.
\end{proof}
{Since $D_1(v\Vert q)$ is convex in $v$, so is $\Delta_1$, and the infimum in \Cref{prop:bandit-ni-converse} is a convex program.}

For arms with the same parameters the optimal reference is explicit. Let
\begin{equation}
 \Lambda(\mu,\eps):=\min_{v\in[0,1]}\max\{D_1(v\Vert\mu),D_1(v\Vert\eps)\}=       \sup_{0<\lambda<1} -\log\bigl[\mu^\lambda\eps^{1-\lambda}+(1-\mu)^\lambda(1-\eps)^{1-\lambda}\bigr]
 \label{eq:chernoff-information}
\end{equation}
be the Chernoff information between $\mathrm{Bern}(\mu)$ and $\mathrm{Bern}(\eps)$, where the equality is from the minimax theorem.

\begin{corollary}[{Closed-form noninteraction bounds}]
\label{cor:bandit-ni-symmetric}
If $\mu_x=\mu$ and $\eps_x=\eps$ for all $x\in[K]$ and $K\geq2$, then $\overline E_{\mathrm b}(P)\leq2\Lambda(\mu,\eps)$. If $\bm\eps=\bm0$ and $K\geq2$, then $\overline E_{\mathrm b}(P)\leq-\log(1-\mu_{(1)})-\log(1-\mu_{(2)})$, where $\mu_{(1)}\geq\mu_{(2)}$ are the two largest means.
\end{corollary}
\begin{proof}
With $v_x\equiv v$, \eqref{eq:bandit-ni-radius} is $2\max\{D_1(v\Vert\mu),D_1(v\Vert\eps)\}$ for every $J\geq2$, and minimizing over $v$ gives \eqref{eq:chernoff-information}. If $\eps_x=0$, then ${\mathsf s_x}=\infty$ unless $v_x=0$, and $v=0$ gives ${\mathsf a_x}=-\log(1-\mu_x)$ and ${\mathsf s_x}=0$, so that \eqref{eq:bandit-ni-radius} is the sum of the two largest values of $-\log(1-\mu_x)$.
\end{proof}

The reference $\bm v$ is the reward law of a lone player that cannot tell whether it shares its arm. The radius is small when collisions are hard to detect, $\eps_x$ close to $\mu_x$, and it vanishes when $\eps_x=\mu_x$ for all $x$. For $\bm\eps=\bm0$ the noninteraction bound is $-\log(1-\mu_{(1)})-\log(1-\mu_{(2)})$, which diverges as $\mu_{(1)}\to1$, although a perfect collision detector has a finite exponent (\Cref{cor:bandit-co-closed}). The following reference does not have this defect.

\paragraph{A common-output reference}
Assume that a collision does not increase the reward probability, i.e., $\eps_x\leq\mu_x$. The idea of the reference is to keep only the trajectories on which the two selected users see the same outputs, so that their views differ only through their private seeds. In the notation of \Cref{sec:meta-converse}, this is a kernel $B$ of the form~\eqref{eq:latent-factorized-kernel} with $m=2$ and the normalized completion (N), with the physical completion of \Cref{rem:physical-completion}. For a kernel with $B\leq P$, the order-infinity case of \Cref{thm:latent-symmetry-meta-converse} gives $P_{\mathrm b}^{(n)}\geq\gamma_{2,J}\,\kappa_2(B)^n$, where $\kappa_2(B)$ is the persistence coefficient \eqref{eq:partial-persistence-kernel-form}; see \eqref{eq:latent-infinity-optimization}. We therefore look for a kernel $B\leq P$ with the largest possible persistence, and compute $\kappa_2(B)$ for the kernel below. Put
\begin{equation}
 \bar\mu_x:=1-\mu_x,\qquad \tau_x:=(1-\eps_x)^2-(1-\mu_x)^2\geq0,\qquad x\in\cX,
 \label{eq:co-parameters}
\end{equation}
so that a silent arm has $\bar\mu_x=1$ and $\tau_x=0$, and for a law $p$ on $\cX$ let
\begin{equation}
 \kappa(p):=\Bigl(\sum_xp_x\bar\mu_x \Bigr)^2+\Bigl(\sum_xp_x\eps_x\Bigr)^2+\sum_x\tau_x\,p_x^2,
 \qquad \kappa_{\mathrm{CO}}:=\min_{p\in\mathcal P(\cX)}\kappa(p)>0.
 \label{eq:co-kappa}
\end{equation}
The function $\kappa$ is a convex quadratic on the simplex.

\begin{proposition}[Common-output converse]
\label{prop:bandit-co-converse}
If $\eps_x\leq\mu_x$ for all $x\in[K]$, then for every $J\geq2$, every $n$ and every symmetric causal protocol (an instance of \Cref{thm:latent-symmetry-meta-converse} with $m=2$ and $\alpha=\infty$),
\begin{equation}
 P_{\mathrm b}^{(n)}\geq\gamma_{2,J}\,\kappa_{\mathrm{CO}}^{\,n},
 \qquad
 \overline E_{\mathrm b}(P)\leq-\log\kappa_{\mathrm{CO}}.
 \label{eq:bandit-co-converse}
\end{equation}
\end{proposition}
\begin{proof}
Define for a pair of selected users
\[
 B^2(y_1,y_2\mid x_1,x_2):=\ind\{y_1=y_2=0\}\Bigl[\bar\mu_{x_1}\bar\mu_{x_2}+\ind\{x_1=x_2\}\tau_{x_1}\Bigr]+\ind\{y_1=y_2=1\}\,\eps_{x_1}\eps_{x_2},
\]
and $B(y_{[J]}\mid x_{[J]}):=B^2(y_1,y_2\mid x_1,x_2)\,P(y_{3:J}\mid x_{[J]})$, where the last factor is the physical law of the unselected outputs, which are independent of $y_1,y_2$. This kernel has a normalized completion, so it satisfies condition~(N). Its selected part is $\sum_{r\in\{\mathsf u\}\cup[K]\cup\{\mathsf e\}}a_r(x_1,y_1)a_r(x_2,y_2)$, with the components $a_{\mathsf u}(x,y)=\bar\mu_x\ind\{y=0\}$, $a_{x'}(x,y)=\sqrt{\tau_{x'}}\,\ind\{x=x',y=0\}$ for $x'\in[K]$, and $a_{\mathsf e}(x,y)=\eps_x\ind\{y=1\}$.

We show $B\leq P$. If $x_1=x_2=x$, both selected users share their arm, and $B^2(0,0\mid x,x)=\bar\mu_x^2+\tau_x=(1-\eps_x)^2$ and $B^2(1,1\mid x,x)=\eps_x^2$ are the physical probabilities. Otherwise, in every placement of the unselected users, each selected user is rewarded with probability $\mu_{x_j}$ or $\eps_{x_j}$, both at least $\eps_{x_j}$ and at most $\mu_{x_j}$ since $\eps_{x_j}\leq\mu_{x_j}$. Hence its probability of a zero is at least $\bar\mu_{x_j}$ and of a one at least $\eps_{x_j}$, and $B^2\leq P_2$ for all placements. Therefore $\rho_{\infty,2}(B;P)\leq1$, and $K_{\infty,2}(B;P)\geq\kappa_2(B)$. Finally, $A_r(x)=\sum_ya_r(x,y)$ equals $\bar\mu_x$, $\sqrt{\tau_{x'}}\ind\{x=x'\}$ and $\eps_x$ for the three kinds of component, and \eqref{eq:partial-persistence-kernel-form} gives $\kappa_2(B)=\min_p\kappa(p)$. \Cref{thm:latent-symmetry-meta-converse} with $\alpha=\infty$ and $b_\infty=1$ concludes.
\end{proof}

The minimization in \eqref{eq:co-kappa} is a quadratic program. For equal parameters it has a closed form.

\begin{corollary}[Equal parameters]
\label{cor:bandit-co-closed}
Let $k\geq1$ arms have the parameters $(\mu,\eps)$ with $\eps\leq\mu$, suppose that the other $K-k\geq0$ arms are silent, and put $\tau:=(1-\eps)^2-(1-\mu)^2$ and $a^\star:=\mu/(\mu^2+\eps^2+\tau/k)$. If some arm is silent and $a^\star\leq1$, the optimal law puts mass $a^\star$ uniformly on the $k$ arms and the rest on a silent arm, and $\kappa_{\mathrm{CO}}=1-\mu a^\star$. Otherwise the optimal law is uniform on the $k$ arms, and $\kappa_{\mathrm{CO}}=(1-\mu)^2+\eps^2+\tau/k$. For $\bm\eps=\bm0$ and a silent arm, $a^\star\leq1$ if and only if $k-2\leq(k-1)\mu$, and then
\begin{equation}
 \kappa_{\mathrm{CO}}=\frac{2-\mu}{2+(k-1)\mu},
 \qquad
 \overline E_{\mathrm b}(P)\leq\log\frac{2+(k-1)\mu}{2-\mu};
 \label{eq:bandit-co-closed}
\end{equation}
for $\bm\eps=\bm0$ and no silent arm, $\kappa_{\mathrm{CO}}=(1-\mu)^2+(2\mu-\mu^2)/k$; and for $k=1$ and a silent arm, $\kappa_{\mathrm{CO}}=1-\mu^2/[2(\mu-\eps(1-\eps))]$ if $2\eps(1-\eps)\leq\mu$, and $\kappa_{\mathrm{CO}}=(1-\eps)^2+\eps^2$ otherwise.
\end{corollary}
\begin{proof}
Since $\kappa$ is convex and invariant under permutations of the $k$ arms, and of the silent arms, an optimal law spreads its mass uniformly on the $k$ arms, and only its total mass on the silent arms matters. With $a$ the mass on the $k$ arms, $\kappa(a)=(1-\mu a)^2+\eps^2a^2+\tau a^2/k$, a convex quadratic minimized over the real line at $a=a^\star$, where $\kappa(a^\star)=1-\mu a^\star$ by the first-order condition; $a=1$ is forced if no arm is silent, and is optimal if $a^\star>1$. For $\bm\eps=\bm0$, $\tau=2\mu-\mu^2$ and $a^\star=k/(2+(k-1)\mu)$, and substitution gives \eqref{eq:bandit-co-closed}; for $k=1$, $\mu^2+\eps^2+\tau=2(\mu-\eps(1-\eps))$.
\end{proof}

For $\mu\to1$ the exponent bound tends to $\log(k+1)$, its value at $\mu=1$; for $\mu<1$ it is smaller. Unlike the noninteraction bound of \Cref{cor:bandit-ni-symmetric}, it stays finite as $\mu\to1$.

\paragraph{Combined statement, optimized references, and the dependence on $J$}
Both propositions hold at once, so that $P_{\mathrm b}^{(n)}\geq\max\{\gamma_{2,J}^{b_\alpha}e^{-n\Delta_\alpha(\bm v)},\ \gamma_{2,J}\kappa_{\mathrm{CO}}^n\}$ for every $\bm v$ and $\alpha$, and $\overline E_{\mathrm b}(P)\leq\min\{\inf_{\bm v}\Delta_1(\bm v),-\log\kappa_{\mathrm{CO}}\}$~(the second terms are present if $\eps_x\leq\mu_x$ for all $x$). The noninteraction bound is the best for noisy collisions, where the two laws to be told apart are close, and the common-output bound for erasures, where the noninteraction bound is large. Those are two kernels of the family of \Cref{thm:combined-protocol-converse}, and neither is optimal in it. A reference $B^2=\sum_ra_r\otimes a_r$ with several components, optimized numerically under the constraint $B\leq P$ or, for a finite $\alpha$, by minimizing $\Psi_{\alpha,2}(B;P)$ under $\kappa_2(B)\geq1$, is a valid reference as soon as it is feasible, and the numerical values in \Cref{tab:bandit-comparison} are of this kind. They improve on the smaller of the two closed forms by up to $12\%$ in the exponent, at the intermediate noise levels where the two closed forms cross, and coincide with them at $\eps=0$ and $\eps=0.2$ for $\mu=0.8$. How they are computed, and how precise they are, is described in \Cref{app:table-numerics}.

Both propositions are instances of the kernel converse with $m=2$. For noninteracting references $m=2$ is optimal, by data processing; for general kernels there is no such monotonicity, and we have not evaluated kernels with $m>2$. As they stand, the exponents of the bounds depend on the number of users only through the distinction between $J=2$ and $J\geq3$ in \Cref{prop:bandit-ni-converse}, and not at all in \Cref{prop:bandit-co-converse}: only the two selected users are charged, and the unselected users are adversarial. \Cref{subsec:erasure-comparison} discusses the dependence on $J$.

\subsection{Block Coding}
\label{subsec:bandit-block}

For block coding the Gallager function of \Cref{thm:multi-user-gallager} is explicit. Fix an input law $Q$ on $\cX$ with $q_x=Q(x)$, an error subset of size $s$, and $c=J-1-s$ correct foreign codewords. Let
\begin{multline}
 G_{\rho,s}^{(J)}(Q;\bm\mu,\bm\eps):=\sum_{x\in\cX}q_x\Biggl\{1-(1-q_x)^c
 +(1-q_x)^c\sum_{y\in\{0,1\}}\Bigl[(1-q_x)^s\,W_{\mu_x}(y)^{\frac1{1+\rho}}\\
 +\bigl(1-(1-q_x)^s\bigr)W_{\eps_x}(y)^{\frac1{1+\rho}}\Bigr]^{1+\rho}\Biggr\},
 \label{eq:multi-bandit-g-function}
\end{multline}
where $W_\nu(1):=\nu$ and $W_\nu(0):=1-\nu$.

\begin{proposition}[Noisy bandit Gallager function]
\label{prop:multi-bandit-exponent}
For the channel of \Cref{def:noisy-bandit}, $E_{0,s}(\rho,Q)=-\log G_{\rho,s}^{(J)}(Q;\bm\mu,\bm\eps)$, and hence
\begin{equation}
 E_{\mathrm r,s}(R,Q)=\max_{0\leq\rho\leq1}\{-\log G_{\rho,s}^{(J)}(Q;\bm\mu,\bm\eps)-\rho sR\}.
 \label{eq:multi-bandit-random-exponent}
\end{equation}
\end{proposition}

The proof is in \Cref{app:erasure-gallager}. Together with \Cref{thm:multi-user-gallager} this gives an explicit error bound for $J$ users. The two terms in the braces of \eqref{eq:multi-bandit-g-function} are the following: if a correct foreign codeword collides with the receiver, the output law does not depend on the candidate and contributes $1$; if no correct codeword collides, the candidate tuple avoids the receiver's arm with probability $(1-q_x)^s$, and the likelihood is $W_{\mu_x}$ or $W_{\eps_x}$ accordingly. \Cref{prop:multi-bandit-exponent} does not require $\eps_x\leq\mu_x$. At $\bm\eps=\bm0$ it is the function of the noiseless bandit channel.

\paragraph{Two users}
For $J=2$ the only error subset is $s=1$, $c=0$. With $\{j,j'\}=[2]$, the receiver's output depends on $X_{j'}$ only through the indicator of a collision, which has probability $q_x$ given $X_j=x$, so that
\begin{align}
 I_Q(X_{j'};Y_j\mid X_j)&=\sum_{x\in\cX}q_x\Bigl[\hb\bigl((1-q_x)\mu_x+q_x\eps_x\bigr)-(1-q_x)\hb(\mu_x)-q_x\hb(\eps_x)\Bigr],
 \label{eq:two-user-bandit-rate}\\
 E_{0,1}(\rho,Q)&=-\log\sum_{x\in\cX}q_x\sum_{y\in\{0,1\}}\Bigl[(1-q_x)W_{\mu_x}(y)^{\frac1{1+\rho}}+q_xW_{\eps_x}(y)^{\frac1{1+\rho}}\Bigr]^{1+\rho}.
 \label{eq:two-user-bandit-ezero}
\end{align}
For $\bm\eps=\bm0$, \eqref{eq:two-user-bandit-ezero} and \eqref{eq:two-user-bandit-rate} are the exponent function of \cite[Cor.~1]{zhang2025} and the rate of \cite[Th.~4]{zhang2025}, and those of \cite[Ths.~1 and~2]{zhang2026} with a single block. With $R$ the name rate, the collision of the two names contributes $e^{-nR}$, so that for $J=2$ the exponent \eqref{eq:multi-user-sync-exponent} is
\begin{equation}
 E_{\mathrm{bc}}(Q)=\sup_{R>0}\min\{R,E_{\mathrm r,1}(R,Q)\}=\max_{0\leq\rho\leq1}\frac{E_{0,1}(\rho,Q)}{1+\rho},
 \label{eq:two-user-bc-exponent}
\end{equation}
because $E_{\mathrm r,1}(R,Q)\geq R$ exactly when $R\leq E_{0,1}(\rho,Q)/(1+\rho)$ for some $\rho$.

For $J$ users the full-set information, which decides the rate threshold in \Cref{thm:multi-user-inner}, is obtained in the same way, with $\theta_x:=(1-q_x)^{J-1}$ the probability that no foreign input equals $x$:
\begin{equation}
 I_Q(X_{-j};Y_j\mid X_j)=\sum_{x\in\cX}q_x\Bigl[\hb\bigl(\theta_x\mu_x+(1-\theta_x)\eps_x\bigr)-\theta_x\hb(\mu_x)-(1-\theta_x)\hb(\eps_x)\Bigr].
 \label{eq:multi-bandit-information}
\end{equation}
It is positive if and only if $\mu_x\neq\eps_x$ for some $x$ with $q_x\in(0,1)$, by the strict concavity of $\hb$.

\begin{corollary}[{Dichotomy}]
\label{cor:bandit-dichotomy}
For every $J\geq2$, $\underline E_{\mathrm b}(P)>0$ if $K\geq2$ and $\mu_x\neq\eps_x$ for some $x\in[K]$, and $\epsilon_n^\star(P)\geq\gamma_J=1-J!/J^J$ for every $n$ if $\mu_x=\eps_x$ for all $x\in[K]$ or if $K=1$.
\end{corollary}
\begin{proof}
If $K\geq2$ and $\mu_x\neq\eps_x$, take $Q$ supported on $\{x,x'\}$ for some $x'\neq x$, with $q_x\in(0,1)$, so that the term of arm $x$ of \eqref{eq:multi-bandit-information} is positive and the others are nonnegative, and apply \Cref{thm:multi-user-inner} and \Cref{thm:multi-user-gallager}. If $\mu_x=\eps_x$ for all $x$, or if $K=1$, so that the users always share the single arm, the output of each user depends only on its own input: $P(y_{[J]}\mid x_{[J]})=\prod_jW(y_j\mid x_j)$ for a one-user channel $W$. The complete local views are then i.i.d.\ given $C$, and so are the identifiers, and \Cref{lem:iid-identifier-collision} with $m=J$ gives the claim.
\end{proof}

Two arms therefore suffice to break the symmetry, one of which may be silent, as soon as a collision changes the reward law of an arm, however little.

\subsection{Interactive Schemes}
\label{subsec:bandit-interactive}

\paragraph{Generalized musical chairs}
Take $\bm\eps=\bm0$ and use $J$ of the arms as the roles, with $\chi_a=a$ and $\mathcal C_a=\{1\}$: a rewarded user is certain to have been alone. \Cref{def:gmc} then gives generalized musical chairs with exclusion probability $\beta$, and \Cref{cor:gmc-exponent} gives a lower bound on its exponent as a spectral radius, which we compute exactly for small $J$ (\Cref{tab:bandit-comparison,tab:gmc-users}); in all these cases it is the exact exponent (\Cref{app:table-numerics}). The parameter $\beta=0$ is musical chairs. Musical chairs is also a confidence controller (\Cref{def:confidence-controller}), with $Q_a=\delta_a$, score $+1$ on a positive output and $-1$ on a zero, $\xi_+^a=\xi_-^a=1$, and the uniform redraw kernel and initial law: a user commits at its first reward and draws a new role after every zero. The two schemes differ only in the identifier of a user that is still searching at time $n$: the last arm it played in musical chairs, and a fresh draw in the confidence controller; this does not change the exponent. For $\beta=0$ the exponent has a closed form.

\begin{proposition}[Musical chairs, $\beta=0$]
\label{prop:bandit-musical-chairs}
On the noiseless bandit channel with $J$ arms of means $\mu_1,\ldots,\mu_J>0$ used as roles, musical chairs satisfies
\begin{equation}
 \lim_{n\to\infty}-\frac1n\log P_{\mathrm b}^{(n)}=-\log\Bigl(1-\frac{\min_x\mu_x}{J}\Bigr).
 \label{eq:musical-chairs-exponent}
\end{equation}
\end{proposition}
\begin{proof}
Let $S$ be the set of committed arms. Uncommitted users redraw uniformly, so the transition of $S$ depends only on $S$, and $S$ is a Markov chain that only grows. Let $s_S$ be the probability that the state $S$ is not left in one slot, that is, that no uncommitted user is rewarded. The substochastic matrix of the chain on the states $S\neq[J]$, ordered by size, is triangular with diagonal entries $s_S$, so that $\PP(S_n\neq[J])$ is at most a constant times $n^{J-1}\max_{S\neq[J]}s_S^{\,n-J+1}$, since a path that has not reached $[J]$ at time $n$ leaves states at most $J-1$ times, hence has probability at most $\max_{S\neq[J]}s_S^{\,n-J+1}$, and there are at most a constant times $n^{J-1}$ such paths, and, since every state is reachable, at least a constant times $\max_{S\neq[J]}s_S^{\,n}$. The maximum is $1-\min_x\mu_x/J$. It is attained by $S=[J]\setminus\{x\}$ with a single uncommitted user, which can be rewarded only if it draws $x$, an event of probability $1/J$, and then is rewarded with probability $\mu_x$, so that $s_S=1-\mu_x/J$. For any other $S$ with $w\leq J$ uncommitted users and any free arm $x$, the probability that exactly one of them draws $x$ is $\frac wJ(1-\frac1J)^{w-1}\geq\frac1J$, since $w(1-\frac1J)^{w-1}$ is nondecreasing in $w\leq J$ and equals $1$ at $w=1$; that user is rewarded with probability $\mu_x$. Hence $s_S\leq1-\mu_x/J$ for every free arm $x$, and in particular $s_S\leq1-\min_x\mu_x/J$.

This gives the exponent of the survival probability, which upper-bounds $P_{\mathrm b}^{(n)}$ by \Cref{prop:persistent-state-exponent}. For the lower bound, let $x$ be an arm of smallest mean and $S^\star=[J]\setminus\{x\}$. The chain reaches $S^\star$ and stays there until the end of slot $n-1$ with probability at least a constant times $s_{S^\star}^{\,n-1}$. In slot $n$ the single uncommitted user draws the arm of a committed user with probability $(J-1)/J$, and its identifier, the last role it drew, then coincides with that of a committed user. Hence $P_{\mathrm b}^{(n)}\geq c\,\frac{J-1}{J}\,s_{S^\star}^{\,n-1}$.
\end{proof}

The exponent is that of the last uncommitted user, which must find its own arm, and it decays like $\mu/J$.

For $\beta>0$ the last uncommitted user no longer searches blindly. In musical chairs it draws among all $J$ arms although $J-1$ of them are held by committed users and a failure reveals it, whereas an excluding user searches the remaining arms. The exponent is not monotone in $\beta$, and it is discontinuous at $\beta=0$. Indeed, let $\beta>0$. For $J=2$, the joint state in which both users search with the same excluded arm is reachable (both draw this arm, collide, and exclude it); it is a failure state, which persists with probability $(1-\beta)^2$, since both users then draw the other arm and collide. For $J\geq3$, the joint state in which $J-1$ users are committed and the last user searches with all arms excluded but one, held by a committed user, is reachable; it is a failure state, which persists with probability $1-\beta$. Hence the exponent of $P_{\mathrm b}^{(n)}$ is at most $-2\log(1-\beta)$ for $J=2$ and $-\log(1-\beta)$ for $J\geq3$, which tends to zero as $\beta\downarrow0$, whereas the exponent for $\beta=0$ is \eqref{eq:musical-chairs-exponent}. For $\mu=1$ and $\beta=0.05$ these upper bounds, $0.103$ for $J=2$ and $0.051$ for $J=3$, coincide with the lower bounds of \Cref{cor:gmc-exponent}, against $0.693$ and $0.405$ for musical chairs, so that $\beta$ has to be tuned (the best values found are in \Cref{tab:gmc-users}). With the best $\beta$ found on a grid, generalized musical chairs improves on musical chairs for every $J\geq2$ in \Cref{tab:gmc-users}, except for $\mu=1$ and $J=2$, where musical chairs already meets the converse, so that $\beta=0$ is optimal; $\beta=1/2$ attains the same exponent.

\paragraph{One-arm musical chairs}
For $J=2$, one informative arm together with a silent arm is enough to break symmetry. Call them arm $1$, with mean $\mu_1$, and arm $2$, which is silent. Role $1$ is a \emph{claimer} that pulls the arm at every use and never releases; role $2$ is a \emph{prober} that pulls arm $1$ with probability $1/2$ and plays the silent arm otherwise. Both users start as probers. A prober that pulls and is rewarded knows that it was alone on the arm; it becomes a claimer and takes the arm for good, after which the other prober is never rewarded again. As a confidence controller, the claimer has $\xi_+^1=0$, so that it is committed on entry, $\xi_-^1=1$ and score $s_1\equiv0$; the prober has $Q_2$ uniform on $\{1,2\}$, $\xi_+^2=\infty$ and $\xi_-^2=1$, with score $-1$ if it played arm $1$ and was rewarded, and $0$ otherwise; the kernel is $R(2,\cdot)=\delta_1$ and $\pi_0=\delta_2$.

\begin{theorem}[One arm and a silent arm]
\label{thm:bandit-one-arm-tight}
Let $J=2$, $K=2$, $\bm\eps=\bm0$, and let arm $2$ be silent, i.e., $\mu_2=0$. One-arm musical chairs has $P_{\mathrm b}^{(n)}=(1-\mu_1/2)^n$, and every symmetric causal protocol satisfies $P_{\mathrm b}^{(n)}\geq\frac12(1-\mu_1/2)^n$. Hence
\[
 \tfrac12\bigl(1-\tfrac{\mu_1}2\bigr)^n\leq\epsilon_n^\star(P)\leq\bigl(1-\tfrac{\mu_1}2\bigr)^n,\qquad
 \overline E_{\mathrm b}(P)=-\log\bigl(1-\tfrac{\mu_1}2\bigr).
\]
\end{theorem}
\begin{proof}
The controller fails exactly when both users are probers at time $n$, since a claimer and a prober have distinct roles, and the users leave the state (prober, prober) in a slot only if exactly one pulls, probability $1/2$, and is rewarded, probability $\mu_1$. Thus $P_{\mathrm b}^{(n)}=(1-\mu_1/2)^n$. The converse is \Cref{prop:bandit-co-converse} with $K=2$, where $\kappa_{\mathrm{CO}}=1-\mu_1/2$ by \Cref{cor:bandit-co-closed} with $k=1$, and $\gamma_{2,2}=\frac12$.
\end{proof}

With exactly one informative arm and a silent arm ($K=2$), this matches the converse. For $J=2$, $\bm\eps=\bm0$ and any channel with a silent arm, using only the best arm, one-arm musical chairs achieves $-\log(1-\mu_{(1)}/2)$; with more informative arms this is only achievability, and the converse can be larger. The controller relies on the certificate, and the silent arm is what lets a prober be alone.

\paragraph{Confidence controllers with noisy collisions}

Take $J=2$, two arms as roles with $\eps_a<\mu_a$, and $Q_a=\delta_a$. A user in role $a$ has score $+u_a$ on a reward and $-d_a$ otherwise, with integers $u_a,d_a\geq1$; its thresholds are $\xi_+^a=\lceil\nu n\rceil$ and $\xi_-^a=\ell$; it redraws its role with the kernel $R(a,\bar a)=w$. When the roles coincide, each user is rewarded with probability $\eps_a$, so that the counter drifts down if $\eps_au_a<(1-\eps_a)d_a$; when they are distinct, it drifts up if $\mu_au_a>(1-\mu_a)d_a$. Integers $u_a,d_a$ satisfying both inequalities exist exactly when $\eps_a<\mu_a$, since they require $(1-\mu_a)/\mu_a<u_a/d_a<(1-\eps_a)/\eps_a$. We do not have a proof of a positive exponent for these controllers. We compute the error probability exactly (\Cref{prop:confidence-exact}) for a chain of $O(n^2)$ states, up to $n=960$ ($1280$ for $\eps=0.2$), and read the exponent off the slope of $-\log P_{\mathrm b}^{(n)}$. Since the threshold $\xi_+^a$ grows with $n$, this exponent is a limit and not a spectral radius, and the values we report are estimates, marked $^\ddagger$ in \Cref{tab:bandit-comparison}.         A truncated chain gives a rigorous upper bound on it, which exceeds the extrapolated value by less than $5\cdot10^{-4}$ for $\eps\leq0.1$ and by less than $3\cdot10^{-3}$ for $\eps=0.2$~(\Cref{app:table-numerics}). A controller of this class is called \emph{uniform} if both roles have the same parameters and the kernel and initial law are uniform ($w=\frac12$, $\pi_0=\mathrm{Unif}([2])$); the \emph{optimized} controller also optimizes the redraw probability $w$, and the score and threshold parameters. The optimized controller at $\eps=0.2$ has $w=1$; its error probability is evaluated exactly in the same way.

In \Cref{tab:bandit-comparison} the parameters $(u,d,\ell,\nu)$ of the uniform controller are $(11,6,5,2)$, $(10,7,3,2)$ and $(4,4,1,1)$ for $\eps=0.05$, $0.1$ and $0.2$, and those $(u,d,\ell,\nu;w,\pi_0)$ of the optimized controller are $(12,6,5,2;0.7,\mathrm{Unif}([2]))$, $(10,7,4,2;0.8,\mathrm{Unif}([2]))$ and $(4,4,1,1;1,\delta_1)$. The initial law changes only the constant, not the exponent: replacing $\delta_1$ by the uniform law at $\eps=0.2$ multiplies $P_{\mathrm b}^{(n)}$ by about $0.6$ and changes the slopes by less than $10^{-4}$, so the gain comes from $w=1$.

\subsection{Comparison}
\label{subsec:erasure-comparison}

\Cref{tab:bandit-comparison} compares the converse of \Cref{subsec:bandit-converse}, block coding, and the interactive schemes for $J=2$ and $K=3$ arms, with $\bm\mu=(0.8,0.8,0)$ and $\bm\eps=(\eps,\eps,0)$, $\eps\in\{0,0.05,0.1,0.2\}$, so that arm $3$ is silent. The converse is for the channel with the three arms; the schemes use only the two positive arms, which can only reduce their exponents. \Cref{tab:gmc-users} adds more users, with as many arms as users and no silent arm.

\begin{table}[t]
\centering

\caption{Exponents (nats per use) for $\bm\mu=(0.8,0.8,0)$, $\bm\eps=(\eps,\eps,0)$, $J=2$. Bold: the smallest converse, and the largest exponent of a scheme (proved for $\eps=0$, estimated for $\eps>0$), in each column. $^\dagger$: valid bound found by a numerical search, which a better search can only improve; $^\ddagger$: numerical estimate, not a proven bound (\Cref{app:table-numerics}). The name-decoding ceiling bounds the name-decoding exponent of every scheme (\Cref{cor:name-decoding-cap}); it is neither a converse nor an achievable exponent.}
\label{tab:bandit-comparison}
\begin{tabular}{lcccc}
\toprule
 & $\eps=0$ & $0.05$ & $0.1$ & $0.2$\\
\midrule
Noninteraction converse (\cref{prop:bandit-ni-converse}) & $3.219$ & $0.915$ & $0.695$ & $\mathbf{0.446}$\\
Common-output converse (\cref{prop:bandit-co-converse}) & $\mathbf{0.847}$ & $0.906$ & $0.963$ & $1.059$\\
Optimized reference (\cref{thm:combined-protocol-converse}) & $\mathbf{0.847}$ & $\mathbf{0.798}^\dagger$ & $\mathbf{0.652}^\dagger$ & $\mathbf{0.446}$\\
\midrule
Name-decoding ceiling (\cref{prop:bandit-name-cap}) & $0.429$ & $0.334$ & $0.276$ & $0.193$\\
Block coding, $\max_QE_{\mathrm{bc}}(Q)$ \eqref{eq:two-user-bc-exponent} & $0.162$ & $0.100$ & $0.079$ & $0.053$\\
\midrule
Musical chairs, $\beta=0$ (\cref{prop:bandit-musical-chairs}) & $0.511$ & --- & --- & ---\\
Generalized musical chairs, {$\beta=0.652$ (\cref{cor:gmc-exponent})} & $\mathbf{0.631}^\dagger$ & --- & --- & ---\\
Confidence controller, {uniform} (\cref{def:confidence-controller}) & --- & $0.369^{{\ddagger}}$ & $0.304^{{\ddagger}}$ & $0.20^{{\ddagger}}$\\
Confidence controller, optimized (\cref{def:confidence-controller}) & --- & $\mathbf{0.406}^{{\ddagger}}$ & $\mathbf{0.337}^{{\ddagger}}$ & $\mathbf{0.22}^{{\ddagger}}$\\
\bottomrule
\end{tabular}
\end{table}

\begin{table}[t]
\centering

\caption{Exponents (nats per use) for $J=K$ users and arms, all with the same mean $\mu$, and $\bm\eps=\bm0$; maximizing $\beta$ of generalized musical chairs in parentheses. Bold: the largest exponent of a scheme in each column. $^\dagger$: valid bound found by a numerical search over $\beta$, exact at the stated $\beta$ (\Cref{app:table-numerics}). The name-decoding ceiling is as in \Cref{tab:bandit-comparison}.}
\label{tab:gmc-users}
\setlength{\tabcolsep}{3pt}
\begin{tabular}{lcccccc}
\toprule
 & \multicolumn{3}{c}{$\mu=1$} & \multicolumn{3}{c}{$\mu=0.8$}\\
\cmidrule(lr){2-4}\cmidrule(lr){5-7}
$J$ & $2$ & $3$ & $4$ & $2$ & $3$ & $4$\\
\midrule
Common-output converse (\cref{cor:bandit-co-closed}) & $0.693$ & $1.099$ & $1.386$ & $0.654$ & $1.022$ & $1.273$\\
\midrule
Name-decoding ceiling (\cref{prop:bandit-name-cap}) & $0.693$ & $0.347$ & $0.231$ & $0.429$ & $0.214$ & $0.143$\\
Block coding, uniform $Q$ (\cref{prop:multi-bandit-exponent}) & $0.347$ & $0.176$ & $0.118$ & $0.162$ & $0.089$ & $0.062$\\
\midrule
Musical chairs, $\beta=0$ \eqref{eq:musical-chairs-exponent} & $\mathbf{0.693}$ & $0.405$ & $0.288$ & $0.511$ & $0.310$ & $0.223$\\
Generalized musical chairs (\cref{cor:gmc-exponent}) & $\mathbf{0.693}$ $(0)$ & $\mathbf{1.031}^\dagger$ $(0.643)$ & $\mathbf{0.950}^\dagger$ $({0.614})$ & $\mathbf{0.631}^\dagger$ $({0.652})$ & $\mathbf{0.536}^{{\dagger}}$ $(1)$ & $\mathbf{0.402}^{{\dagger}}$ $(1)$\\
\bottomrule
\end{tabular}
\end{table}

\paragraph{Interaction beats name decoding}
For a name-decoding scheme (\Cref{subsec:name-decoding-limits}), \Cref{cor:name-decoding-cap} caps the name-decoding exponent, the exponent of $\PP(\cN_n)$, at $I(X_{-j};Y_j\mid X_j)/(J-1)$ maximized over joint input laws, whether the scheme is block or interactive. For the noisy bandit channel this maximum is explicit.

\begin{proposition}[Name-decoding ceiling]
\label{prop:bandit-name-cap}
Let $\mathsf C(\mu,\eps):=\max_{c\in[0,1]}\bigl[\hb(c\eps+(1-c)\mu)-c\,\hb(\eps)-(1-c)\hb(\mu)\bigr]$ be the capacity of the binary-input channel with output laws $\mathrm{Bern}(\eps)$ and $\mathrm{Bern}(\mu)$. For every sequence of name-decoding schemes, the exponent of $\PP(\cN_n)$ is at most $\max_{x\in[K]}\mathsf C(\mu_x,\eps_x)/(J-1)$. For $\eps\neq\mu$, $\mathsf C(\mu,\eps)=\log(1+e^{-\varsigma})+\mu\varsigma-\hb(\mu)$, where $\varsigma:=(\hb(\eps)-\hb(\mu))/(\eps-\mu)$; in particular, $\mathsf C(\mu,0)=\log(1+e^{-\hb(\mu)/\mu})$, the capacity of a Z-channel; for $J=2$ and arms with equal means, this is the upper bound on the capacity of the bandit channel in \cite[Cor.~2]{zhang2025}.
\end{proposition}
\begin{proof}
Given $X_j=x$, the output $Y_j$ depends on $X_{-j}$ only through the indicator $A$ of a collision, and the conditional law of $Y_j$ is $\mathrm{Bern}(\eps_x)$ if $A=1$ and $\mathrm{Bern}(\mu_x)$ otherwise. Hence $I(X_{-j};Y_j\mid X_j)=I(A;Y_j\mid X_j)\leq\max_x\mathsf C(\mu_x,\eps_x)$, since given $X_j=x$ the channel from $A$ to $Y_j$ has capacity $\mathsf C(\mu_x,\eps_x)$, and the claim follows from \Cref{cor:name-decoding-cap}. The closed form follows by setting to zero the derivative of the concave objective, which gives $c\eps+(1-c)\mu=1/(1+e^{\varsigma})$. 
\end{proof}

The ceiling is governed by the best arm alone, and it does not depend on any other arm. It bounds the name-decoding exponent of every scheme, and hence the random-coding exponent of block coding, which by \eqref{eq:block-exponent-threshold} is even at most $\max_x\mathsf C(\mu_x,\eps_x)/J$. In \Cref{tab:bandit-comparison} the interactive schemes exceed it. For $\eps=0$ the comparison is proved: the exponent of musical chairs is a closed form, and that of generalized musical chairs is computed exactly. For $J$ arms of mean $\mu\in(0,1)$, musical chairs exceeds the ceiling for every $J\geq2$. Indeed, $(J-1)\bigl[-\log(1-\mu/J)\bigr]$ is nondecreasing in $J$, since its derivative in $J$ is at least $\mu(1-\mu)/(J(J-\mu))\geq0$ by $-\log(1-u)\geq u$; and with $t=1-\mu$, the inequality $-\log(1-\mu/2)>\mathsf C(\mu,0)$ reads $\varphi(t):=t\log t+(1-t)\log(1+t)<0$, which holds on $(0,1)$ because $\varphi(0^+)=\varphi(1)=0$ and $\varphi''(t)=(1-t)/(t(1+t)^2)>0$. For $\mu=1$ the two exponents are equal for $J=2$, and musical chairs is larger for $J\geq3$. For $\eps>0$ the exponents of the controllers are numerical estimates; if they are accurate, the optimized controller exceeds the name-decoding ceiling by $14\%$ to $22\%$. The comparison is least certain at $\eps=0.2$, where the margin is smallest and only two digits of the estimate are reliable. Directly contending for roles through the channel is not a form of name decoding, and it does better.

\paragraph{Gap to the converse}
For $\eps>0$ the converse is about twice the exponent of the optimized controller. Both sides are numerical at $\eps\in\{0.05,0.1\}$: a better search for the reference could lower the converse, and better parameters could raise the exponent of the controller, so the gap may be smaller. The optimized controller improves on the uniform one by about $10\%$, through the redraw probability and the score and threshold parameters. For $\eps=0$ and $J=2$, \Cref{thm:bandit-one-arm-tight} closes the gap when there is exactly one positive arm and a silent arm. The gap is also closed for two arms of mean $\mu=1$ without a silent arm (first column of \Cref{tab:gmc-users}): musical chairs has $P_{\mathrm b}^{(n)}=2^{-n}$, since the users leave their initial state exactly when they draw distinct arms, and \Cref{prop:bandit-co-converse} gives $\epsilon_n^\star(P)\geq\frac12\,2^{-n}$, so that the optimal error probability is again determined within a factor of two. With two positive arms and a silent one a gap remains between the converse and musical chairs, one-arm musical chairs, which has the same exponent at equal means, and generalized musical chairs on the two positive arms. Without the silent arm, the converse for $\mu=0.8$ is $0.654$, against $0.631$ for generalized musical chairs (\Cref{tab:gmc-users}), so that most of the gap at $\eps=0$ in \Cref{tab:bandit-comparison} is due to the silent arm, which raises the converse but which the schemes on the two positive arms do not use. Whether schemes that use both positive arms and the silent arm close the gap is open.

\paragraph{Scaling with the number of users}
The $m=2$ converse bounds of \Cref{subsec:bandit-converse} do not depend on $J$ beyond $J\geq3$. The exponent of musical chairs is $-\log(1-\mu/J)\sim\mu/J$, the name-decoding ceiling is at most $\log2/(J-1)$, and the random-coding exponent of block coding is at most $\log2/J$. These three exponents decay like $1/J$, but the decay is not intrinsic to interactive schemes: generalized musical chairs, which does not decode names, stays above the ceiling, and far above musical chairs, at $J=3$ and $J=4$ in \Cref{tab:gmc-users}. Against the $m=2$ converse the gap is much smaller than the decay of musical chairs suggests, though it still widens from $J=3$ to $J=4$ as the converse grows with the number of arms. Whether the true exponent decays with $J$, and whether kernels with $m>2$ show it, is open.

\section{Conclusion}
\label{sec:conclusion}

We studied how fast the error probability of anonymous symmetry breaking can decay with the blocklength over symmetric multiway channels. A R\'enyi change of measure gives a converse for every symmetric causal strategy, and for every population vector with at least two positive entries. The name-decoding exponent, that is, the exponent of the probability that two names coincide or are decoded wrongly, is limited by a cut-set bound, and so is the random-coding exponent of block coding; controllers that contend for identifiers without decoding names do not face this limit. On the noisy bandit channel, interaction pays: musical chairs already provably exceeds the limit of name decoding when collisions are noiseless and the arms have a common mean below one, and generalized musical chairs, introduced here, improves substantially on musical chairs. For two users, noiseless collisions, one informative arm and a silent arm, the converse and one-arm musical chairs determine the optimal error probability up to a factor of two, and so do the converse and musical chairs for two arms of mean one.

Several questions remain open.
\begin{itemize}
 \item \emph{Gap between converse and schemes.} With noisy collisions the converse is about twice the exponent of the best controller we found, and with two positive arms and a silent arm it exceeds the exponent of generalized musical chairs. It is unknown whether the converse or the schemes are loose. We have no proof that the confidence controllers have a positive exponent, and it is open whether a controller with a bounded state set and time-varying maps, which can drive the error probability to zero on the noisy bandit channel, can achieve a positive exponent when every output has positive probability.
 \item \emph{Dependence on the number of users.} The converse bounds we evaluated involve $m=2$ selected users and do not depend on $J$ beyond $J\geq3$. The exponents of musical chairs and block coding decay like $1/J$, and so does the ceiling on the name-decoding exponent, but the exponents of generalized musical chairs that we evaluated, for $J=3$ and $J=4$, are much larger and too few to show a rate. How the optimal exponent depends on $J$ is open, and so is whether kernels that involve $m>2$ users, which we did not evaluate, capture it.
 \item \emph{Sharper converses.} The references used here are built from kernels with a product structure over the selected users. Posterior-dependent references, and kernels with several users, may give tighter bounds.
 \item \emph{Population vectors.} Every achievability result holds for any population vector, in particular for leader election, and so do the converses, with the same kernels and a range of $m$ that depends on the vector, when it has at least two positive entries (\Cref{prop:population-converse}); otherwise the task is trivial. Whether the intermediate tasks, and leader election in particular, have strictly larger optimal exponents than the task $\bm N=(1,\ldots,1)$ is open.
 \item \emph{Unknown channels and complexity.} We assumed that the channel and the number of users are known, and that the users are synchronized. Schemes that learn the channel parameters while breaking the symmetry, universal decoding for block coding, and low-complexity implementations are of practical interest, as is the use of symmetry breaking as the initialization phase of multiplayer bandit algorithms: with horizon $T$, an initialization of length about $(\log T)/E$, for a scheme of exponent $E$, keeps the probability of a failed initialization of order $1/T$, so that $1/E$ multiplies $\log T$ in the regret.
\end{itemize}

\appendices
\crefalias{section}{appendix}
\section{Proof of the Kernel-Reference Lemmas}
\label{app:kernel-reference}
\label{app:CO-iid}
\label{app:CO-radius}

We prove Lemma~\ref{lem:product-form} and \Cref{lem:factorized-reference-structure}. Fix $\pi_n$, $c$, $2\leq m\leq J$, and $B\in\mathfrak A_{m,J}(P)$; for \Cref{lem:product-form} assume $\beta_B^{\pi_n,c}>0$, and for \Cref{lem:factorized-reference-structure} assume $\kappa_m(B)>0$. Throughout, $x_j[i]=f_i(c,s_j,y_j^{i-1})$, $u[i]=x_{m+1:J}[i]$ and $v[i]=y_{m+1:J}[i]$. The seeds may take values in a general space; the outputs use counting measure.

\paragraph{Branch decomposition}
For $0\leq t\leq n$, let $R^{(t)}$ denote the measure \eqref{eq:kernel-trajectory-measure} with $n$ replaced by $t$, so that $R^{(n)}=R_B^{\pi_n,c}$ and $R^{(0)}=P_S^{\otimes J}$. Expanding each factor $B(y[i]\mid x[i])$ according to \eqref{eq:latent-factorized-kernel} gives $R^{(t)}=\sum_{r^t\in\mathcal R^t}R_{r^t}$ with
\begin{equation}
 R_{r^t}(ds_{[J]},dy_{[J]}^t):=\prod_{j=1}^JP_S(ds_j)\prod_{i=1}^t\left[\prod_{j=1}^ma_{r_i}\bigl(x_j[i],y_j[i]\bigr)\right]c_{r_i}\bigl(v[i]\bigm| x[i],y_{[m]}[i]\bigr).
 \label{eq:branch-measure}
\end{equation}
Define the one-user measure
\begin{equation}
 \lambda_{r^t}(ds,dy^t):=P_S(ds)\prod_{i=1}^ta_{r_i}\bigl(f_i(c,s,y^{i-1}),y[i]\bigr),
 \qquad \ell(r^t):=\lambda_{r^t}(\text{total}).
 \label{eq:partial-selected-seed-measure}
\end{equation}

\emph{Claim.} The marginal of $R_{r^t}$ on the selected views $(s_{[m]},y_{[m]}^t)$ equals $\zeta(r^t)\prod_{j=1}^m\lambda_{r^t}(ds_j,dy_j^t)$ for a constant $\zeta(r^t)\geq0$, and $\zeta\equiv1$ under (N).

\emph{Under (I).} The factors $c_{r_i}(u[i],v[i])$ involve only unselected seeds, inputs and outputs. The unselected inputs are functions of the unselected seeds and outputs, and the selected inputs are functions of the selected seeds and outputs. Hence $R_{r^t}$ is exactly the product of $\prod_{j\leq m}\lambda_{r^t}(ds_j,dy_j^t)$ and
\[
 \eta_{r^t}(ds_{m+1:J},dv^t):=\prod_{j>m}P_S(ds_j)\prod_{i=1}^tc_{r_i}(u[i],v[i]),
\]
and the claim holds with $\zeta(r^t):=\eta_{r^t}(\text{total})$.

\emph{Under (N).} Integrate out the unselected outputs in the order $v[t],v[t-1],\ldots,v[1]$, then the unselected seeds. Suppose $v[k]$ has already been integrated out for every $k>i$. The factors $c_{r_k}$ for $k>i$ have then disappeared, and they were the only factors containing the unselected inputs $u[k]$ for $k>i$. These inputs are the only other places where $v[i]$ appears. The one remaining factor containing $v[i]$ is therefore $c_{r_i}(v[i]\mid x[i],y_{[m]}[i])$, which sums to one by (N). The selected factors never involve unselected variables. After all unselected variables are integrated out, what remains is $\prod_{j\leq m}\lambda_{r^t}(ds_j,dy_j^t)$.

The claim does not assert that the unselected views are independent of the selected ones under (N), and this is not needed. In both cases the total mass of $R_{r^t}$ is
\begin{equation}
 w(r^t):=\zeta(r^t)\,\ell(r^t)^m.
 \label{eq:partial-branch-masses}
\end{equation}

\paragraph{{Proof of \Cref{lem:product-form}}}
Let $\widehat Q$ be the probability law on $(r^n,h_{[J]})$ proportional to $R_{r^n}(dh_{[J]})$. Its $h$-marginal is $Q_B^{\pi_n,c}$. Conditionally on $r^n$ with $w(r^n)>0$, the claim shows that the selected views have law $\prod_{j\leq m}\lambda_{r^n}/\ell(r^n)$, so they are i.i.d. Moreover, since the seed law and the causal maps are common to both measures, $B\ll P$ gives $dR_B^{\pi_n,c}/dP^{\pi_n,c}=\prod_iB(Y[i]\mid X[i])/P(Y[i]\mid X[i])$, hence \eqref{eq:causal-trajectory-likelihood-ratio}, and in particular $Q_B^{\pi_n,c}\ll P^{\pi_n,c}$. Hence $Q_B^{\pi_n,c}\in\mathfrak L_{m,J}^{(n)}(\pi_n,c;P)$ with $\Lambda=r^n$.

\paragraph{{Proof of \Cref{lem:factorized-reference-structure}, first item}}
Let $M_t:=\sum_{r^t}w(r^t)$, so $M_0=1$ and $M_n=\beta_B^{\pi_n,c}$. These are masses of unnormalized prefix measures, not marginals of $Q_B^{\pi_n,c}$. Fix a prefix $r^t$ with $w(r^t)>0$. Let $p_{r^t}$ be the law of the next input $f_{t+1}(c,S,Y^t)$ of one selected user under $\lambda_{r^t}/\ell(r^t)$. Summing $a_r(x,y[t+1])$ over $y[t+1]$ gives
\begin{equation}
 \ell(r^t,r)=\ell(r^t)\sum_xp_{r^t}(x)A_r(x).
 \label{eq:partial-selected-child-mass}
\end{equation}
Under (N), $\zeta(r^t,r)=1=\zeta(r^t)$. Under (I), let $q_{r^t}$ be the law of the next unselected input tuple under $\eta_{r^t}/\zeta(r^t)$; then $\zeta(r^t,r)=\zeta(r^t)\sum_uq_{r^t}(u)C_r(u)$. Setting $C_r\equiv1$ and letting $q_{r^t}$ be arbitrary under (N), both cases give
\begin{align}
 \sum_rw(r^t,r)
 &=w(r^t)\sum_r\left[\sum_xp_{r^t}(x)A_r(x)\right]^m\left[\sum_uq_{r^t}(u)C_r(u)\right]\notag\\
 &=w(r^t)\,\EE_{p_{r^t}^{\otimes m}q_{r^t}}[b_B(X_{[J]})]
 \geq w(r^t)\,\kappa_m(B).
 \label{eq:partial-mass-extension}
\end{align}
The last step uses \eqref{eq:partial-persistence-kernel-form} and the linearity of the expectation in $q_{r^t}$. The inequality \eqref{eq:partial-mass-extension} also holds when $w(r^t)=0$: then $R_{r^t}$ is the zero measure, so every extension $(r^t,r)$ has $w(r^t,r)=0$. Summing over $r^t$ gives $M_{t+1}\geq\kappa_m(B)M_t$, and iterating proves $\beta_B^{\pi_n,c}\geq\kappa_m(B)^n$.

\paragraph{{Proof of \Cref{lem:factorized-reference-structure}, second item}}
Let $g_i:=B(Y[i]\mid X[i])/P(Y[i]\mid X[i])$ and $F_n:=\prod_{i=1}^ng_i$, which is $dR_B^{\pi_n,c}/dP^{\pi_n,c}$, so that $dQ_B^{\pi_n,c}/dP^{\pi_n,c}=F_n/\beta_B^{\pi_n,c}$ by \eqref{eq:causal-trajectory-likelihood-ratio}. Under $P^{\pi_n,c}$, given the seeds and all past outputs, the current input tuple $X[i]$ is fixed and $Y[i]\sim P(\cdot\mid X[i])$. Hence, for $1<\alpha<\infty$,
\begin{equation}
 \EE\bigl[g_i^\alpha\bigm| S_{[J]},Y_{[J]}^{i-1}\bigr]=\sum_{y:P(y\mid X[i])>0}B(y\mid X[i])^\alpha P(y\mid X[i])^{1-\alpha}\leq\Psi_{\alpha,m}(B;P).
 \label{eq:partial-density-one-step-moment}
\end{equation}
By the tower property, $\EE[F_n^\alpha]\leq\Psi_{\alpha,m}(B;P)^n$. The first item then gives
\begin{align}
 D_\alpha(Q_B^{\pi_n,c}\Vert P^{\pi_n,c})
 &=\frac{\log\EE[F_n^\alpha]-\alpha\log\beta_B^{\pi_n,c}}{\alpha-1}\notag\\
 &\leq\frac{n\log\Psi_{\alpha,m}(B;P)-\alpha n\log\kappa_m(B)}{\alpha-1}
 =n\,d_{\alpha,m}^{\mathrm{ref}}(B\Vert P).
\end{align}
At order infinity, $F_n\leq\rho_{\infty,m}(B;P)^n$, and hence
\[
 \frac{dQ_B^{\pi_n,c}}{dP^{\pi_n,c}}\leq\left[\frac{\rho_{\infty,m}(B;P)}{\kappa_m(B)}\right]^n.
\] The bounds $\EE[F_n^\alpha]\leq\Psi_{\alpha,m}(B;P)^n$ and $F_n\leq\rho_{\infty,m}(B;P)^n$ use neither (I) nor (N); admissibility enters only through \Cref{lem:product-form} and the first item.

\section{{Proof of the Block-Coding Exponent}}
\label{app:multi-user-gallager}

{We prove \Cref{thm:multi-user-gallager}, conditioning on distinct names $M_1,\ldots,M_J$. If receiver $j$ errs, some unordered set $\cA\neq\{M_k:k\neq j\}$ satisfies $L_j(\cA)\geq L_j(\{M_k:k\neq j\})$ in \eqref{eq:unordered-ml-likelihood}. The users whose names are missing from $\cA$ form a nonempty set $\cS\subseteq[J]\setminus\{j\}$, and $\cA$ consists of the names of the users in $\cS^c\setminus\{j\}$ together with $s=|\cS|$ names outside $\{M_1,\ldots,M_J\}$. Every error is therefore of the following form.}

Fix receiver $j$, a nonempty set $\cS\subseteq[J]\setminus\{j\}$ of users whose names are replaced, that is, wrongly decoded, and $s=|\cS|$. The receiver input and the $J-1-s$ correctly retained foreign codewords form $X_{\cS^c}^n$. The number of unordered competing sets differing in exactly these $s$ names is
\begin{equation}
 \binom{M-J}{s}\leq(M-J)^s.
 \label{eq:multi-candidate-count}
\end{equation}
Channel symmetry {(\Cref{lem:single-local-law})} makes any assignment of the new codewords to the positions in $\cS$ equivalent.

With the likelihood
\begin{equation}
 W_j^n(y_j^n\mid x_{[J]}^n) =\prod_{i=1}^nW_j(y_j[i]\mid x_{[J]}[i])
 \label{eq:multi-product-likelihood}
\end{equation}
and $0\leq\rho\leq1$, the indicator of an ML error on $\cS$, the event that some candidate set that differs from the true one exactly in the names of $\cS$ has a likelihood at least that of the true set, satisfies
\begin{equation}
 \ind\{\text{ML error on }\cS\} \leq \Bigg( \sum_{\widetilde m_\cS} \bigg[ \frac{W_j^n\bigl(Y_j^n\mid \ins{\cS}{X_\cS^n(\widetilde m_\cS)}{X_{\cS^c}^n}\bigr)}{W_j^n(Y_j^n\mid X_{[J]}^n)} \bigg]^{\frac1{1+\rho}} \Bigg)^\rho,
 \label{eq:multi-gallager-indicator}
\end{equation}
where the sum is taken over one representative tuple $\widetilde m_\cS$ of $s$ distinct unused names per candidate set. Indeed, on the error event at least one summand is at least one; off the event the right side is nonnegative.

Concavity of $t\mapsto t^\rho$ gives the factor $(M-J)^{s\rho}${: conditioned on the transmitted codewords and on $Y_j^n$, the competing codewords are i.i.d.\ with law $Q_X^n$, because distinct names index independently generated codewords, so Jensen's inequality and \eqref{eq:multi-candidate-count} move the expectation inside $(\cdot)^\rho$}. Averaging the candidate tuple, the true tuple, the conditioned correct codewords, and the output yields the one-letter quantity
\begin{equation}
 \sum_{x_{\cS^c},y_j}\Bigl(\prod_{k\in\cS^c}Q_X(x_k)\Bigr)\Bigl[\sum_{x_\cS}\Bigl(\prod_{k\in\cS}Q_X(x_k)\Bigr)W_j(y_j\mid x_{[J]})^{\frac1{1+\rho}}\Bigr]^{1+\rho}.
 \label{eq:multi-one-letter-gallager-proof}
\end{equation}
This is $e^{-{E_{0,s}}(\rho,Q_X)}$. Product coding and memorylessness raise it to the $n${-}th power. There are $\binom{J-1}{s}$ choices of $\cS$, so {the union bound over $\cS$, with $\rho$ optimized separately for each $s$,} proves \eqref{eq:multi-user-gallager-bound}.

For $M=\lceil e^{nR}\rceil$, optimize each term over $\rho$, sum over the $J$ receivers, and choose a deterministic codebook no worse than the ensemble average. Adding the collision term from \eqref{eq:birthday-bound} proves \eqref{eq:multi-user-sync-bound}. {Taking $-\frac1n\log$ and optimizing over $R$ gives part (iii).}

For completeness, differentiating \eqref{eq:multi-user-ezero} at zero can be done by writing its inner sum as an exponential moment of $\log W_j(Y_j\mid X_{[J]})$. At $\rho=0$ the normalizing expression is one; the derivative of its negative logarithm is
\begin{equation}
 \sum_{x_{[J]},y_j}Q_X^J(x_{[J]})W_j(y_j\mid x_{[J]})\log\frac{W_j(y_j\mid x_{[J]})}{\sum_{\widetilde x_\cS}Q_X^s(\widetilde x_\cS)W_j(y_j\mid\ins{\cS}{\widetilde x_\cS}{x_{\cS^c}})}=I(X_\cS;Y_j\mid X_{\cS^c}),
 \label{eq:multi-ezero-derivative-proof}
\end{equation}
which verifies \eqref{eq:multi-ezero-derivative}.

\section{Proof of the Monotonicity of Normalized Subset Information}
\label{app:normalized-information}

We prove \Cref{lem:normalized-information-decreasing}.  Let
\begin{equation}
 I_s=I(X_{[s]};Y\mid X_{[m]\setminus[s]},T),
 \qquad 1\leq s\leq m.
 \label{eq:appendix-is-definition}
\end{equation}
Fix $s<m$ and put
\begin{equation}
 U=(Y,T,X_{s+2},\ldots,X_m).
 \label{eq:appendix-u-definition}
\end{equation}
Because the inputs are i.i.d. and independent of $T$,
\begin{align}
 I_s &=sH(X_1)-H(X_1,\ldots,X_s\mid X_{s+1},U),
 \label{eq:appendix-is-entropy}\\
 I_{s+1} &=(s+1)H(X_1)-H(X_1,\ldots,X_{s+1}\mid U).
 \label{eq:appendix-isplus-entropy}
\end{align}
Let $A=H(X_1,\ldots,X_s\mid X_{s+1},U)$.  By the chain rule,
\begin{equation}
 H(X_1,\ldots,X_{s+1}\mid U) =H(X_{s+1}\mid U)+A.
 \label{eq:appendix-chain-a}
\end{equation}
Moreover,
\begin{align}
 A &\leq\sum_{i=1}^sH(X_i\mid X_{s+1},U) \notag\\
 &\leq\sum_{i=1}^sH(X_i\mid U) =sH(X_{s+1}\mid U).
 \label{eq:appendix-a-bound}
\end{align}
The final equality in \eqref{eq:appendix-a-bound} follows from exchangeability. Indeed, the inputs are i.i.d., and the conditional law of $Y$ is invariant under permutations of $X_1,\ldots,X_{s+1}$. Combining
\eqref{eq:appendix-chain-a} and \eqref{eq:appendix-a-bound} yields
\begin{equation}
 \frac{H(X_1,\ldots,X_{s+1}\mid U)}{s+1}
 \geq\frac{A}{s}.
 \label{eq:appendix-normalized-entropy}
\end{equation}
Subtract the two sides from $H(X_1)$ using
\eqref{eq:appendix-is-entropy}--\eqref{eq:appendix-isplus-entropy} to obtain
$I_s/s\geq I_{s+1}/(s+1)$.

\section{Proof of the Multiway Cut-Set Bound}
\label{app:cutset}

We prove \Cref{prop:multi-user-cutset}. Let $M_1,\ldots,M_J$ be the independent uniform names, each on a set of size $\lceil e^{nR}\rceil$, and $\Gamma_j=(M_j,\Theta_j)$ the name together with the private randomness of user~$j$. Then $\Gamma_1,\ldots,\Gamma_J$ are independent, and independent of the common randomness $C$, and $X_j[i]$ is a function of $(C,\Gamma_j,Y_j^{i-1})$. Until \eqref{eq:cutset-unequal-rates}, all entropies and mutual informations are conditional on $C$, which we omit from the notation. Fix a nonempty proper subset $\cS\subset[J]$, and let $\mathfrak M_\cS$ be the multiset of the names of the users in $\cS$. Every terminal knows its own name and estimates the multiset of all names, so that the terminals of $\cS^c$, viewed collectively, know $M_{\cS^c}$ and can recover $\mathfrak M_\cS$ from $(\Gamma_{\cS^c},Y_{\cS^c}^n)$ with vanishing error probability. The multiset loses at most the order of the $|\cS|$ names, so that
\begin{equation}
 H(\mathfrak M_\cS\mid \Gamma_{\cS^c})\geq |\cS|\log\lceil e^{nR}\rceil-\log|\cS|!\geq|\cS|\,nR-\log|\cS|!.
\end{equation}
Fano's inequality gives
\begin{equation}
 H(\mathfrak M_\cS\mid \Gamma_{\cS^c},Y_{\cS^c}^n)\leq n\varepsilon_n,
 \qquad \varepsilon_n\to0.
 \label{eq:cutset-fano}
\end{equation}
Since $\mathfrak M_\cS$ is a function of $\Gamma_\cS$,
\begin{align}
 n|\cS|R-\log|\cS|!
 &\leq I(\Gamma_\cS;Y_{\cS^c}^n\mid \Gamma_{\cS^c})+n\varepsilon_n
 \notag\\
 &=\sum_{i=1}^n
 I(\Gamma_\cS;Y_{\cS^c}[i]\mid
 \Gamma_{\cS^c},Y_{\cS^c}^{i-1})+n\varepsilon_n.
 \label{eq:cutset-chain-rule}
\end{align}
We bound each term by a difference of entropies. The input $X_{\cS^c}[i]$ is a function of $(\Gamma_{\cS^c},Y_{\cS^c}^{i-1})$, and conditioning reduces entropy, so that
\[
 H(Y_{\cS^c}[i]\mid \Gamma_{\cS^c},Y_{\cS^c}^{i-1})\leq H(Y_{\cS^c}[i]\mid X_{\cS^c}[i]).
\]
Also $X_{[J]}[i]$ is a function of $(\Gamma_{[J]},Y_{[J]}^{i-1})$, and the channel is memoryless, so that
\[
 H(Y_{\cS^c}[i]\mid \Gamma_{[J]},Y_{\cS^c}^{i-1})\geq H(Y_{\cS^c}[i]\mid \Gamma_{[J]},Y_{[J]}^{i-1})=H(Y_{\cS^c}[i]\mid X_{[J]}[i]).
\]
Subtracting,
\begin{equation}
 I(\Gamma_\cS;Y_{\cS^c}[i]\mid
 \Gamma_{\cS^c},Y_{\cS^c}^{i-1})
 \leq I(X_\cS[i];Y_{\cS^c}[i]\mid X_{\cS^c}[i]),
 \label{eq:cutset-single-time}
\end{equation}
where the right-hand side is evaluated under the joint law of $X_{[J]}[i]$ induced by the protocol; this law is arbitrary, since the inputs may be dependent through the past outputs.

Let $U$ be uniform on $[n]$, independent of the experiment, and define $(X_{[J]},Y_{[J]})=(X_{[J]}[U],Y_{[J]}[U])$. Divide \eqref{eq:cutset-chain-rule} by $n$, use \eqref{eq:cutset-single-time}, and make the conditioning on $C$ explicit again, to obtain, for the scheme of blocklength $n$,
\begin{equation}
 |\cS|R
 \leq I(X_\cS;Y_{\cS^c}\mid X_{\cS^c},U,C)+\varepsilon_n+\frac{\log|\cS|!}{n}.
 \label{eq:cutset-unequal-rates}
\end{equation}
Since $(U,C)-X_{[J]}-Y_{[J]}$ is a Markov chain, $I(X_\cS;Y_{\cS^c}\mid X_{\cS^c},U,C)=H(Y_{\cS^c}\mid X_{\cS^c},U,C)-H(Y_{\cS^c}\mid X_{[J]})\leq I(X_\cS;Y_{\cS^c}\mid X_{\cS^c})$, where the right-hand side is evaluated under the law $\bar P_n$ of $X_{[J]}$, averaged over $U$ and $C$. Hence $|\cS|R\leq I_{\bar P_n}(X_\cS;Y_{\cS^c}\mid X_{\cS^c})+\eta_n$ for every $\cS$, with $\eta_n\to0$. The laws on $\cX^J$ form a compact set, and for a fixed channel the mutual information is continuous in the input law, so that along a subsequence $\bar P_n$ converges to a law $P_{X_{[J]}}$ that satisfies \eqref{eq:multiway-cutset} for every $\cS$. For $\cS=[J]\setminus\{j\}$, \eqref{eq:single-receiver-cutset} follows, and $I(X_{-j};Y_j\mid X_j)\leq H(Y_j)\leq\log|\cY|$.

\section{Bounded Memory with Time-Varying Maps}
\label{app:bounded-memory}

The floor of \Cref{prop:finite-controller-floor} needs stationarity. The following controller uses the device that Cover used for hypothesis testing between two Bernoulli laws with finite memory~\cite{cover1969}.
\begin{proposition}[Bounded memory with time-varying maps]
\label{prop:bounded-memory-clock}
On the noisy bandit channel of \Cref{def:noisy-bandit} with $J$ arms of common parameters $0\leq\eps<\mu\leq1$, in particular with $0<\eps<\mu<1$, where every output has positive probability, there is a role controller with $4J$ states for every $n$, whose maps vary with time, and whose error probability tends to zero as $n\to\infty$.
\end{proposition}
\begin{proof}
The state of a user is $(a,A,F)\in[J]\times\{0,1\}^2$, initially $(1,1,0)$; the user plays arm $a$, and its identifier is $a$. Time is divided into epochs, and epoch $k$ consists of $N_k:=\lceil k\mu^{-k}\rceil$ windows of $k$ channel uses. Within a window, the flag $A$ (for \emph{all}) records whether all outputs so far equal $1$, and the flag $F$ (for \emph{found}) records whether a window of ones has been found in the current epoch; at the end of a window, $F\leftarrow\max\{F,A\}$ and $A\leftarrow1$; at the end of an epoch, a user with $F=0$ draws a new role uniformly in $[J]$, and $F\leftarrow0$. The maps depend on the time index only through this schedule, and the roles do not change during an epoch. A user alone on its arm finds no window of ones with probability $(1-\mu^k)^{N_k}\leq e^{-k}$, and a user that shares its arm finds one with probability at most $\delta_k:=N_k\eps^k\leq k(\eps/\mu)^k+\eps^k$. If the roles of epoch $k$ are distinct, they remain unchanged with probability at least $1-Je^{-k}$. Otherwise, with probability at least $1-J(\delta_k+e^{-k})$ exactly the users that share a role redraw, and the $c$ users that redraw take the $c$ free roles with probability $c!/J^c\geq J!/J^J$. Hence the probability $p_k$ that the roles of epoch $k$ are not distinct satisfies $p_{k+1}\leq p_k\bigl[1-\frac{J!}{J^J}(1-J\delta_k-Je^{-k})\bigr]+Je^{-k}$, so that $p_k\to0$, and $P_{\mathrm b}^{(n)}\leq\max\{p_k,p_{k+1}\}$ when time $n$ lies in epoch $k$.
\end{proof}
In this construction the error probability decays only subexponentially in $n$, polynomially when $\mu<1$. Indeed, for $\mu<1$ a user alone on its arm misses every window of epoch $k$ with probability of order $e^{-k}$ and then collides with probability at least $1/J$, while epoch $k$ ends at a time of order $k^2\mu^{-k}$; for $\mu=1$ all users start on the same role, a redraw of $c\geq2$ users fails with probability at least $1-c!/J^c\geq1/2$, and epoch $k$ ends at a time of order $k^3$. The floor of \Cref{prop:finite-controller-floor} is therefore a property of stationary finite-state controllers: a controller with a bounded state set may use the time index as a free clock, which \Cref{rem:clock-in-state} counts as memory.

\section{Evaluation of the Noisy Bandit Gallager Function}
\label{app:erasure-gallager}

We prove \Cref{prop:multi-bandit-exponent}. Fix the receiver $j$, an error subset of size $s$, and a realization $X_j=x$. Let $c=J-1-s$ correct foreign codewords remain fixed. By \eqref{eq:noisy-bandit-law}, the local law $W(\cdot\mid x,x_{-j})$ depends on the foreign inputs only through whether at least one of them equals $x$: it is $W_{\mu_x}$ if none does and $W_{\eps_x}$ otherwise, where $W_\nu(1)=\nu$ and $W_\nu(0)=1-\nu$; for a silent arm both laws are $\delta_0$.

Each of the $c$ correct inputs equals $x$ with probability $q_x$. If at least one does, which has probability
\begin{equation}
 1-(1-q_x)^c,
 \label{eq:erasure-proof-correct-collision}
\end{equation}
the likelihood is $W_{\eps_x}(y)$ for every candidate tuple, so that the inner sum in \eqref{eq:multi-user-ezero} is $W_{\eps_x}(y)^{1/(1+\rho)}$, its $(1+\rho)$-th power summed over $y\in\{0,1\}$ is $W_{\eps_x}(0)+W_{\eps_x}(1)=1$, and the outer average contributes \eqref{eq:erasure-proof-correct-collision}.

Suppose none of the correct inputs equals $x$, which has probability $(1-q_x)^c$. A random candidate tuple avoids $x$ with probability
\begin{equation}
 \theta_x=(1-q_x)^s.
 \label{eq:erasure-proof-ax}
\end{equation}
If it does, the likelihood of the output $y$ is $W_{\mu_x}(y)$; otherwise it is $W_{\eps_x}(y)$. The inner sum for the output $y$ is therefore
\begin{equation}
 \theta_xW_{\mu_x}(y)^{\frac1{1+\rho}}+(1-\theta_x)W_{\eps_x}(y)^{\frac1{1+\rho}}.
 \label{eq:erasure-proof-output}
\end{equation}
Raise \eqref{eq:erasure-proof-output} to the power $1+\rho$, sum over $y\in\{0,1\}$, multiply by $(1-q_x)^c$, add \eqref{eq:erasure-proof-correct-collision}, and finally average over $x\sim Q$. The result is exactly $G_{\rho,s}^{(J)}(Q;\bm\mu,\bm\eps)$ in \eqref{eq:multi-bandit-g-function}. Taking the negative logarithm proves \Cref{prop:multi-bandit-exponent}. For $J=2$, $s=1$ and $c=0$ it is \eqref{eq:two-user-bandit-ezero}. Differentiating at $\rho=0$ and using \eqref{eq:multi-ezero-derivative} gives \eqref{eq:two-user-bandit-rate} and \eqref{eq:multi-bandit-information}: for $s=J-1$ and $c=0$, \eqref{eq:erasure-proof-output} is the mixture, with weight $\theta_x$ on $W_{\mu_x}$, of the two output laws of a receiver that is alone or in collision, and the derivative is the mutual information between the collision indicator and the output given $X_j=x$, averaged over $x\sim Q$.

\section{Numerical Evaluation of the Tables}
\label{app:table-numerics}
The entries of \Cref{tab:bandit-comparison,tab:gmc-users} are of three kinds.

\paragraph{Exact entries (unmarked)}
The common-output converse, musical chairs, and the converse and the name-decoding ceiling of \Cref{tab:gmc-users} are closed forms; the noninteraction converse and the ceiling of \Cref{tab:bandit-comparison} are one-dimensional convex problems, solved to $10^{-12}$. Block coding is maximized over $\rho$ and over the laws on the two positive arms in \Cref{tab:bandit-comparison}, where the maximum is at the uniform law and $\rho=1$, and restricted to uniform laws in \Cref{tab:gmc-users}. At a given $\beta$, the spectral radius of generalized musical chairs is computed to about $10^{-10}$ on a chain of at most $6560$ states, and $-\log\rho(K_{\mathcal B})$ is then the exponent of $P_{\mathrm b}^{(n)}$ itself. For $\mu=1$, a user that is still searching after the first channel use collided at its last draw, so that every joint state outside $\mathcal G$ is a failure state. For $\mu=0.8$, the class that attains the spectral radius contains failure states in each of its cyclic classes, as we checked on the chain whose states include the last roles drawn, and \Cref{prop:persistent-state-exponent} applies.

\paragraph{Bounds found by a numerical search ($^\dagger$)}
These are valid bounds, and a better search can only improve them. Each optimized reference is an explicit feasible kernel: of order $\infty$ with $8$ components at $\eps=0.05$, rescaled so that $B\leq P$ holds exactly, and of order $1.25$ with $6$ components at $\eps=0.1$; $\Psi_{\alpha,2}(B;P)$ and $\kappa_2(B)$ are recomputed from the kernel, $\kappa_2(B)$ by an exact minimization over the simplex. The search is not convex. Its restarts agree to $10^{-8}$ at the orders that define the entries, and refining the grid of orders lowered the entry at $\eps=0.1$ from $0.653$ to $0.652$. At $\eps=0$ and $\eps=0.2$ the search does not improve on the closed forms, which are the entries. The best $\beta$ of generalized musical chairs is searched on successively finer grids; refining them to $10^{-4}$ changes no digit shown, and for $\mu=0.8$ and $J\in\{3,4\}$ the exponent increases up to $\beta=1$ on the grid.

\paragraph{Estimates ($^\ddagger$)}
The commit threshold $\lceil\nu n\rceil$ of the confidence controllers grows with $n$, so that \Cref{prop:persistent-state-exponent} does not apply, and we have no proof that their exponent exists. \Cref{prop:confidence-exact} gives $P_{\mathrm b}^{(n)}$ exactly, which we compute up to $n=960$ ($1280$ for $\eps=0.2$); it agrees with Monte Carlo simulation ($4\cdot10^6$ runs at $n=4,8,\ldots,40$) within $2.72$ binomial standard deviations. The slopes of $-\log P_{\mathrm b}^{(n)}$ decrease beyond $n=200$, and least-squares fits of $-\log P_{\mathrm b}^{(n)}=En+c\log n+a+b/n$ extrapolate them; these fits assume the form of the correction, so they give estimates. For an upper bound, let $K_L^\circ$ be the transition matrix of the controller in which no user commits and a path is killed when a counter exceeds $L$, restricted to the joint states reachable from the initial ones. Both rewards have probabilities in $(0,1)$, so that the support graph of $K_L^\circ$ is the direct product of two copies of that of one user, which is strongly connected, and every component of this product is closed and strongly connected. For every $L$ we used, the joint states reachable from the initial ones form one such component, so that $K_L^\circ$ is irreducible~(with period $2$ for the optimized controller at $\eps=0.2$, aperiodic otherwise), and it has failure states. For $n>L/\nu$ the paths on which no counter exceeds $L$ have the same probability in both chains, so that $\liminf_{n\to\infty}-\frac1n\log P_{\mathrm b}^{(n)}\leq-\log\rho(K_L^\circ)$ by the Perron--Frobenius theorem. We bracket $\rho(K_L^\circ)$ by the Collatz--Wielandt inequalities, with a positive vector obtained by power iteration, to a relative width below $10^{-6}$; standard eigenvalue solvers fail on this far-from-normal chain. \Cref{tab:bandit-comparison} shows the extrapolation, rounded, and the upper bound exceeds it by less than $5\cdot10^{-4}$ for $\eps\leq0.1$ and by less than $3\cdot10^{-3}$ for $\eps=0.2$ (\Cref{tab:controller-precision}), where both still decrease at the largest $n$ and $L$. The parameters of the controllers were tuned numerically, so that a better controller of this form may exist.

\begin{table}[t]
\centering
\caption{{Confidence controllers of \Cref{tab:bandit-comparison}: slope of $-\log P_{\mathrm b}^{(n)}$ on $[240,320]$ and on the last window ($[800,960]$, or $[960,1280]$ for $\eps=0.2$), range of the extrapolations, upper bound $-\log\rho(K_L^\circ)$ at the largest $L$, and entry of \Cref{tab:bandit-comparison}.}}
\label{tab:controller-precision}
\begin{tabular}{llcccccc}
\toprule
 & $\eps$ & slope $[240,320]$ & slope, last window & extrapolation & upper bound & $L$ & entry\\
\midrule
{uniform} & $0.05$ & $0.36950$ & $0.36946$ & $0.36945$--$0.36947$ & $0.36947$ & $800$ & $0.369$\\
 & $0.1$ & $0.30488$ & $0.30443$ & $0.30424$--$0.30432$ & $0.30470$ & $800$ & $0.304$\\
 & $0.2$ & $0.20763$ & $0.20543$ & $0.20439$--$0.20457$ & $0.20491$ & $1600$ & $0.20$\\
optimized & $0.05$ & $0.40568$ & $0.40567$ & $0.40567$--$0.40567$ & $0.40567$ & $1600$ & $0.406$\\
 & $0.1$ & $0.33747$ & $0.33736$ & $0.33730$--$0.33731$ & $0.33765$ & $800$ & $0.337$\\
 & $0.2$ & $0.22623$ & $0.22385$ & $0.22306$--$0.22314$ & $0.22597$ & $1600$ & $0.22$\\
\bottomrule
\end{tabular}
\end{table}

\section*{Acknowledgment}
The authors thank Henrique Miyamoto for helpful discussions on the problem formulation and the error-exponent analysis, and for his careful reading of the manuscript and his constructive suggestions.

\bibliographystyle{IEEEtran}
\bibliography{references}

@article{aldawsari2020multichannel,
  title = {Naming in Multichannel with Beeps in the Strong Model},
  author = {Aldawsari, Layla S. and Altman, Tom},
  journal = {Appl. Sci.},
  volume = {10},
  number = {20},
  pages = {7164},
  year = {2020},
  publisher = {MDPI},
  doi = {10.3390/app10207164},
}

@ARTICLE{Anandkumar2011,
  author = {Anandkumar, Animashree and Michael, Nithin and Tang, Ao Kevin and Swami, Ananthram},
  journal = {IEEE J. Sel. Areas Commun.},
  title = {Distributed Algorithms for Learning and Cognitive Medium Access with Logarithmic Regret},
  year = {2011},
  volume = {29},
  number = {4},
  pages = {731--745},
  doi = {10.1109/JSAC.2011.110406},
}

@article{ashkenazi2022noisy,
  author = {Ashkenazi, Yagel and Gelles, Ran and Leshem, Amir},
  title = {Noisy Beeping Networks},
  journal = {Inf. Comput.},
  year = {2022},
  volume = {289},
  note = {Art. no.~104925},
  doi = {10.1016/j.ic.2022.104925},
}

@article{berg2023survey,
  author = {Berg, Tomer and Ordentlich, Or and Shayevitz, Ofer},
  title = {Statistical Inference With Limited Memory: A Survey},
  journal = {IEEE J. Sel. Areas Inf. Theory},
  year = {2024},
  volume = {5},
  pages = {623--644},
  doi = {10.1109/JSAIT.2024.3481296},
}

@inproceedings{boursier_sic-mmab_2019,
  author = {Boursier, Etienne and Perchet, Vianney},
  booktitle = {Proc. Adv. Neural Inf. Process. Syst. (NeurIPS 2019)},
  title = {{SIC}-{MMAB}: Synchronisation Involves Communication in Multiplayer Multi-Armed Bandits},
  volume = {32},
  year = {2019},
}

@article{boursier_survey_2024,
  author = {Etienne Boursier and Vianney Perchet},
  title = {A Survey on Multi-player Bandits},
  journal = {J. Mach. Learn. Res.},
  year = {2024},
  volume = {25},
  number = {137},
  pages = {1--45},
}

@article{chlebus2017naming,
  author = {Chlebus, Bogdan S. and De Marco, Gianluca and Talo, Muhammed},
  title = {Naming a Channel with Beeps},
  journal = {Fundam. Inform.},
  volume = {153},
  number = {3},
  pages = {199--219},
  year = {2017},
  doi = {10.3233/FI-2017-1537},
}

@article{cover1969,
  author = {Cover, T. M.},
  journal = {Ann. Math. Statist.},
  title = {Hypothesis Testing with Finite Statistics},
  year = {1969},
  volume = {40},
  number = {3},
  pages = {828-835},
}

@article{cover1970,
  author = {Cover, T. M. and Hellman, M. E.},
  journal = {IEEE Trans. Inf. Theory},
  title = {The Two-Armed-Bandit Problem With Time-Invariant Finite Memory},
  year = {1970},
  volume = {16},
  number = {2},
  pages = {185-195},
}

@article{cuff2010coordination,
  author = {Cuff, Paul and Permuter, Haim H. and Cover, Thomas M.},
  title = {Coordination Capacity},
  journal = {IEEE Trans. Inf. Theory},
  volume = {56},
  number = {9},
  pages = {4181--4206},
  year = {2010},
  doi = {10.1109/TIT.2010.2054651},
}

@article{dickinson2014,
  author = {Dickinson, Peter J. C. and Gijben, Luuk},
  title = {On the Computational Complexity of Membership Problems for the Completely Positive Cone and its Dual},
  journal = {Comput. Optim. Appl.},
  year = {2014},
  volume = {57},
  number = {2},
  pages = {403--415},
  doi = {10.1007/s10589-013-9594-z},
}

@article{dueck1979capacity,
  title = {The capacity region of the two-way channel can exceed the inner bound},
  author = {Dueck, Gunter},
  journal = {Inf. Control},
  volume = {40},
  number = {3},
  pages = {258--266},
  year = {1979},
  publisher = {Elsevier},
}

@book{el_gamal_network_2012,
  author = {El Gamal, Abbas and Kim, Young-Han},
  title = {Network Information Theory},
  publisher = {Cambridge Univ. Press},
  address = {Cambridge, U.K.},
  year = {2011},
}

@book{gallager1968,
  author = {Robert G. Gallager},
  title = {Information Theory and Reliable Communication},
  publisher = {Wiley},
  address = {New York, NY, USA},
  year = {1968},
}

@article{gallager1985perspective,
  author = {Gallager, Robert G.},
  title = {A Perspective on Multiaccess Channels},
  journal = {IEEE Trans. Inf. Theory},
  year = {1985},
  volume = {31},
  number = {2},
  pages = {124--142},
  doi = {10.1109/TIT.1985.1057022},
}

@article{garivier2019explore,
  author = {Garivier, Aur{\'e}lien and M{\'e}nard, Pierre and Stoltz, Gilles},
  title = {Explore First, Exploit Next: The True Shape of Regret in Bandit Problems},
  journal = {Math. Oper. Res.},
  year = {2019},
  volume = {44},
  number = {2},
  pages = {377--399},
  doi = {10.1287/moor.2017.0928},
}

@inproceedings{ghaffari2016leader,
  author = {Ghaffari, Mohsen and Newport, Calvin},
  title = {Leader Election in Unreliable Radio Networks},
  booktitle = {Proc. 43rd Int. Colloq. Automata, Lang., Program. (ICALP)},
  series = {LIPIcs},
  volume = {55},
  pages = {138:1--138:14},
  year = {2016},
  publisher = {Schloss Dagstuhl -- Leibniz-Zentrum f{\"u}r Informatik},
  doi = {10.4230/LIPIcs.ICALP.2016.138},
}

@article{han1984general,
  title = {A general coding scheme for the two-way channel},
  author = {Han, Te},
  journal = {IEEE Trans. Inf. Theory},
  volume = {30},
  number = {1},
  pages = {35--44},
  year = {1984},
  publisher = {IEEE},
}

@book{hardy1952inequalities,
  author = {Hardy, G. H. and Littlewood, J. E. and P{\'o}lya, G.},
  title = {Inequalities},
  publisher = {Cambridge Univ. Press},
  address = {Cambridge, U.K.},
  edition = {2nd},
  year = {1952},
}

@article{haroutunian1977,
  author = {Haroutunian, E. A.},
  title = {Lower Bound for Error Probability in Channels with Feedback},
  journal = {Probl. Inf. Transm.},
  year = {1977},
  volume = {13},
  number = {2},
  pages = {107--114},
  note = {Translated from \emph{Probl. Peredachi Inf.}, vol.~13, no.~2, pp.~36--44, 1977},
}

@inproceedings{hayashi1999initialization,
  author = {Hayashi, Tatsuya and Nakano, Koji and Olariu, Stephan},
  title = {Randomized Initialization Protocols for Packet Radio Networks},
  booktitle = {Proc. 13th Int. Parallel Process. Symp. and 10th Symp. Parallel Distrib. Process. (IPPS/SPDP)},
  pages = {544--548},
  year = {1999},
  doi = {10.1109/IPPS.1999.760530},
}

@article{hellman1970,
  author = {Hellman, M. E. and Cover, T. M.},
  journal = {Ann. Math. Statist.},
  title = {Learning with Finite Memory},
  year = {1970},
  volume = {41},
  number = {3},
  pages = {765-782},
  doi = {10.1214/aoms/1177696958},
}

@InProceedings{huang22a,
  title = {Towards Optimal Algorithms for Multi-Player Bandits without Collision Sensing Information},
  author = {Huang, Wei and Combes, Richard and Trinh, Cindy},
  booktitle = {Proc. 35th Conf. Learn. Theory (COLT)},
  pages = {1990--2012},
  year = {2022},
  volume = {178},
}

@article{itai1990symmetry,
  author = {Itai, Alon and Rodeh, Michael},
  title = {Symmetry Breaking in Distributed Networks},
  journal = {Inf. Comput.},
  year = {1990},
  volume = {88},
  number = {1},
  pages = {60--87},
  doi = {10.1016/0890-5401(90)90004-2},
}

@article{kaufmann2016complexity,
  author = {Kaufmann, Emilie and Capp{\'e}, Olivier and Garivier, Aur{\'e}lien},
  title = {On the Complexity of Best-Arm Identification in Multi-Armed Bandit Models},
  journal = {J. Mach. Learn. Res.},
  year = {2016},
  volume = {17},
  number = {1},
  pages = {1--42},
}

@article{lim2011noisy,
  author = {Sung Hoon Lim and Young-Han Kim and Abbas {El Gamal} and
             Sae-Young Chung},
  title = {Noisy Network Coding},
  journal = {IEEE Trans. Inf. Theory},
  year = {2011},
  volume = {57},
  number = {5},
  pages = {3132--3152},
  doi = {10.1109/TIT.2011.2119930},
}

@article{lipton1990processor,
  author = {Lipton, Richard J. and Park, Arvin},
  title = {The Processor Identity Problem},
  journal = {Inf. Process. Lett.},
  year = {1990},
  volume = {36},
  number = {2},
  pages = {91--94},
  doi = {10.1016/0020-0190(90)90103-5},
}

@article{lugosi2022multiplayer,
  author = {Lugosi, G{\'a}bor and Mehrabian, Abbas},
  title = {Multiplayer Bandits Without Observing Collision Information},
  journal = {Math. Oper. Res.},
  year = {2022},
  volume = {47},
  number = {2},
  pages = {1247--1265},
  doi = {10.1287/moor.2021.1168},
}

@article{magesh2022decentralized,
  author = {Magesh, Akshayaa and Veeravalli, Venugopal V.},
  title = {Decentralized Heterogeneous Multi-Player Multi-Armed Bandits With Non-Zero Rewards on Collisions},
  journal = {IEEE Trans. Inf. Theory},
  year = {2022},
  volume = {68},
  number = {4},
  pages = {2622--2634},
  doi = {10.1109/TIT.2021.3136095},
}

@article{massey1985collision,
  author = {Massey, James L. and Mathys, Peter},
  title = {The Collision Channel Without Feedback},
  journal = {IEEE Trans. Inf. Theory},
  year = {1985},
  volume = {31},
  number = {2},
  pages = {192--204},
  doi = {10.1109/TIT.1985.1057010},
}

@article{nakano2000initialization,
  author = {Nakano, Koji and Olariu, Stephan},
  title = {Energy-Efficient Initialization Protocols for Single-Hop Radio
    Networks with No Collision Detection},
  journal = {IEEE Trans. Parallel Distrib. Syst.},
  volume = {11},
  number = {8},
  pages = {851--863},
  year = {2000},
  doi = {10.1109/71.877942},
}

@article{nazari2014error,
  author = {Nazari, Ali and Anastasopoulos, Achilleas and Pradhan, S. Sandeep},
  title = {Error Exponent for Multiple-Access Channels: Lower Bounds},
  journal = {IEEE Trans. Inf. Theory},
  year = {2014},
  volume = {60},
  number = {9},
  pages = {5095--5115},
  doi = {10.1109/TIT.2014.2331952},
}

@inproceedings{newport2014radio,
  author = {Newport, Calvin},
  title = {Radio Network Lower Bounds Made Easy},
  booktitle = {Proc. 28th Int. Symp. Distrib. Comput. (DISC)},
  editor = {Kuhn, Fabian},
  series = {Lecture Notes in Computer Science},
  volume = {8784},
  publisher = {Springer},
  address = {Berlin, Heidelberg, Germany},
  year = {2014},
  pages = {258--272},
  doi = {10.1007/978-3-662-45174-8_18},
}

@inproceedings{polyanskiy2010arimoto,
  author = {Polyanskiy, Yury and Verd{\'u}, Sergio},
  title = {{Arimoto} Channel Coding Converse and {R\'enyi} Divergence},
  booktitle = {Proc. 48th Annu. Allerton Conf. Commun., Control, Comput.},
  year = {2010},
  pages = {1327--1333},
  doi = {10.1109/ALLERTON.2010.5707067},
}

@article{polyanskiy2010channel,
  author = {Polyanskiy, Yury and Poor, H. Vincent and Verd{\'u}, Sergio},
  title = {Channel Coding Rate in the Finite Blocklength Regime},
  journal = {IEEE Trans. Inf. Theory},
  year = {2010},
  volume = {56},
  number = {5},
  pages = {2307--2359},
  doi = {10.1109/TIT.2010.2043769},
}

@inproceedings{polyanskiy2017perspective,
  title = {A perspective on massive random-access},
  author = {Polyanskiy, Yury},
  booktitle = {Proc. 2017 IEEE Int. Symp. Inf. Theory (ISIT)},
  pages = {2523--2527},
  year = {2017},
}

@inproceedings{rosenski_multi-player_nodate,
  author = {Rosenski, Jonathan and Shamir, Ohad and Szlak, Liran},
  title = {Multi-Player Bandits -- a Musical Chairs Approach},
  booktitle = {Proc. 33rd Int. Conf. Mach. Learn. (ICML)},
  pages = {155--163},
  year = {2016},
  volume = {48},
}

@inproceedings{shannon_two-way_1961,
  author = {Shannon, Claude E.},
  title = {Two-Way Communication Channels},
  booktitle = {Proc. 4th Berkeley Symp. Math. Statist. Probab.},
  editor = {Neyman, Jerzy},
  volume = {1},
  publisher = {Univ. California Press},
  address = {Berkeley, CA, USA},
  year = {1961},
  pages = {611--644},
}

@inproceedings{shi_decentralized_2020,
  author = {Shi, Chengshuai and Xiong, Wei and Shen, Cong and Yang, Jing},
  title = {Decentralized Multi-player Multi-armed Bandits with No Collision Information},
  booktitle = {Proc. 23rd Int. Conf. Artif. Intell. Statist. (AISTATS)},
  pages = {1519--1528},
  year = {2020},
  volume = {108},
}

@article{vanerven2014,
  author = {Tim van Erven and Peter Harremo\"es},
  title = {R\'enyi Divergence and {Kullback--Leibler} Divergence},
  journal = {IEEE Trans. Inf. Theory},
  year = {2014},
  volume = {60},
  number = {7},
  pages = {3797--3820},
  doi = {10.1109/TIT.2014.2320500},
}

@inproceedings{wang2021contention,
  author = {Wang, Dingyu},
  title = {Optimal Protocols for 2-Party Contention Resolution},
  booktitle = {Proc. 23rd Int. Symp. Stabilization, Safety, Security Distrib. Syst. (SSS)},
  editor = {Johnen, Colette and Schiller, Elad Michael and Schmid, Stefan},
  series = {Lecture Notes in Computer Science},
  volume = {13046},
  publisher = {Springer},
  address = {Cham, Switzerland},
  year = {2021},
  pages = {456--468},
  doi = {10.1007/978-3-030-91081-5_30},
}

@inproceedings{wang_optimal_2020,
  author = {Wang, Po-An and Proutiere, Alexandre and Ariu, Kaito and Jedra, Yassir and Russo, Alessio},
  title = {Optimal Algorithms for Multiplayer Multi-Armed Bandits},
  booktitle = {Proc. 23rd Int. Conf. Artif. Intell. Statist. (AISTATS)},
  pages = {4120--4129},
  year = {2020},
  volume = {108},
}

@article{willard1986,
  author = {Willard, Dan E.},
  title = {Log-Logarithmic Selection Resolution Protocols in a Multiple Access Channel},
  journal = {SIAM J. Comput.},
  year = {1986},
  volume = {15},
  number = {2},
  pages = {468--477},
  doi = {10.1137/0215032},
}

@inproceedings{zhang2025,
  author = {Zhang, Raymond and Combes, Richard and Yang, Sheng},
  title = {The Bandit Channel},
  booktitle = {Proc. 2025 IEEE Inf. Theory Workshop (ITW)},
  year = {2025},
  pages = {1--7},
}

@INPROCEEDINGS{zhang2026,
  author = {Zhang, Raymond and Miyamoto, Henrique and Combes, Richard and Yang, Sheng},
  booktitle = {Proc. 2026 IEEE Int. Symp. Inf. Theory (ISIT)},
  title = {An Efficient Synchronization Scheme for Distributed Bandits},
  year = {2026},
  volume = {},
  number = {},
  pages = {1-6},
  doi = {10.1109/ISIT62367.2026.11654092},
}

\end{document}